\documentclass[11pt]{article}
\usepackage[body={9.5in},left=1in,right=1in]{geometry}
\usepackage{amsmath,amssymb,mathrsfs,amsfonts,graphicx,amsthm,nicefrac,mathtools,color}
\usepackage{arydshln}
\usepackage{subcaption}
\usepackage{enumitem}
\usepackage{hyperref}
\usepackage{setspace}
\usepackage{lipsum}
\usepackage{titlesec}
\usepackage{relsize}
\usepackage{tabu}
\usepackage{bbm}
\usepackage{tikz}
\usepackage{booktabs}
\usepackage{array}
\usepackage{float}
\usepackage{natbib}
\usepackage{setspace}
\usepackage{graphicx}

\newcommand\mbb[1]{\mathbb{#1}}

\newcommand\abs[1]{\left|#1\right|}

\def\indic#1{\mbb{I}\left({#1}\right)} % Indicator function
\providecommand{\abs}{\mathop\mathrm{abs}}

\def\E{\mathbb{E}} % Expectation symbol
\def\Earg#1{\E\left[{#1}\right]}

\def\P{\mathbb{P}} % Probability symbol
\def\Parg#1{\P\left({#1}\right)}

\theoremstyle{defn}

\newtheorem{thm}{Theorem}

\newtheorem{prop}{Proposition}
\newtheorem{cor}{Corollary}

\newtheorem{assumption}{Assumption}

\hypersetup{
  colorlinks   = true, % colors links instead of boxes
  urlcolor     = black, % color for external hyperlinks
  linkcolor    = black, % color of internal links
  citecolor   = black % color of citations
}
\setcitestyle{open={(},close={)}}

\newcommand*{\thisdraft}{This draft: September 25, 2025}
\newcommand*{\firstdraft}{First draft: September 10, 2018} 

\usepackage[margin=10pt,labelfont=bf]{caption}

\newcommand\tcapfig[1]{\captionsetup{position=top, font=normalsize, labelfont=bf, textfont=normalfont, justification=centering, margin=0mm, aboveskip=2mm, belowskip=0mm, labelsep=colon, singlelinecheck=false}\caption{#1}}
\newcommand\bnotefig[1]{\captionsetup{position=bottom, font=footnotesize,  textfont=normalfont, margin=1mm, skip=2mm, justification=justified, singlelinecheck=false}\caption*{#1}}

\begin{document}
 	\title{Change-Point Testing for Risk Measures in Time Series\thanks{\scriptsize We thank seminar and conference participants at Stanford and the NBER-NSF Time-Series Conference for helpful comments.}}

\date{\thisdraft \\ \firstdraft}
%\date{\today}
				\author{Lin Fan\thanks{ \scriptsize Northwestern University, Kellogg School of Management, Email: \texttt{lin.fan@kellogg.northwestern.edu}.}
		\and
			Junting Duan\thanks{ \scriptsize Stanford University, Department of Management Science \& Engineering, Email: \texttt{duanjt@stanford.edu}.} 
               \and
              Peter Glynn\thanks{ \scriptsize Stanford University, Department of Management Science \& Engineering, Email: \texttt{glynn@stanford.edu}.}
                \and
		Markus Pelger\thanks{\scriptsize Stanford University, Department of Management Science \& Engineering, Email: \texttt{mpelger@stanford.edu}.}
	}
	
	\onehalfspacing

	\begin{titlepage}
		\maketitle
		\thispagestyle{empty}
		
		\begin{abstract}

%We investigate methods of change-point testing for nonparametric estimators of expected shortfall and related risk measures in weakly dependent time series. A key aspect of our work is the ability to detect general multiple structural changes in the tails of time series marginal distributions.
%Unlike extant approaches for detecting tail structural changes using quantities such as tail index, our approach does not require parametric modeling of the tail and detects more general changes in the tail. Additionally, our methods are based on the recently introduced self-normalization technique for time series, allowing for statistical analysis without the issues of consistent standard error estimation. The theoretical foundation for our methods are functional central limit theorems, which we develop under weak assumptions. An empirical study of S\&P 500 returns and US 30-Year Treasury bonds illustrates the practical use of our methods in detecting and quantifying market instability via the tails of financial time series during times of financial crisis.

We propose novel methods for change-point testing for nonparametric estimators of expected shortfall and related risk measures in weakly dependent time series. We can detect general multiple structural changes in the tails of marginal distributions of time series under general assumptions. Self-normalization allows us to avoid the issues of standard error estimation. The theoretical foundations for our methods are functional central limit theorems, which we develop under weak assumptions. An empirical study of S\&P 500 and US Treasury bond returns illustrates the practical use of our methods in detecting and quantifying instability in the tails of financial time series.

			\vspace{1cm}
			
				
		\end{abstract}
\end{titlepage}

\section{Introduction} \label{S1}

The quantification of risk is a central topic of study in finance. The need to guard against unforeseeable events with adverse and potentially catastrophic consequences has led to an extensive and burgeoning literature on risk measures. 
The two most popular financial risk measures, the value-at-risk (VaR) and the expected shortfall (ES), are extensively used for risk and portfolio management as well as regulation in the finance industry \citep{mcneil2015quantitative}.
Parallel to the developments in risk estimation, there has been longstanding interest in detecting changes in financial time series, especially changes in the tail structure, which is essential for effective risk and portfolio management.
Indeed, empirical findings strongly suggest that financial time series frequently exhibit changes in their underlying statistical structure due to changes in economic conditions, e.g. monetary policies, or critical social events.
Although there is an established literature on structural change detection for parametric time series models, including monitoring of proxies for risk such as tail index, there is little work concerning monitoring of general tail structure, or of risk measures in particular.
To underscore this point, the existing literature on risk measure estimation assumes stationarity of time series observations over a time period of interest, with the stationarity of the risk measure being key for the estimation to make sense.
However, it is a challenge to verify this assumption.

We provide tools to detect general and potentially multiple changes in the tail structure of time series, and in particular, tools for monitoring for changes in risk measures such as ES and related measures such as conditional tail moments (CTM) \citep{methni_etal2014} over time periods of interest.
Specifically, we develop retrospective change-point tests to detect changes in ES and related risk measures. 
Additionally, we offer new ways of constructing confidence intervals for these risk measures.
Our methods are applicable to a wide variety of time series models as they depend on functional central limit theorems for the risk measures, which we develop under weak assumptions.
As described below, our methods complement and extend the existing literature in several ways.

As mentioned previously, the literature lacks tools to monitor general tail structure or risk measures such as ES or CTM over time.
This deficiency appears to be two-fold.
(1) Although there are studies on VaR change-point testing, for example, \cite{qu2008}, often one is interested in characterizations of tail structure that are more informative than simple location measures.
Indeed, the introduction of ES as an alternative risk measure to VaR was, to a great extent, driven by the need to quantify tail structure, particularly, the expected magnitude of losses conditional on losses being in the tail.
Aligned with this goal, a popular measure of tail structure is the tail index, which describes tail thickness and governs distributional moments.
In tail index estimation, an extreme value theory approach is typically taken with the assumption of so-called regularly-varying Pareto-type tails. However, tail index estimation is very sensitive to the choice of which fraction of sample observations is classified to be ``in the tail''.\footnote{The Hill estimator \citep{hill1975} is widely used and requires the user to choose the fraction of sample observations deemed to be ``in the tail'' to use for estimation. 
However, generally, there appears to be no best way to select this fraction.
In the specific setting of change-point testing for tail index, recommendations for this fraction in the literature range from the top 20th percentile to the top 5th percentile of observations \citep{kim_etal2009,kim_etal2011,hoga2017}.
Such choice heavily influences the quality of tail index estimation and change detection and must be made on a case-by-case basis \citep{rocco2014}.
It is a delicate matter as choosing too small of a fraction results in high estimation variance and choosing too large of a fraction often results in high bias due to misspecification of where the tail begins.}
Moreover, the typical regularly-varying Pareto-type tail assumptions may not even be valid in some situations. 
And even if they are valid, the tail index is invariant to changes of the location and scale types, and thus these types of structural changes in the tail would remain undetected using tail index-based change-point tests.
(2) As mentioned before, all previous studies on risk measure estimation implicitly assume the risk measure is constant over some time period of interest---otherwise, risk measure estimation and confidence interval construction could behave erratically.
For instance, if there is a sudden change in VaR (at some level) in the middle of a time series, naively estimating VaR using the entire time series could result in a wrong estimate of VaR or ES.
Hence, given the importance of ES and related risk measures, a simple test for ES change over a time period of interest is a useful first step prior to follow-up statistical analysis.

To simultaneously address both deficiencies, we propose retrospective change-point testing for ES and related risk measures such as CTM.
We introduce, in particular, a consistent test for a potential ES change at an unknown time based on a variant of the widely used cumulative sum (CUSUM) process.\footnote{The cumulative sum (CUSUM) process was first introduced by \cite{page1954} and is discussed in detail in \cite{csorgo_etal2011}.}
We subsequently generalize this test to the case of multiple potential ES change points, leveraging recent work by \cite{zhang_etal2018}, and unlike existing change-point testing methodologies in the literature, our test does not require the number of potential change points to be specified in advance.
Our use of a risk measure such as ES is attractive in many ways.
First, the fraction of observations used in ES estimation, for example, the upper 5th percentile of observations, directly has meaning, and is often comparable to the fraction of observations used in tail index estimation, as discussed previously.
Second, the use of ES does not require parametric-type tail assumptions, which is not the case with use of the tail index.
Third, ES can detect much more general structural changes in the tail such as location or scale changes, while such changes go undetected when using tail index.
Moreover, our change-point tests can be used to check in a statistically principled way whether or not ES and related risk measures are constant over a time period of interest, and provide additional validity when applying existing estimation methods for these risk measures.

A key detail that has been largely ignored in previous studies is standard error estimation.
Here, the issue is two-fold.
(1) In the construction of confidence intervals for risk measures, consistent estimation of standard errors is nontrivial due to standard errors involving the entire time series correlation structure at all integer lags, and thus being infinite-dimensional in nature.
A few studies in which standard error estimation has been addressed (or bypassed) are \cite{chen_etal2005,chen2008,wang_etal2008,xu2016}.
However, confidence interval construction in these studies all require user-specified tuning parameters such as the bandwidth in periodogram kernel smoothing or window width in the blockwise bootstrap and empirical likelihood methods.
Although the choice of these tuning parameters significantly influences the quality of the resulting confidence intervals, it is not always clear how to best to select them.
(2) 
Choosing the tuning parameters is not only difficult, but data-driven approaches may result in non-monotonic test power, as pointed out in numerous studies \citep{vogelsang1999,crainiceanu_etal2007,deng_etal2008,shao_etal2010}.\footnote{For general change-point tests with time series observations, consistent estimation of standard errors is typically done using periodogram kernel smoothing, and the performance of such tests is heavily influenced by the choice of kernel bandwidth.} 

We offer the following solutions to the above issues. (1) To address the issue of often problematic selection of tuning parameters for confidence interval construction with time series data, we make use of ratio statistics to cancel out unknown standard errors and form pivotal quantities, thereby avoiding the often difficult estimation of such nuisance parameters. We examine confidence interval construction using a technique originally referred to in the simulation literature as sectioning \citep{asmussen_etal2007}, which involves splitting the data into equal-size non-overlapping sections, separately evaluating the estimator of interest using the data in each section, and relying on a normal approximation to form an asymptotically pivotal t-statistic.
We also examine its generalization, referred to in the simulation literature as standardized time series \citep{schruben1983,glynn_etal1990} and in the time series literature as self-normalization \citep{lobato2001,shao2010}, which uses functionals different from the t-statistic to create asymptotically pivotal quantities.
(2) In the context of change-point testing using ES and related risk measures, to avoid potentially troublesome standard error estimation, we follow \cite{shao_etal2010} and \cite{zhang_etal2018} and apply the method of self-normalization for change-point testing by dividing CUSUM-type processes by corresponding processes designed to cancel out the unknown standard error.
The processes we divide by take into account potential change point(s) and thus avoids the problem of non-monotonic power which often plagues change-point tests that rely on consistent standard error estimation, as discussed previously.

The outline of our paper is as follows. We describe the setting and assumptions in Section \ref{S2p1}. In Section \ref{S2p2}, we develop the asymptotic theory for VaR and ES, specifically, functional central limit theorems, which provide the theoretical basis for the proposed confidence interval construction and change-point testing methodologies.
In introducing our statistical methods, we first discuss the simpler task of confidence interval construction for risk measures in Section \ref{S3p1}.
Then, with several fundamental ideas in place, we take up testing for a single potential change point in VaR and ES in Section \ref{S3p2}.
We extend the change-point testing methodology to an unknown, possibly multiple, number of change points in Section \ref{S3p3}.
In Section \ref{S4}, we demonstrate the good finite-sample performance of our proposed methods through simulations.
We conclude with an empirical study in Section \ref{S5} of returns data for the S\&P 500 index and US Treasury bond returns. The proofs of our theoretical results as well as additional simulation and empirical results can be found in the Appendix.

% END OF INTRODUCTION

\section{Asymptotic Theory} \label{S2}
\subsection{Model Setup} \label{S2p1}
Suppose the random variable $X$ and the stationary sequence of random variables $X_1, \dots, X_n$ have the marginal distribution function $F$.
For some level $p \in (0,1)$, we want to estimate the risk measures VaR (value-at-risk) and ES (expected shortfall) defined by:\footnote{It is reasonable to define VaR and ES as upper tail characteristics when considering a loss process.} 
\begin{eqnarray*}
& VaR = \inf \{ x \in \mathbb{R} : F(x) \ge p\} \\
& ES = \Earg{X \mid X \ge VaR}.
\end{eqnarray*}
Let $\widehat{F}_n(\cdot) = n^{-1} \sum_{i=1}^n \indic{X_i \le \cdot}$ be the sample distribution function.
We consider the following plug-in sample-based nonparametric estimators:
\begin{eqnarray} \label{E1}
& \widehat{VaR}_n = \inf \{x \in \mathbb{R} : \widehat{F}_n(x) \ge p\} \\
\label{E2}
& \widehat{ES}_n = \frac{1}{1-p} \frac{1}{n} \sum_{i=1}^n X_i \indic{X_i \ge \widehat{VaR}_n}.
\end{eqnarray}
For the change-point testing, we estimate these risk measures on subsets of the data. Hence, for $m > l$, we will also consider the following nonparametric estimators based on samples $X_l,\dots, X_m$, with $\widehat{F}_{l:m}(\cdot) = (m-l+1)^{-1} \sum_{i=l}^m \indic{X_i \le \cdot}$:
\begin{eqnarray}
\label{E8}
& \widehat{VaR}_{l:m} = \inf \{x \in \mathbb{R} : \widehat{F}_{l:m}(x) \ge p\} \\
\label{E9}
& \widehat{ES}_{l:m} = \frac{1}{1-p} \frac{1}{m-l+1} \sum_{i=l}^m X_i \indic{X_i \ge \widehat{VaR}_{l:m}}.
\end{eqnarray}
We derive the asymptotic theory under general assumptions for the stochastic process. Let $\mathcal{F}_l^m$ denote the $\sigma$-algebra generated by $X_l, \dots, X_m$, and let $\mathcal{F}_l^\infty$ denote the $\sigma$-algebra generated by $X_l, X_{l+1}, \dots$.
The $\alpha$-mixing coefficient, as introduced by \cite{rosenblatt1956}, is defined as
$$\alpha(k) = \sup_{A \in \mathcal{F}_1^j, \, B \in \mathcal{F}_{j+k}^\infty, \, j \ge 1} \abs{\Parg{A}\Parg{B} - \Parg{A,B}},$$
and a sequence is said to be $\alpha$-mixing if $\lim_{k \to \infty} \alpha(k) = 0$.
The dependence described by $\alpha$-mixing is the least restrictive as it is implied by the other types of mixing; see \cite{doukhan1994} for a comprehensive discussion.
In what follows, $D[0,1]$ denotes the space of real-valued functions on $[0,1]$ that are right-continuous and have left limits, and convergence in distribution on this space is defined with respect to the Skorohod topology \citep{billingsley1999}.
Also, for some index set $\mathcal{I}$, $\ell^\infty(\mathcal{I})$ denotes the space of real-valued bounded functions on $\mathcal{I}$, and convergence in distribution on this space is defined with respect to the uniform topology.
We denote the integer part of a real number $x$ by $[x]$ and the positive part by $[x]_+$.
A standard Brownian motion on the real line is denoted by $W$.
Our theoretical results are developed using the following assumption.
%\begin{assumption} \label{A1}
%There exists $a > 1$ such that the following hold.
%~\begin{enumerate}
%\item[(i)] $X_1, X_2, \dots$ is $\alpha$-mixing with %$\alpha(k) = O(k^{-a})$
%\item[(ii)] $X$ has positive and continuous density $f$ in a neighborhood of $VaR$, and for each $k \ge 2$, $(X_1,X_k)$ has joint density in a neighborhood of $(VaR,VaR)$.
%\end{enumerate}
%\end{assumption}
%\begin{assumption} \label{A2}
%There exists $\delta > 0$ such that Assumption \ref{A1} holds with $a > (2+\delta)/\delta$ along with $\Earg{\abs{X}^{2+\delta}} < \infty$.
%\end{assumption}
\begin{assumption} \label{A3}
For some constants $\gamma > 0$ and $a > \max(3,(2+\gamma)/\gamma)$:
%For some $\epsilon > 0$:
\begin{enumerate}
\item[(i)] $X_1, X_2, \dots$ is $\alpha$-mixing with $\alpha(k) = O(k^{-a})$
\item[(ii)] $\Earg{\abs{X}^{2+\gamma}} < \infty$
\item[(iii)] $X$ has a positive and differentiable density $f$ in a neighborhood of $VaR$, and for each $k \ge 2$, $(X_1,X_k)$ has joint density in a neighborhood of $(VaR,VaR)$.
\end{enumerate}
\end{assumption}
Assumption \ref{A3} condition (i) is a form of asymptotic independence, which ensures that the time series is not too serially dependent so that non-degenerate limit distributions are possible.
A very wide range of commonly used financial time series models such as GARCH models and diffusion models satisfy this condition.
Condition (ii) indicates a trade-off between the strength of the moment condition of the underlying marginal distribution and the rates of $\alpha$-mixing, with weaker $\alpha$-mixing conditions requiring stronger moment conditions and vice versa.
%The moment condition (ii) accompanies 
Condition (iii) is a standard assumption in VaR and ES estimation, and rules out pathological situations in which there are many ties among the observations near $VaR$.

We point out, in particular, that while our results are illustrated for the most widely used risk measures, VaR and ES, they can be adapted to many other important functionals of the underlying marginal distribution.
One straightforward adaptation (by assuming stronger $\alpha$-mixing and moment conditions) is to CTM: $\Earg{X^\beta \mid X > VaR}$ for some level $p \in (0,1)$ and some $\beta > 0$ \citep{methni_etal2014}.
Our results also easily extend to multivariate time series, but for simplicity of illustration, we focus on univariate time series.

\subsection{Functional Central Limit Theorems} \label{S2p2}

We develop functional central limit theorems jointly for VaR and ES under weak assumptions. These functional central limit theorems allow the construction of general change point statistics.

% OLD VERSION OF THEOREM 1
%\begin{thm} \label{thm1}
%Under Assumption \ref{A1} with the modifications: $a \ge 3$ and $X$ has positive and differentiable %density at $VaR(p)$, the process
%\begin{align} \label{E3}
%\{\sqrt{n}t(\widehat{VaR}_{[nt]}(p) - VaR(p)), \: t \in [0,1]\}
%\end{align}
%converges in distribution in $D[0,1]$ to $\sigma_{VaR} W$, where $$\sigma_{VaR}^2 = %\frac{1}{f^2(VaR(p))}\left(\Earg{g_p^2(X_1)} + 2 \sum_{i=2}^\infty \Earg{g_p(X_1)g_p(X_i)}\right)$$ %with $g_p(X) = \indic{X \le VaR(p)} - p$. 
%Under Assumption \ref{A2}, the process
%\begin{align} \label{E4}
%\left\{\sqrt{n}t(\widehat{ES}_{[nt]}(p) - ES(p)), \: t \in [0,1]\right\}
%\end{align}
%converges in distribution to $\sigma_{ES} W$ in $D[0,1]$, where $$\sigma_{ES}^2 = %\frac{1}{(1-p)^2}\left(\Earg{h_p^2(X_1)} + 2 \sum_{i=2}^\infty \Earg{h_p(X_1)h_p(X_i)}\right)$$ with %$h_p(X) = \max(X, VaR(p)) - \Earg{\max(X, VaR(p))}$.
%\end{thm}

\begin{thm} \label{thm1}
Under Assumption \ref{A3}, the process
\begin{align} \label{E3}
\left\{ \sqrt{n} t\begin{bmatrix}\widehat{VaR}_{[nt]} - VaR \\ \widehat{ES}_{[nt]} - ES \end{bmatrix} : t \in [0,1]\right\}
\end{align}
converges in distribution in $D[0,1] \times D[0,1]$ to $\Sigma^{1/2} (W_1,W_2)$, where $W_1$ and $W_2$ are independent standard Brownian motions, and $\Sigma \in \mathbb{R}^{2 \times 2}$ is symmetric positive-semidefinite with components
\begin{align}
& \Sigma_{11} = \frac{1}{f^2(VaR)}\left(\Earg{g^2(X_1)} + 2 \sum_{i=2}^\infty \Earg{g(X_1)g(X_i)}\right) \label{sigma_11} \\
& \Sigma_{12} = \frac{1}{f(VaR)(1-p)}\left(\Earg{g(X_1)h(X_1)} + \sum_{i=2}^\infty \Bigl(\Earg{g(X_1)h(X_i)} + \Earg{h(X_1)g(X_i)}\Bigr)\right) \label{sigma_12} \\
& \Sigma_{22} = \frac{1}{(1-p)^2}\left(\Earg{h^2(X_1)} + 2 \sum_{i=2}^\infty \Earg{h(X_1)h(X_i)}\right), \label{sigma_22}
\end{align}
where $g(X) = \indic{X \le VaR} - p$ and $h(X) = [X-VaR]_+ - \Earg{[X-VaR]_+}$.
\end{thm}
%From this, we immediately have the following central limit theorems for VaR and ES.
%\begin{cor} \label{cor2}
%Under Assumption \ref{A3},
%$$\sqrt{n}\begin{bmatrix}\widehat{VaR}_n - VaR \\ \widehat{ES}_n - ES \end{bmatrix} \overset{d}{\to} \mathcal{N}(0,\Sigma),$$
%where $\Sigma$ is the same as in Theorem \ref{thm1}.
%\end{cor}
Our Assumption $\ref{A3}$ conditions (i)-(ii) are weaker than those of existing central limit theorems in the literature (c.f. \cite{chen_etal2005,chen2008}), which require $\alpha$-mixing coefficients to decay exponentially fast, as well as additional regularity of the marginal and pairwise joint densities of the observations.

%The above functional central limit theorem for VaR is a modification of a result due to \cite{sen1972}.
%The condition $a \ge 3$ for the VaR functional central limit theorem is due to our use of a so-called Bahadur representation for $\widehat{VaR}_n$ by \cite{wendler2011}.
%Although such a condition can likely be weakened, we do not pursue that here.
%Moreover, the above functional central limit theorem for ES does not require such a condition; Assumption \ref{A2} is all that is needed.
In deriving the ES functional central limit theorem, we used the following Bahadur representation for $\widehat{ES}_n$.
\begin{prop} \label{prop1}
%Suppose $X_1,X_2,\dots$ is $\alpha$-mixing with $\alpha(k) = O(k^{-a})$ for some $a > (2+\gamma)/\gamma$, $\Earg{\abs{X}^{2+\gamma}} < \infty$ for some $\gamma > 0$, and Assumption \ref{A3} (iii) holds. 
Suppose Assumption \ref{A3} holds for some $\gamma > 0$ and $a > (2+\gamma)/\gamma$.
Then, for any $\gamma' > 0$ satisfying $-1/2 + 1/(2a) + \gamma' < 0$, 
$$\widehat{ES}_n - \biggl(VaR + \frac{1}{1-p} \frac{1}{n} \sum_{i=1}^n [X_i - VaR]_+ \biggr) = o_{a.s.}(n^{-1 + 1/(2a) + \gamma'} \log n).$$
\end{prop}
\cite{sun_etal2010} developed such a Bahadur representation in the setting of independent, identically-distributed data, but to the best of our knowledge, no such representation exists in the stationary, $\alpha$-mixing setting.
Such a Bahadur representation is generally useful for developing limit theorems in many different contexts.

We also have the following extension of the functional central limit theorems for VaR and ES in Theorem \ref{thm1}, where now there are two ``time'' indices instead of just one.
The standard functional central limit theorems are useful for detecting a single change point in a time series, but the following extension will allow us to detect an unknown, possibly multiple, number of change points, as we will discuss later.
We point out that this result does not follow automatically from Theorem \ref{thm1} and an application of the continuous mapping theorem, because the estimators in (\ref{E8})-(\ref{E9}) are not additive, for instance, for $m > l$, $\widehat{ES}_{l:m} \ne \widehat{ES}_{1:m} - \widehat{ES}_{1:l-1}$.

\begin{thm} \label{thm2}
Fix any $\delta > 0$ and consider the index set $\Delta = \{ (s,t) \in [0,1]^2 : t-s \ge \delta \}$.
Under Assumption \ref{A3}, with the modification that $(X_1, X_k)$ has a joint density for all $k \ge 2$, the process
\begin{align} \label{E14}
\left\{ \sqrt{n} (t-s)\begin{bmatrix}\widehat{VaR}_{[ns]:[nt]} - VaR \\ \widehat{ES}_{[ns]:[nt]} - ES \end{bmatrix} : (s,t) \in \Delta \right\}
\end{align}
converges in distribution in $\ell^\infty(\Delta) \times \ell^\infty(\Delta)$ to the process
\begin{align}
\left\{\Sigma^{1/2} \begin{bmatrix} W_1(t) - W_1(s) \\ W_2(t) - W_2(s) \end{bmatrix} : (s,t) \in \Delta \right\}, \label{bivariate_limit}
\end{align}
where $\Sigma$ is from Theorem \ref{thm1}, and $W_1$ and $W_2$ are independent standard Brownian motions.
\end{thm}

\section{Statistical Methods}

\subsection{Confidence Intervals} \label{S3p1}
In time series analysis, confidence interval construction for an unknown quantity is often difficult, due to dependence.
Indeed, from Theorem \ref{thm1}, we see that the standard errors appearing in the normal limiting distributions depend on the time series autocovariance at all integer lags.
To construct confidence intervals using Theorem \ref{thm1}, these standard errors must be estimated.
One approach, taken in \cite{chen_etal2005,chen2008}, is to estimate using kernel smoothing the spectral density at zero frequency of the transformed time series $g(X_1),g(X_2),\dots$ and $h(X_1),h(X_2),\dots$, where $g$ and $h$ are from Theorem \ref{thm1}.
Although it is known that under certain moment and correlation assumptions, spectral density estimators are consistent for stationary processes \citep{brockwell_etal1991,anderson1971}, in practice it is nontrivial to obtain robust and precise estimates due to the need to select tuning parameters for the kernel smoothing-based approach.
As for other approaches, under certain conditions, resampling methods such as the moving block bootstrap \citep{kunsch1989,liu_etal1992} and the subsampling method for time series \citep{politis_etal1999} bypass direct standard error estimation and yield confidence intervals that asymptotically have the correct coverage probability.
However, these approaches also require user-chosen tuning parameters such as the block length in the moving block bootstrap and the window width in subsampling.
We introduce an alternative way to construct confidence intervals with asymptotically correct coverage, where the quality of the resulting confidence intervals is less sensitive to the choice of tuning parameters and is therefore more robust.

We first examine a technique called sectioning from the simulation literature \citep{asmussen_etal2007}, which can be used to construct confidence intervals for general estimators.
%Although the method can be used to construct confidence intervals for general risk measures, in light of our results from Section \ref{S2}, we apply the method to VaR and ES in particular.
The method is as follows.
Let $Y_1(\cdot), Y_2(\cdot), \dots$ be a sequence of random bounded real-valued functions on $[0,1]$.
For some user-specified integer $m \ge 2$, suppose we have the joint convergence in distribution:
\begin{align} \label{E5}
\biggl(Y_n(1/m) - Y_n(0), Y_n(2/m) - Y_n(1/m), \dots, Y_n(1) - Y_n((m-1)/m) \biggr) \overset{d}{\to} \frac{\sigma}{m^{1/2}} \biggl( \mathcal{N}_1, \mathcal{N}_2, \dots, \mathcal{N}_m \biggr),
\end{align}
where $\sigma > 0$ and $\mathcal{N}_1, \dots, \mathcal{N}_m$ are independent standard normal random variables.
Taking $Y_n(\cdot)$ to be the processes in (\ref{E3}) for VaR or ES, this is guaranteed by Theorem \ref{thm1}.
Then, with
\begin{align*}
 \overline{Y}_n &= m^{-1} \sum_{i=1}^m (Y_n(i/m) - Y_n((i-1)/m)) \\
 S_n &= \left((m-1)^{-1} \sum_{i=1}^m (Y_n(i/m) - Y_n((i-1)/m) - \overline{Y}_n)^2 \right)^{1/2},
\end{align*}
by the Continuous Mapping Theorem, as $n \to \infty$ with $m$ fixed, $m^{1/2}\overline{Y}_n/S_n$ converges in distribution to the Student's $t$-distribution with $m-1$ degrees of freedom.
So using the $t$-distribution critical values, we can obtain confidence intervals for VaR and ES with asymptotically correct coverage. 
Note that this method requires the number $n/m$ to be sufficiently large for the asymptotic distribution in (\ref{E5}) to hold. Following \cite{asmussen_etal2007}, we recommend selecting the integer $m$ within the range of $[2,10]$ in practice. In Section \ref{S4}, we demonstrate through simulations that the sectioning method is very robust to choices of $m$ within this range.
%With the limiting Student's $t$-distribution, we may construct confidence intervals for VaR or ES by taking the random functions $Y_n(\cdot)$ to be the processes in (\ref{E3}).
%The distributional convergence result in (\ref{E5}) is easily obtained by applying Theorem \ref{thm1}.

Next, we examine a generalization of sectioning, called self-normalization.
This technique has been studied recently in the time series literature \citep{lobato2001,shao2010}, as well as earlier in the simulation literature, where it is known as standardization \citep{schruben1983,glynn_etal1990}.
As with sectioning, the method applies to confidence interval construction for general estimators.
With self-normalization, the idea is to use a ratio-type statistic where the unknown standard error appears in both the numerator and the denominator and thus cancels, resulting in a pivotal limiting distribution.
As before, let $Y_1(\cdot), Y_2(\cdot), \dots$ be a sequence of random bounded real-valued functions on $[0,1]$.
Suppose we have the distributional convergence in $D[0,1]$: $Y_n(\cdot) \overset{d}{\to} \sigma W(\cdot)$, where $\sigma > 0$.
Again, taking $Y_n(\cdot)$ to be the processes in (\ref{E3}) for VaR or ES, this distributional convergence is guaranteed by Theorem \ref{thm1}.
Consider some positive homogeneous functional $T : D[0,1] \mapsto \mathbb{R}$, i.e., satisfying $T(\sigma Y) = \sigma T(Y)$ for $\sigma > 0$ and $Y \in D[0,1]$. 
If the Continuous Mapping Theorem can applied, then
\begin{eqnarray} \label{E6}
\frac{Y_n(1)}{T(Y_n)} \overset{d}{\to} \frac{W(1)}{T(W)}.
\end{eqnarray}
The right side of (\ref{E6}) is a pivotal limiting distribution.
So using its critical values, which may be computed via simulation, we can obtain confidence intervals for VaR and ES with asymptotically correct coverage.
%The distributional convergence result in (\ref{E6}) follows directly from Theorem \ref{thm1}.
%As an example, taking the functional to be $T(Y) = \left(\int_0^1(Y(t) - tY(1))^2 dt\right)^{1/2}$, we obtain the following result for ES:
%\begin{eqnarray} \label{E7}
%\frac{\widehat{ES}_n(p) - ES(p)}{\left(\int_0^1 t^2 \left(\widehat{ES}_{[nt]}(p) - \widehat{ES}_n(p)\right)^2 dt\right)^{1/2}} \overset{d}{\to} \frac{W(1)}{(\int_0^1 (W(t) - t W(1))^2 dt)^{1/2}}.
%\end{eqnarray}

The choice of form for the functional $T$ is up to the user.
One convenient choice is $T(Y) = \left(\int_0^1(Y(t) - tY(1))^2 dt\right)^{1/2}$, using which we obtain the following result for ES:
\begin{eqnarray} \label{E7}
\frac{\widehat{ES}_n(p) - ES(p)}{\left(\int_0^1 t^2 \left(\widehat{ES}_{[nt]}(p) - \widehat{ES}_n(p)\right)^2 dt\right)^{1/2}} \overset{d}{\to} \frac{W(1)}{(\int_0^1 (W(t) - t W(1))^2 dt)^{1/2}}.
\end{eqnarray}
This form for $T$, as advocated for by \cite{lobato2001}, \cite{shao2010} and \cite{shao_etal2010} among others, ensures that the numerator and denominator of the right side of (\ref{E7}) are quantities based on independent normal random variables, which makes quantiles for their ratio particularly easy to compute.

\subsection{Testing for a Single Change Point} \label{S3p2}
As is the case with confidence interval construction with time series data, change-point testing in time series based on statistics constructed from functional central limit theorems and the continuous mapping theorem is often nontrivial due to the need to estimate standard errors.
Motivated by the maximum likelihood method in the parametric setting, variants of the \cite{page1954} CUSUM statistic are commonly used for nonparametric change-point tests \citep{csorgo_etal2011}, and generally rely on asymptotic approximations (via functional central limit theorems and the continuous mapping theorem) to supply critical values of pivotal limiting distributions under the null hypothesis of no change.
As discussed in \cite{vogelsang1999,shao_etal2010,zhang_etal2018}, testing procedures where standard errors are estimated directly, for example, by estimating the spectral density of transformed time series via a kernel-smoothing approach, can be biased under the change-point alternative. Such bias can result in nonmonotonic power, i.e., power can decrease in some ranges as the alternative deviates from the null.
To avoid this issue, \cite{shao_etal2010} and \cite{zhang_etal2018} propose using self-normalization techniques to general change-point testing. We adopt this idea to our specific problem of detecting changes in tail risk measures.

As motivated in the Introduction (Section \ref{S1}), it is important to perform hypothesis tests for abrupt changes of risk measures in the time series setting. 
We introduce the methodology for joint testing of VaR and ES.
In this section, we consider the case of at most one change point.
We consider the case of an unknown number of change points in the next section, using the approach in \cite{zhang_etal2018}.
For a time series $X_1, \dots, X_n$, let $(VaR_{X_i}, ES_{X_i})$ be the VaR and ES at a fixed level $p$ for the marginal distribution of $X_i$.
We test the following null and alternative hypotheses.
\begin{align*}
& \mathcal{H}_0: X_1, \dots, X_n \text{ is stationary, and in particular, } \begin{bmatrix} VaR_{X_1} \\ ES_{X_1} \end{bmatrix} = \dots = \begin{bmatrix} VaR_{X_n} \\ ES_{X_n} \end{bmatrix}. \\
& \mathcal{H}_1: \text{There is } t^* \in (0,1) \text{ such that } \\
& \qquad  \begin{bmatrix} VaR_{X_1} \\ ES_{X1} \end{bmatrix} = \dots = \begin{bmatrix} VaR_{X_{[nt^*]}} \\ ES_{X_{[nt^*]}} \end{bmatrix}  \ne \begin{bmatrix} VaR_{X_{[nt^*]+1}} \\ ES_{X_{[nt^*]+1}} \end{bmatrix} = \dots = \begin{bmatrix} VaR_{X_n} \\ ES_{X_n} \end{bmatrix}, \\
& \qquad  \text{ and } X_1, \dots, X_{[nt^*]} \text{ and } X_{[nt^*]+1}, \dots, X_n \text{ are separately stationary.}
%\begin{bmatrix} VaR_{X_1} \\ ES_{X1} \end{bmatrix} = \dots = \begin{bmatrix} VaR_{X_{[nt^*]}} \\ ES_{X_{[nt^*]}} \end{bmatrix} \\
%& \qquad \qquad \qquad \qquad \qquad \qquad \qquad \quad \ne \begin{bmatrix} VaR_{X_{[nt^*]+1}} \\ ES_{X_{[nt^*]+1}} \end{bmatrix} = \dots = \begin{bmatrix} VaR_{X_n} \\ ES_{X_n} \end{bmatrix}, \\
%& \qquad \text{ and } X_1, \dots, X_{[nt^*]} \text{ and } X_{[nt^*]+1}, \dots, X_n \text{ are separately stationary.}
\end{align*}
We base our change-point test on the following variant of the CUSUM process:
\begin{align}
\label{E10}
\left\{\sqrt{n} t(1-t) \begin{bmatrix} \widehat{VaR}_{1:[nt]} - \widehat{VaR}_{[nt]+1:n} \\ \widehat{ES}_{1:[nt]} - \widehat{ES}_{[nt]+1:n} \end{bmatrix} : t \in [0,1] \right\}
\end{align}
Note that we split the above process for all possible break points $t \in (0,1)$ into a difference of two estimators, an estimator using $X_1, \dots, X_{[nt]}$ and an estimator using $X_{[nt]+1}, \dots, X_n$.
Moreover, splitting the process as in (\ref{E10}) avoids potential VaR estimation using a sequence containing the change point, which could have undesirable behavior.
In Proposition \ref{prop2} below, we self-normalize the process in (\ref{E9}) using the approach of \cite{shao_etal2010}.
Note that the self-normalizer of (\ref{E11}) below takes into account the potential change point and is split into two separate integrals involving $X_1, \dots, X_{[nt]}$ and $X_{[nt]+1}, \dots, X_n$.
\begin{prop} \label{prop2}
Suppose Assumption \ref{A3} holds. Under the null hypothesis $\mathcal{H}_0$,
\small
\begin{align}
G_n := \sup_{t \in [0,1]} C_n(t)^T D_n(t)^{-1} C_n(t), \label{E11}
\end{align}
with
\begin{align*}
& C_n(t) := t(1-t) \begin{bmatrix} \widehat{VaR}_{1:[nt]} - \widehat{VaR}_{[nt]+1:n} \\ \widehat{ES}_{1:[nt]} - \widehat{ES}_{[nt]+1:n} \end{bmatrix} \\
& D_n(t) := n^{-1}\sum_{i=1}^{[nt]} \left(\frac{i}{n}\right)^2 \begin{bmatrix} \widehat{VaR}_{1:i} - \widehat{VaR}_{1:[nt]} \\ \widehat{ES}_{1:i} - \widehat{ES}_{1:[nt]} \end{bmatrix}^{\otimes 2} + n^{-1}\sum_{i=[nt]+1}^n \left(\frac{n-i+1}{n}\right)^2 \begin{bmatrix} \widehat{VaR}_{i:n} - \widehat{VaR}_{[nt]+1:n} \\ \widehat{ES}_{i:n} - \widehat{ES}_{[nt]+1:n} \end{bmatrix}^{\otimes 2},
\end{align*}
\normalsize
converges in distribution to
\small
\begin{align}
G := \sup_{t \in [0,1]} C(t)^T D(t)^{-1} C(t), \label{E12}
\end{align}
with
\begin{align*}
& C(t) := \begin{bmatrix} W_1(t) - tW_1(1) \\ W_2(t) - tW_2(1) \end{bmatrix} \\
& D(t) := \int_0^t \begin{bmatrix} W_1(s) - \frac{s}{t}W_1(t) \\ W_2(s) - \frac{s}{t}W_2(t) \end{bmatrix}^{\otimes 2} ds + \int_t^1 \begin{bmatrix} W_1(1) - W_1(s) - \frac{1-s}{1-t}(W_1(1) - W_1(t)) \\ W_2(1) - W_2(s) - \frac{1-s}{1-t}(W_2(1) - W_2(t)) \end{bmatrix}^{\otimes 2} ds.
\end{align*}
Assume the alternative hypothesis $\mathcal{H}_1$ is true with the change point occurring at some fixed (but unknown) $t^* \in (0,1)$.
For any fixed difference $VaR_{X_{[nt^*]}} = c_1 \ne c_2 = VaR_{X_{[nt^*]+1}}$ or $ES_{X_{[nt^*]}} = d_1 \ne d_2 = ES_{X_{[nt^*]+1}}$, we have $G_n \overset{P}{\to} \infty$ as $n \to \infty$.
Furthermore, if the difference varies with $n$ according to $c_1 - c_2 = n^{-1/2 + \epsilon} L$ or $d_1 - d_2 = n^{-1/2 + \epsilon} L$ for some $L \ne 0$ and $\epsilon \in (0,1/2)$, then
$G_n \overset{P}{\to} \infty$ as $n \to \infty$.

\normalsize
\end{prop}
The distribution of $G$ is pivotal, and its critical values may be obtained via simulation.
For testing $\mathcal{H}_0$ versus $\mathcal{H}_1$ at some level, we reject $\mathcal{H}_0$ if the test statistic $G_n$ exceeds some corresponding critical value of $G$.
To obtain critical values, we simulate 5,000 replications, with each replication consisting of 2,000 independent standard normal random variables to approximate standard Brownian motions on $[0,1]$.

In subsequent discussions concerning (\ref{E11}), we will refer to the process appearing in the numerator as the ``CUSUM process'', the process appearing in the denominator as the ``self-normalizer process'', and the entire ratio process as the ``self-normalized CUSUM process''.\footnote{To theoretically evaluate the efficiency of statistical tests, an analysis based on sequences of so-called local limiting alternatives (for example, sequences $c_1 - c_2 = O(n^{-1/2})$ in Proposition \ref{prop2} above) can be considered (see, for example, \cite{vandervaart1998}).
However, such an analysis would be considerably involved, and we leave it for future study.}

\subsection{Extension to Multiple Change Points} \label{S3p3}
We extend our single change-point testing methodology to the case of multiple change points.
Typically, the number of potential change points in the alternative hypothesis must be prespecified.
However, we leverage the recent work of \cite{zhang_etal2018} and introduce joint change-point tests for VaR and ES, that can accommodate an unknown, possibly multiple, number of change points in the alternative hypothesis.
%For illustration, we introduce the method using ES for univariate time series, but the method extends easily to related risk measures and also to multivariate time series.
We fix some small $\delta > 0$ and consider the following null and alternative hypotheses (following the notation from Section \ref{S3p2}).
\begin{align*}
& \mathcal{H}_0 : X_1, \dots, X_n \text{ is stationary, and in particular, } \begin{bmatrix} VaR_{X_1} \\ ES_{X_1} \end{bmatrix} = \dots = \begin{bmatrix} VaR_{X_n} \\ ES_{X_n} \end{bmatrix}. \\
& \mathcal{H}_1 : \text{There are } 0 = t_0^* < t_1^* < \dots < t_k^* < t_{k+1}^* = 1 \text{ with } t_{j}^* - t_{j-1}^* > \delta \text{ for } j=1,\dots,k+1 \\
& \qquad \text{ such that } \begin{bmatrix} VaR_{X_{[nt_j^*]}} \\ ES_{X_{[nt_j^*]}} \end{bmatrix} \ne \begin{bmatrix} VaR_{X_{[nt_j^*]+1}} \\ ES_{X_{[nt_j^*]+1}} \end{bmatrix}, \text{ and } X_{[nt_j^*]+1}, \dots, X_{[nt_{j+1}^*]} \text{ are separately stationary} \\
& \qquad \text{ for } j=0,\dots,k.
\end{align*}
Consider the index set $\Delta = \{ (s,t) \in [\delta,1-\delta]^2 : t-s \ge \delta \}$ and the test statistic
\begin{align*}
H_n & = \sup_{(s_1,s_2) \in \Delta} E_n^f(s_1,s_2)^T F_n^f(s_1,s_2)^{-1} E_n^f(s_1,s_2) + \sup_{(t_1,t_2) \in \Delta} E_n^b(t_1,t_2)^T F_n^b(t_1,t_2)^{-1} E_n^b(t_1,t_2),
\end{align*}
where
\begin{align*}
 E_n^f(s_1,s_2) =& \frac{[ns_1]([ns_2] - [ns_1])}{[ns_2]^{3/2}} \begin{bmatrix} \widehat{VaR}_{1:[ns_1]} - \widehat{VaR}_{[ns_1]+1:[ns_2]} \\ \widehat{ES}_{1:[ns_1]} - \widehat{ES}_{[ns_1]+1:[ns_2]} \end{bmatrix} \\
 F_n^f(s_1,s_2) =& \sum_{i=1}^{[ns_1]} \frac{i^2 ([ns_1] - i)^2}{[ns_2]^2 [ns_1]^2} \begin{bmatrix} \widehat{VaR}_{1:i} - \widehat{VaR}_{i+1:[ns_1]} \\ \widehat{ES}_{1:i} - \widehat{ES}_{i+1:[ns_1]} \end{bmatrix}^{\otimes 2} \\
  &+ \sum_{i=[ns_1]+1}^{[ns_2]} \frac{(i - 1 - [ns_1])^2 ([ns_2] - i + 1)^2}{[ns_2]^2 ([ns_2] - [ns_1])^2} \begin{bmatrix} \widehat{VaR}_{[ns_1]+1:i-1} - \widehat{VaR}_{i:[ns_2]} \\ \widehat{ES}_{[ns_1]+1:i-1} - \widehat{ES}_{i:[ns_2]} \end{bmatrix}^{\otimes 2} \\
%\end{eqnarray*}
%\begin{eqnarray*}
 E_n^b(t_1,t_2) =& \frac{([nt_2] - [nt_1])(n-[nt_2]+1)}{(n-[nt_1]+1)^{3/2}} \begin{bmatrix} \widehat{VaR}_{[nt_2]:n} - \widehat{VaR}_{[nt_1]:[nt_2]-1} \\ \widehat{ES}_{[nt_2]:n} - \widehat{ES}_{[nt_1]:[nt_2]-1} \end{bmatrix} \\
 F_n^b(t_1,t_2) =& \sum_{i=[nt_1]}^{[nt_2]-1} \frac{(i-[nt_1]+1)^2 ([nt_2]-1-i)^2}{(n-[nt_1]+1)^2([nt_2]-[nt_1])^2} \begin{bmatrix} \widehat{VaR}_{[nt_1]:i} - \widehat{VaR}_{i+1:[nt_2]-1} \\ \widehat{ES}_{[nt_1]:i} - \widehat{ES}_{i+1:[nt_2]-1} \end{bmatrix}^{\otimes 2}  \\
&+ \sum_{i=[nt_2]}^{n} \frac{(i-[nt_2])^2 (n-i+1)^2}{(n-[nt_1]+1)^2 (n-[nt_2]-1)^2} \begin{bmatrix} \widehat{VaR}_{i:n} - \widehat{VaR}_{[nt_2]:i-1} \\ \widehat{ES}_{i:n} - \widehat{ES}_{[nt_2]:i-1} \end{bmatrix}^{\otimes 2}.
\end{align*}
Then, under $\mathcal{H}_0$, applying Theorem 3 of \cite{zhang_etal2018}, our Theorem \ref{thm2} above yields the following result.
\begin{cor}
Suppose Assumption \ref{A3} holds. Under the null hypothesis $\mathcal{H}_0$,
\begin{align*}
H_n \overset{d}{\to} & \sup_{(s_1,s_2) \in \Delta} E(0,s_1,s_2)^T F(0,s_1,s_2)^{-1} E(0,s_1,s_2) + \sup_{(t_1,t_2) \in \Delta} E(t_1,t_2,1)^T F(t_1,t_2,1)^{-1} E(t_1,t_2,1) \\
& := H,
\end{align*}
where
\begin{align*}
E(r_1,r_2,r_3) = & \begin{bmatrix} W_1(r_2) - W_1(r_1) - \frac{r_2 - r_1}{r_3 - r_1}(W_1(r_3) - W_1(r_1)) \\ W_2(r_2) - W_2(r_1) - \frac{r_2 - r_1}{r_3 - r_1}(W_2(r_3) - W_2(r_1)) \end{bmatrix} \\
 F(r_1,r_2,r_3) = & \int_{r_1}^{r_2} \begin{bmatrix} W_1(s) - W_1(r_1) - \frac{s - r_1}{r_2 - r_1}(W_1(r_2) - W_1(_1)) \\ W_2(s) - W_2(r_1) - \frac{s - r_1}{r_2 - r_1}(W_2(r_2) - W_2(r_1)) \end{bmatrix}^{\otimes 2} ds \\
 & \quad + \int_{r_2}^{r_3} \begin{bmatrix} W_1(r_3) - W_1(s) - \frac{r_3 - s}{r_3 - r_2}(W_1(r_3) - W_1(r_2)) \\ W_2(r_3) - W_2(s) - \frac{r_3 - s}{r_3 - r_2}(W_2(r_3) - W_2(r_2)) \end{bmatrix}^{\otimes 2} ds.
\end{align*}
Under the alternative $\mathcal{H}_1$, $H_n \overset{P}{\to} \infty$ as $n \to \infty$.
\end{cor}
%We reject $\mathcal{H}_0$ if $H_n$ exceeds the critical value corresponding to a desired test level of the pivotal quantity $H$, which may be obtained via simulation.
%Moreover, under $\mathcal{H}_1$, our test is asymptotically consistent, and the analogous version of Proposition \ref{prop2} holds.

To reduce the computational burden of the method, we use a grid approximation suggested by \cite{zhang_etal2018}, where in the doubly-indexed set $\Delta$, one index is reduced to a coarser grid. 
Specifically, let $\mathcal{G}_\delta = \{(1+k\delta)/2 : k \in \mathbb{Z}\} \cap [0,1]$ and consider the modified statistic
\begin{align}
\widetilde{H}_n & = \sup_{(s_1,s_2) \in \Delta \cap ([0,1] \times \mathcal{G}_\delta)} E_n^f(s_1,s_2)^T F_n^f(s_1,s_2)^{-1} E_n^f(s_1,s_2) \nonumber \\
& \qquad + \sup_{(t_1,t_2) \in \Delta \cap (\mathcal{G}_\delta \times [0,1])} E_n^b(t_1,t_2)^T F_n^b(t_1,t_2)^{-1} E_n^b(t_1,t_2). \label{E15}
\end{align}
As before, under $\mathcal{H}_0$, we have
\begin{align}
\widetilde{H}_n \overset{d}{\to} & \sup_{(s_1,s_2) \in \Delta \cap ([0,1] \times \mathcal{G}_\delta)} E(0,s_1,s_2)^T F(0,s_1,s_2)^{-1} E(0,s_1,s_2) \nonumber \\
& \quad + \sup_{(t_1,t_2) \in \Delta \cap (\mathcal{G}_\delta \times [0,1])} E(t_1,t_2,1)^T F(t_1,t_2,1)^{-1} E(t_1,t_2,1) \nonumber \\
& := \widetilde{H}. \label{E16}
\end{align}
Note that simply using the original doubly-indexed set $\Delta$, for a sample of size $n$ and an arbitrary number of change points, we would need to search for maxima over $O(n^2)$ points.
However, using the grid approximation, we need only search for maxima over $O(n)$ points.
In contrast, if we were to use a direct extension of the single-change point detection methodology of Section \ref{S3p2}, with $m$ change points (which needs to be specified in advance), we would need to search for maxima over $O(n^m)$ points.
Hence, the methodology introduced in this section offers significant computational savings.

We reject $\mathcal{H}_0$ if $\widetilde{H}_n$ exceeds the critical value corresponding to a desired test level of the pivotal quantity $\widetilde{H}$.
To obtain critical values, we simulate 10,000 replications, with each replication consisting of 5,000 independent standard normal random variables to approximate standard Brownian motion on $[0,1]$.

%Figure \ref{F11} shows the approximate distribution based on 10,000 samples of the test statistic from (\ref{E16}), and the estimated 0.95 quantile is 138.19.

%\begin{figure}[t!]
%\tcapfig{Density Function of Test Statistic}
%\centering
%\minipage{0.5\textwidth}
%  \includegraphics[width=\linewidth]{figure11.jpg}
%\endminipage
%\bnotefig{This figure shows the density of the test statistic in equation (\ref{E16}) estimated by simulation with 10,000 samples, where each sample utilizes 5,000 independent standard normal random variables to approximate the standard Brownian motion on $[0,1]$. The 0.95 quantile is 138.19 and is indicated by the vertical red dashed line.}
%\label{F11}
%\end{figure}

\section{Simulations} \label{S4}

We perform an extensive simulation study to investigate the finite-sample performance of ES confidence interval construction using the sectioning and self-normalization methods (Section \ref{S3p1}) as well as upper tail change detection (Sections \ref{S3p2} and \ref{S3p3}) using ES. 
Our simulations are based on a widely used and practically relevant data-generating process for modeling financial time series, namely  Generalized Autoregressive Conditional Heteroskedasticity (GARCH) models. Specifically, in the main text we consider
\[
X_i=\sigma_i\epsilon_i,
\]
\[
\sigma_i^2 = \omega + \lambda_1 X_{i-1}^2+\lambda_2\sigma_{i-1}^2,
\]
where the conditional variance $\sigma_i^2$ follows a GARCH(1,1) process. The innovations $\epsilon_i$ are assumed to be i.i.d. standard normal, and the model parameters are set as $\omega=0.01$, $\lambda_1=0.1$, and $\lambda_2=0.8$. This parameterization yields a highly persistent yet stationary process, as $\lambda_1+\lambda_2$ approaches 1. Our choice of parameters follows \cite{bollerslev1986generalized}, where these parameters are estimated for a GARCH(1,1) model on lower-frequency financial time series data.\footnote{\cite{bollerslev1986generalized} estimates the parameters $(\omega,\lambda_1,\lambda_2)=(0.007,0.135,0.829)$ {using quarterly U.S. inflation data from February 1948 to April 1983. While these parameters are not calibrated based on high-frequency asset returns, they are widely cited in the literature and serve as a benchmark for evaluating model performance.}} The Appendix collects a comprehensive set of robustness results with different parameters for the GARCH model and $t$-distributed innovations. In particular, in Figure \ref{fig:A1} we also show the results for a more persistent GARCH(1,1) process with parameters $\omega=0.01, \lambda_1=0.1$, and $\lambda_2=0.9$, which better reflects the dynamics of higher-frequency financial return series. This choice of $\lambda_1$ and $\lambda_2$ is also adopted in the recent paper \cite{horvath2024sequential}. All our results are robust to these parameter choices.  

% AR(1): $X_{i+1} = \phi X_i + \epsilon_i$
% and ARCH(1):
% $X_{i+1} = \sqrt{\beta + \lambda X_i^2}\epsilon_i$.
% We take the innovations $\epsilon_i$ to be i.i.d. standard normal, and we use parameters $\phi = 0.5$, $\beta = 1$ and $\lambda = 0.3$.
% The stationary distribution of the AR(1) process is mean-zero normal with variance $1/(1-\phi^2)$.
% According to \cite{embrechts_etal1997}, the above choice of parameters for the ARCH(1) process yields a stationary distribution $F$ with right tail $1-F(x) \sim x^{-8.36}$ as $x \to \infty$. The AR model allows us to capture the effect of time-series dependency, while the ARCH model also leads to heavier tails. 

\subsection{Confidence Intervals} \label{S4p1}

\begin{figure}[t!]
\tcapfig{Coverage Probability and Confidence Interval Width for Different Sample Sizes}
\centering
\minipage{0.45\textwidth}
  \includegraphics[trim={0 2.5cm 0 3.5cm},clip,width=\linewidth]{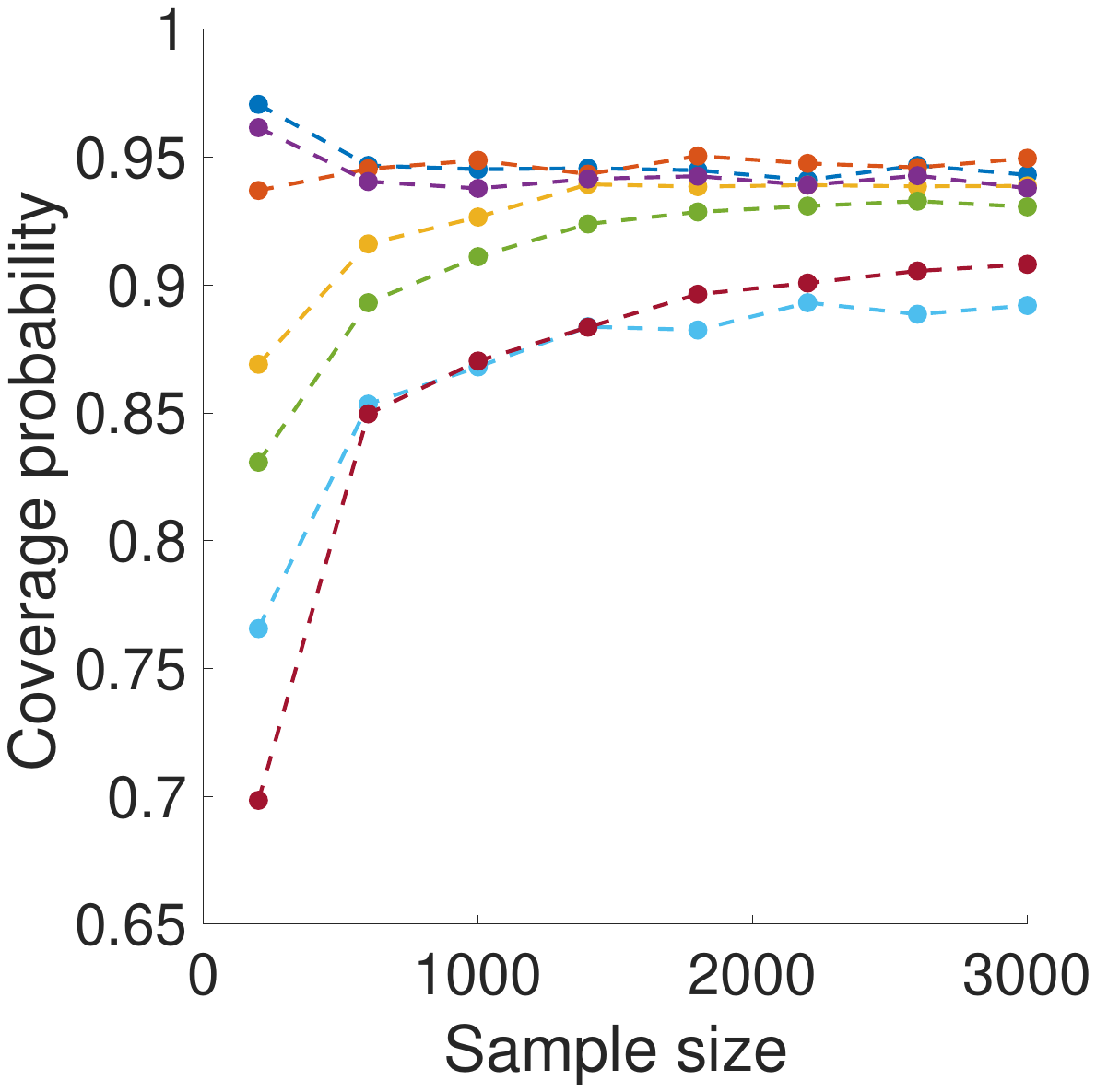}
\endminipage
\minipage{0.45\textwidth}
  \includegraphics[trim={0 2.5cm 0 3.5cm},clip,width=\linewidth]{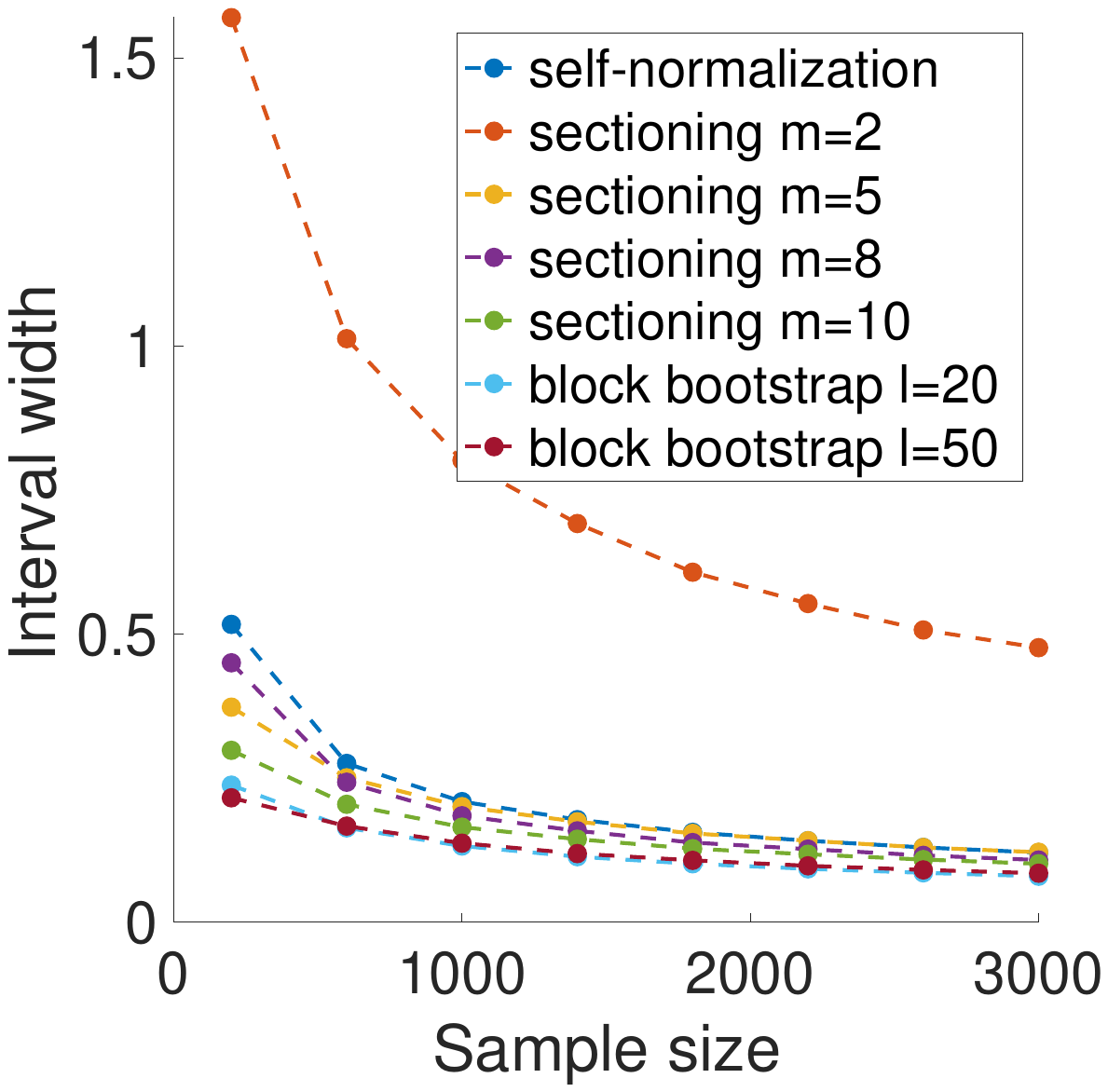}
\endminipage
\bnotefig{This figure shows the coverage probability and confidence interval width for different sample size. We use the GARCH(1,1) process with i.i.d. standard normal innovations, and set the model parameters as $\omega=0.01, \lambda_1=0.1$, and $\lambda_2=0.8$. Left subfigure: relationship between sample size and empirical coverage probability of 95\% confidence intervals for ES in the upper 5th percentile computed for stationary GARCH(1,1) process. Right subfigure: relationship between time series sample size and width of 95\% confidence intervals for ES in the upper 5th percentile computed for a stationary GARCH(1,1) process. Each plotted point is the averaged result over 10,000 replications.}
\label{F1}
\end{figure}

We begin by studying how the empirical coverage probability of confidence intervals depends on the sample size. 
In Figure \ref{F1}, we vary the time series sample size from 200 to 3,000 and examine the widths and empirical coverage probabilities of the 95\% confidence intervals for ES in the upper 5th percentile. These confidence intervals are constructed using the sectioning and self-normalization methods (Section \ref{S3p1}) for the GARCH(1,1) process introduced above. For comparison, we also include confidence intervals constructed using the block bootstrap method, as described in \cite{buhlmann2002bootstraps}.

For the sectioning method, we show that the results are robust to the sectioning parameter $m$. We study the coverage probability and confidence intervals for the sectioning parameter  $m\in \{2,5,8,10\}$. For the self-normalization method, we use the functional $T(Y)$ defined as $T(Y)=( \int_0^1 (Y(t)-tY(1))^2dt)^{1/2}$, as recommended in Section \ref{S3p1}. The block bootstrap method is applied with block lengths $l\in \{20,50\}$, and for this method, confidence intervals are constructed using the 0.025 and 0.975 quantiles of its empirical distribution. We average the estimates run over 10,000 replications. For each replication, we use a burn-in period of 5,000 to ensure the GARCH(1,1) process approximately reaches stationarity. Appendix B collects a comprehensive set of additional robustness results that confirm our findings. We repeat the simulation study for GARCH (1,1) models with parameters $\lambda_2 \in \{0.5, 0.6, 0.7, 0.8\}$ and $t$-distributed innovations $\epsilon_i$ with degrees of freedom $\nu=15$.

%For the sectioning method, we examine its performance with the user-specified parameter $m$ set to $m\in \{2,5,8,10\}$. For the self-normalization method, we use the functional $T(Y)$ defined as $T(Y)=( \int_0^1 (Y(t)-tY(1))^2dt)^{1/2}$, as recommended in Section \ref{S3p1}. The block bootstrap method is applied with block lengths $l\in \{20,50\}$, and for this method, confidence intervals are constructed using the 0.025 and 0.975 quantiles of its empirical distribution. In Figure \ref{F1}, each data point is the averaged result over 10,000 replications. For each replication, we use a burn-in period of 5,000 to ensure the GARCH(1,1) process approximately reaches stationarity. Additional simulation results with GARCH(1,1) processes with different parameters, as well as with i.i.d. $t$-distributed innovations $\epsilon_i$ with degrees of freedom $\nu=15$, are provided in Appendix B. 
% These results show that our findings are robust to the choices of parameters.

Figure \ref{F1} shows that both the self-normalization method and sectioning method with different parameters $m$ achieve good coverage when the sample size is large. In particular, the coverage probability of the confidence intervals constructed by the sectioning method is robust to different choices of $m \in [2,10]$. In contrast, the block bootstrap method yields lower coverage probabilities and performs worse than our proposed methods{, requiring larger sample sizes to approximate the nominal coverage rate}. Additionally, the block bootstrap is also sensitive to the choice of the block length. Furthermore, the sectioning method with a larger $m\in \{5,8,10\}$ produces smaller confidence intervals, but generally yields lower empirical coverage probabilities compared to the self-normalization method. This is especially pronounced for small sample sizes such as 200 or 500.\footnote{{While a sample size of 200 or 500 is statistically small, obtaining such samples at lower frequencies (e.g., quarterly data) requires a long time span.}} As the sample size increases, the performance of the two methods becomes more similar, and the desired coverage probability of 0.95 is approximately reached. While choosing $m=2$ in the sectioning method can lead to wider confidence intervals, this can be easily identified in practice, allowing for the selection of a larger $m$ that leads to a tighter interval width.

\subsection{Detection of Location Change in Tail} \label{S4p2}

We investigate through simulations the detection of abrupt location changes using ES in the upper 5th percentile, as discussed in Section \ref{S3p2}. To assess the robustness of our method in detecting change points at different locations within the time series, we consider changes occurring at $t=rT$ with $r\in \{0.5,0.6,0.7,0.8,0.9\}$, where $T$ is the time horizon.

We consider the GARCH(1,1) process introduced previosuly before the potential distribution change at time $t=rT$. After the change point, we consider the following process, where the magnitude $\mu$ of the location change varies between 0 (the null hypothesis of no change) and 1, that is, $\mu \in [0,1]$:
\[
X_i=\mu+\sigma_i\epsilon_i,
\]
\[
\sigma_i^2 = 0.01 + 0.1 (X_{i-1}-\mu)^2+0.8\sigma_{i-1}^2.
\]

\begin{figure}[t!]
\tcapfig{Empirical Detection Probability for Different Magnitudes of Changes}
\centering
\includegraphics[trim={0 3cm 0 3cm},clip,width = 0.45\textwidth]{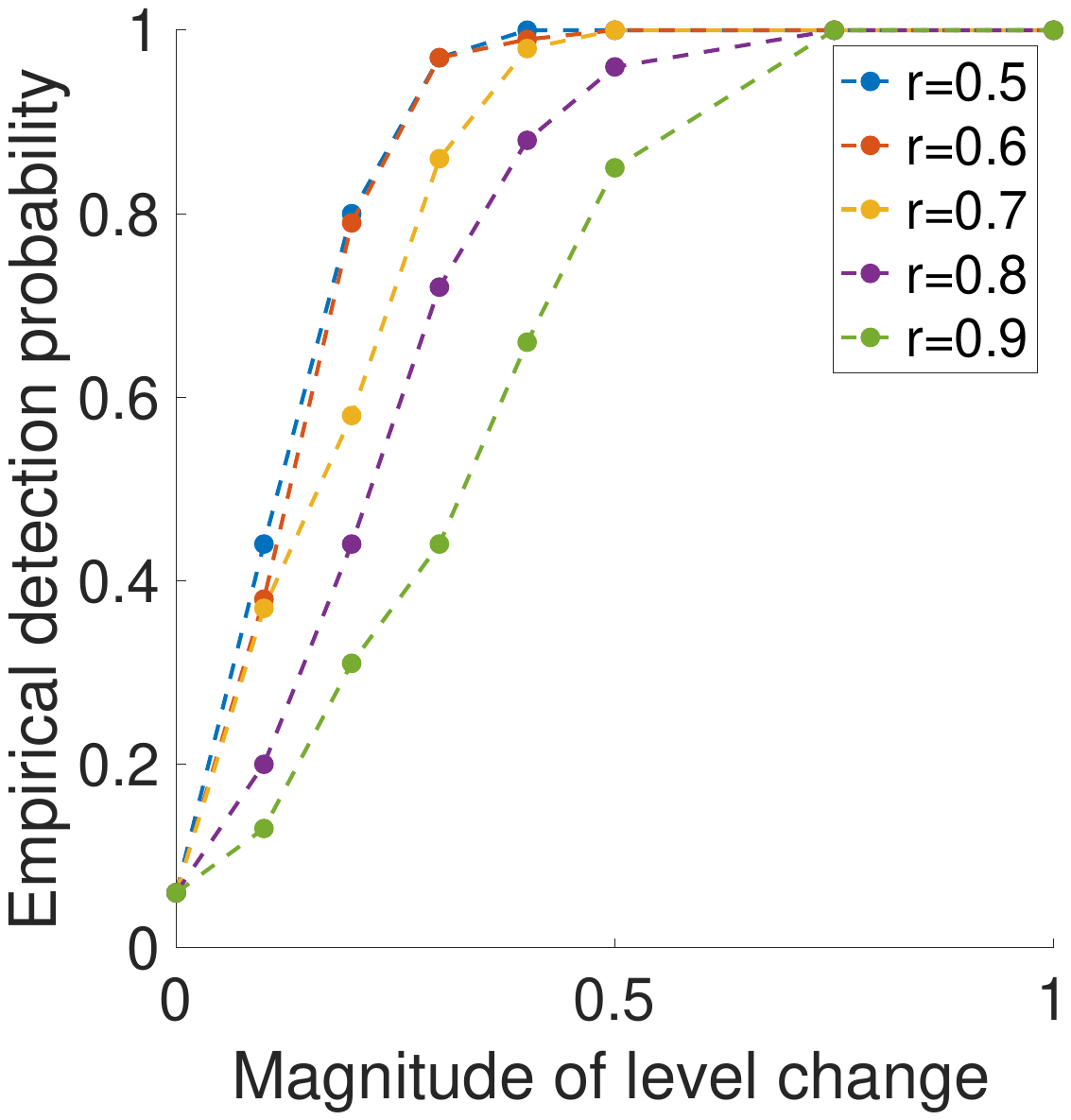}
\bnotefig{This figure shows the relationship between the empirical detection probability and magnitude of location change in the mean for change point detection with 0.05 significance level using ES in the upper 5th percentile. The abrupt location change occurs at $t=rT$ with $r\in \{0.5,0.6,0.7,0.8,0.9\}$ of the time series sequence. The magnitude of the change in the mean is in the interval $[0,1]$. Each plotted point is the average over 100 replications with time series sequences of length 2,000 in each replication.}
\label{F3}
\end{figure}

Figure \ref{F3} shows the approximate power of change-point tests using the self-normalized CUSUM statistic (see (\ref{E11})-(\ref{E12})) at the 0.05 significance level for varying magnitudes of the abrupt location change. For each data point, we perform 100 replications of the change-point testing using times series sequences of length 2,000. For each replication, we use a burn-in period of 5,000 to ensure the GARCH(1,1) process approximately reaches stationarity. 

As expected, the power of the test increases monotonically with the magnitude of the location change. In accordance with the desired 0.05 significance level of our procedure, the probability of false positive detection is approximately 0.05, as indicated by the points with zero magnitude of location change in Figure \ref{F3}. While the location of the change point does affect the detection probability, the test is robust when the change occurs between the $r=0.5$ and $0.8$ fractions of the series. For changes occurring at the very end of the series (at $r=0.9$), the detection probability decreases. However, the test can still achieve high detection power as the magnitude of the change increases.

While our method can detect changes in the mean, it is specifically designed to capture more general changes in the distribution. Therefore, when the change occurs solely in the mean, our method may require a relatively larger magnitude of shift for detection compared to methods designed specifically for mean change detection. In the next section, we focus on more complex structural breaks, where our method is particularly useful and demonstrates strong performance.

\subsection{Detection of General Change in Tail} \label{S4p3}
We also investigate the detection of general structural changes in the upper tail of the underlying marginal distribution occurring at different points within the time interval. Although the relationship between power and the ``magnitude" of the change in the upper tail is not as simple as in the case of pure location change, nevertheless, with Proposition \ref{prop2} we will detect the change with high probability as our sample size increases.
In our simulations, we study the detection of general structural changes in the tail using ES in the upper 5th percentile, as discussed in Section \ref{S3p2}.

\begin{figure}[t!]
\tcapfig{Empirical Detection Probability for Different Changes in Tail Parameters}
\centering
\minipage{0.33\textwidth}
  \includegraphics[trim={0 3cm 0 3cm},clip,width=\linewidth]{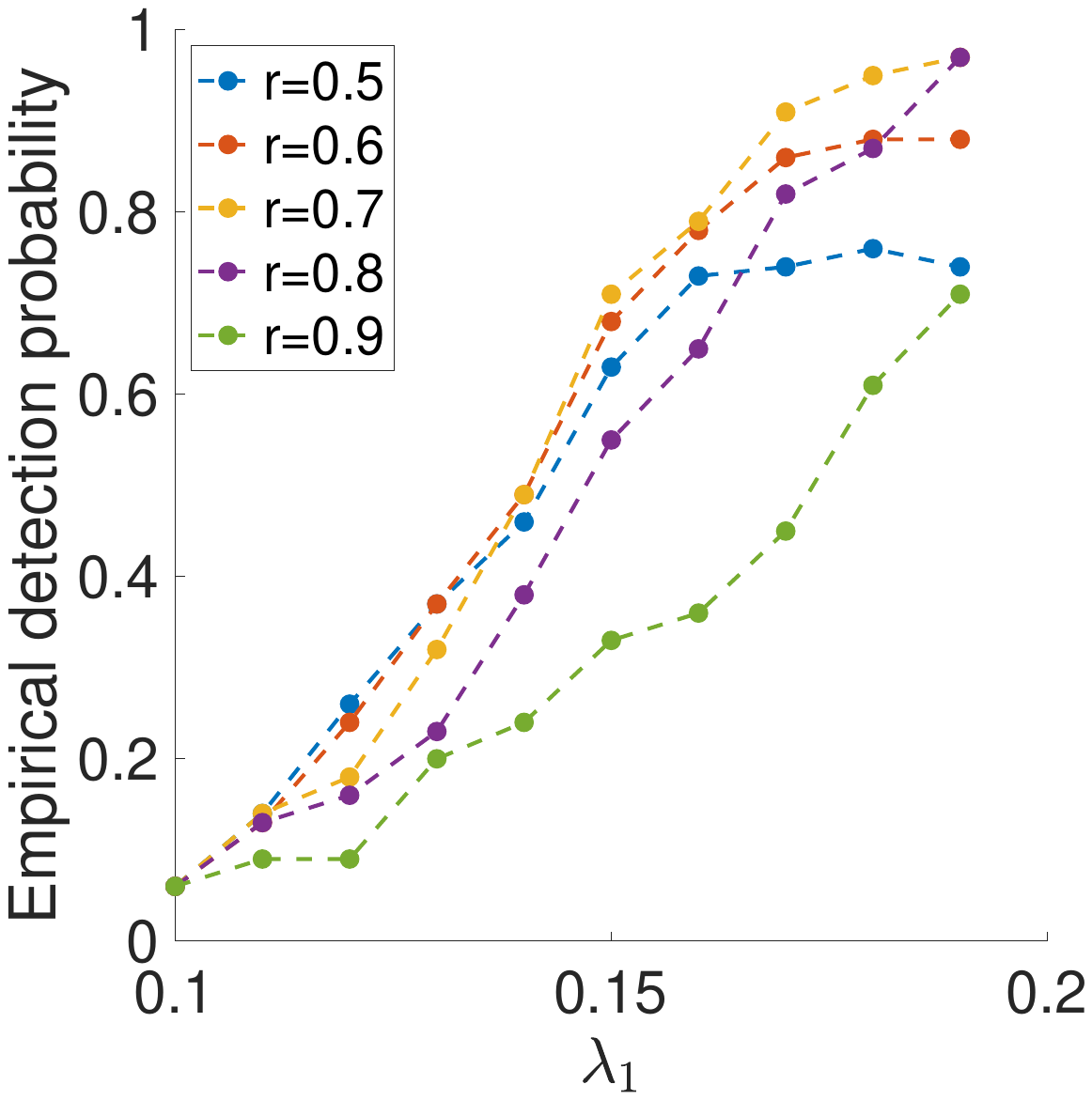}
\endminipage
\minipage{0.33\textwidth}
  \includegraphics[trim={0 3cm 0 3cm},clip,width=\linewidth]{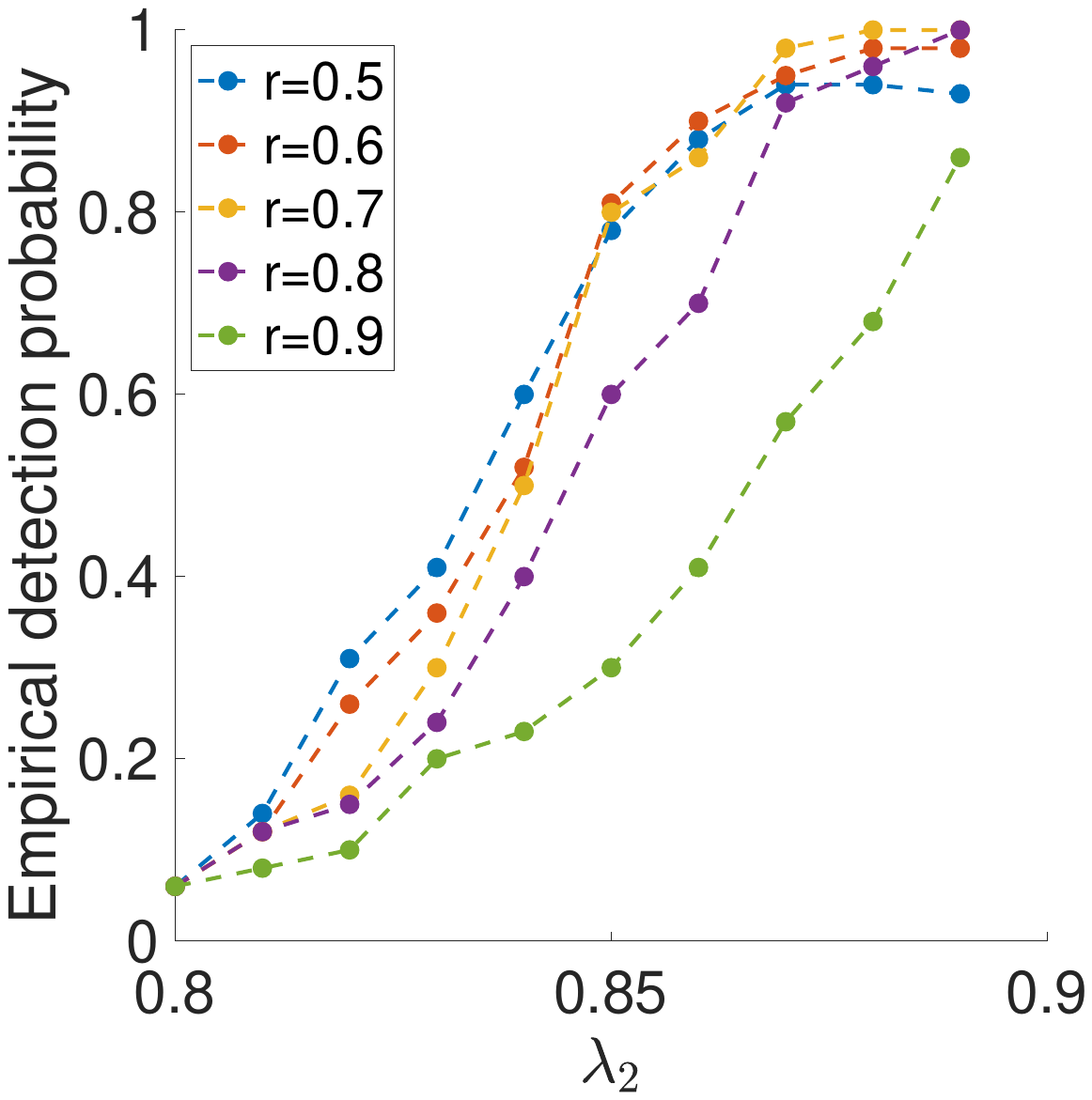}
\endminipage
\minipage{0.33\textwidth}
  \includegraphics[trim={0 3cm 0 3cm},clip,width=\linewidth]{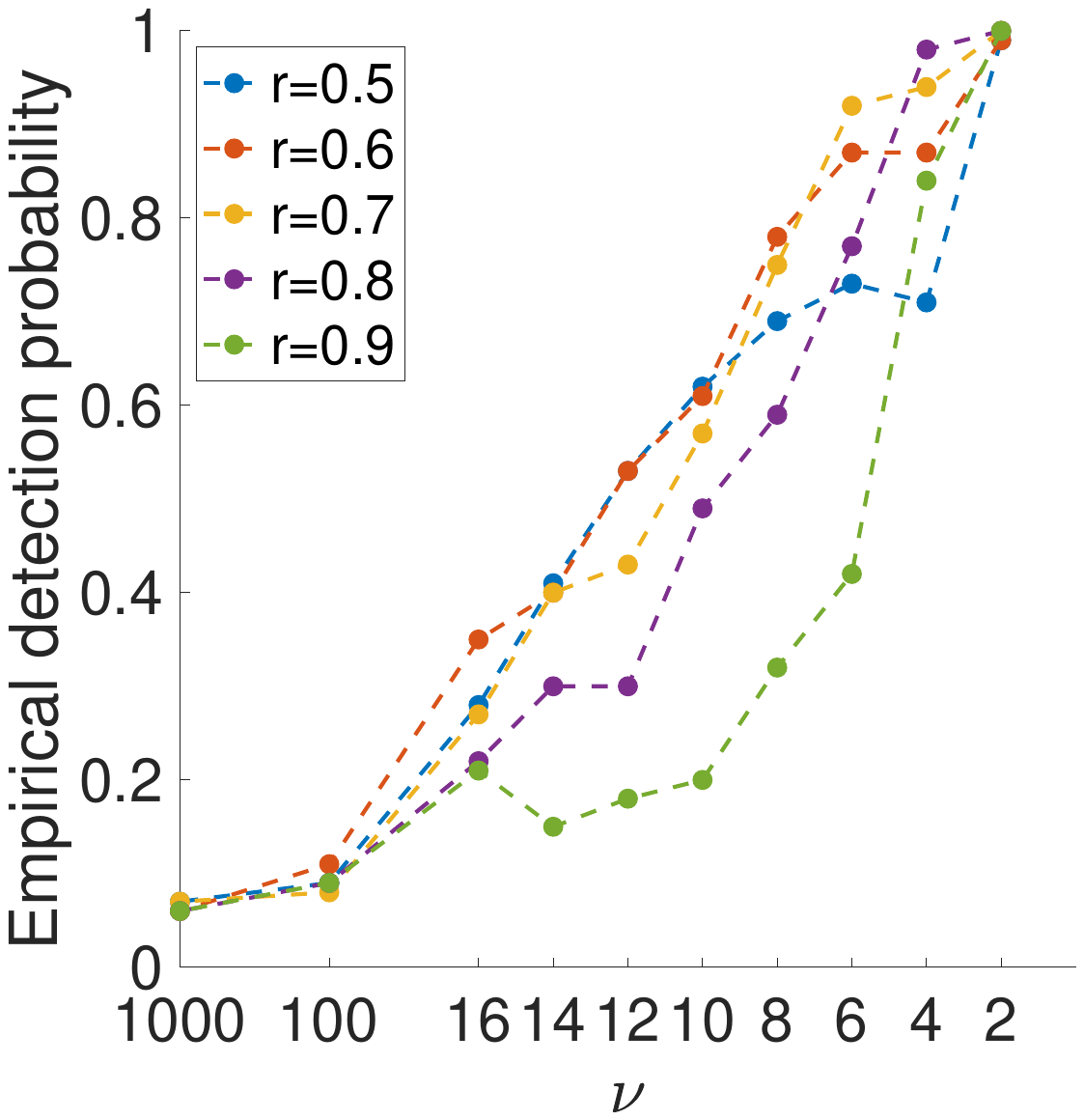}
\endminipage
\bnotefig{This figure shows the empirical detection probability for different changes in parameters that affect the tail behavior. 
Left subfigure: change-point tests for setup 1. Relationship between empirical detection probability and $\lambda_1$ after the change point at $t=rT$. Before the change point, $\lambda_1 = 0.1$.
Middle subfigure: change-point tests for setup 2. Relationship between empirical detection probability and $\lambda_2$ after the change point at $t=rT$. Before the change point, $\lambda_2 = 0.8$.
Right subfigure: change-point tests for setup 3. Relationship between empirical detection probability and the degrees of freedom $\nu$ in $t(\nu)$-distributed innovations after the change point at $t=rT$. Before the change point, the innovations are standard normal distributed.
In all cases, change-point testing is conducted with a 0.05 significance level using ES in the upper 5th percentile. Each plotted point is an average over 100 replications using time series sequences of length 2,000.}
\label{F5}
\end{figure}

As in the previous section, we consider the GARCH(1,1) process with i.i.d. normal innovations and parameters $\omega=0.01$, $\lambda_1=0.1$, and $\lambda_2=0.8$ before the distribution change at time $t=rT$, with $r=\{0.5,0.6,0.7,0.8,0.9\}$.  After the change point, we consider the following processes:
\begin{enumerate}[label=(\arabic*)]
    \item $X_i=\sigma_i\epsilon_i$, $\sigma_i^2 = 0.01 + \lambda_1 X_{i-1}^2+0.8\sigma_{i-1}^2$, $\epsilon_i \overset{i.i.d.}{\sim} \mathcal{N}(0,1)$; 
    \item $X_i=\sigma_i\epsilon_i$, $\sigma_i^2 = 0.01 + 0.1 X_{i-1}^2+\lambda_2\sigma_{i-1}^2$, $\epsilon_i \overset{i.i.d.}{\sim} \mathcal{N}(0,1)$; 
    \item $X_i=\sigma_i\epsilon_i$, $\sigma_i^2 = 0.01 + 0.1 X_{i-1}^2+0.8\sigma_{i-1}^2$, $\epsilon_i \overset{i.i.d.}{\sim} t(\nu)$.
\end{enumerate}
In the first setup, we vary the persistence parameter $\lambda_1$ from $0.1$ to $0.19$. In the second setup, we vary the parameter $\lambda_2$ from $0.8$ to $0.89$. {These ranges of parameters ensure that the process remains stationary.} In the last setup, we consider a change in the heavy-tailedness of the underlying distribution, transitioning from Gaussian innovations to $t$-distributed innovations with {degrees of freedom $\nu\in \{2,4,6,8,10,12,14,16,100,1000\}$. As $\nu$ increases, the $t$-distribution approaches the Gaussian distribution, and hence we expect the detection probability to decline accordingly}. Figure \ref{F5} shows the approximate power of change-point tests using the self-normalized CUSUM statistic (see (\ref{E11})) at the 0.05 significance level. For each data point, we perform 100 replications using times series sequences of length 2,000 and use an initial burn-in period of 5,000 to approximately reach stationarity.

As expected, for all types of changes, the power is approximately a monotonic function of the magnitude of change. Furthermore, our method is generally robust to the location of the change point, provided that it is not located near the very end of the time period. When the change point occurs close to the end of the series, the detection probability decreases slightly, but our method is still able to detect the change with high probability as the magnitude of the change increases.

\subsection{Detection of Multiple Changes in Tail} \label{S4p4}

We additionally investigate the detection of multiple structural changes in the upper tail of the underlying marginal distribution.
Here, we again use ES in the upper 5th percentile and compare the practical performance of the single change-point methodology discussed in Section \ref{S3p2} with the unsupervised multiple change-point methodology discussed in Section \ref{S3p3}.
We consider the following variant of the GARCH(1,1) process introduced previously:
\begin{align} \label{E17}
\text{GARCH(1,1) process:} \qquad X_i = 
\begin{cases} 
\sigma_i\epsilon_i, \epsilon_i\overset{i.i.d.}{\sim} N(0,1) &\text{for } i\leq [n/3] \\ 
\sigma_i u_i, u_i\overset{i.i.d.}{\sim} t(\nu) &\text{for } [n/3]\leq i\leq [2n/3] \\ 
\sigma_i\epsilon_i, \epsilon_i\overset{i.i.d.}{\sim} N(0,1) &\text{for } i\geq [2n/3] 
\end{cases} ,
\end{align}
\[
\qquad \qquad \sigma_i^2 = 0.01 + 0.1 X_{i-1}^2+0.8\sigma_{i-1}^2.
\]
In this GARCH(1,1) process, the innovations initially have relatively light tails, then change to heavier tails with $t(\nu)$-distributed innovations, and finally revert back to the original lighter tails.

\begin{figure}[b!]
\tcapfig{Comparison between Unsupervised Multiple and Single Change-Point Tests}
\centering
%\minipage{0.5\textwidth}
 % \includegraphics[width=0.5\linewidth]{new_plots/figure6.pdf}
  \includegraphics[trim={0 3.5cm 0 4cm},clip,width=0.5\linewidth]{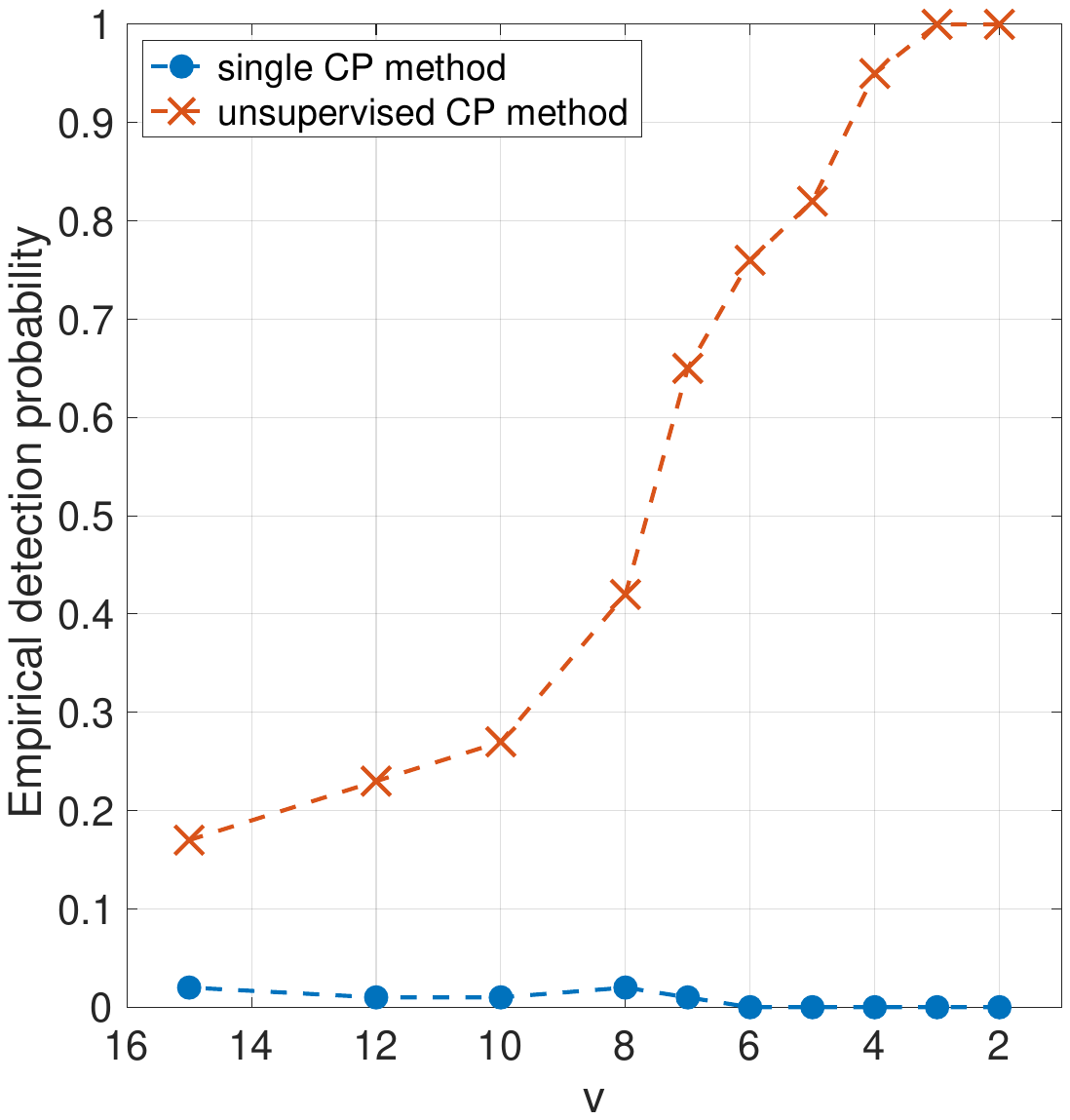}  
%\endminipage
\bnotefig{This figure compares unsupervised multiple change-point tests with single change-point tests at the 0.05 significance level using ES in the upper 5th percentile. We plot the relationship between the empirical detection probability of at least one change point and the degrees of freedom $\nu$ of t-distributed innovations. Each plotted point is an average over 100 replications using time series sequences of length 1,200.}
\label{F10}
\end{figure}

Figure \ref{F10} shows the approximate detection power of at least one change point as $\nu$ varies using the single change-point methodology (see (\ref{E11})-(\ref{E12})) versus the unsupervised multiple change-point methodology (see (\ref{E15})-(\ref{E16})) at the 0.05 significance level. For each data point, we perform 100 replications using times series sequences of length 1,200. For each replication, we use an initial burn-in period of 5,000 to approximately reach stationarity in the GARCH(1,1) process.

We observe that the single change-point testing method is unable to detect a change in the process in (\ref{E17}), even for extremely strong deviations from the null such as the case $\nu = 2$. 
In fact, its power decays to zero as the magnitude of the change increases. This occurs because the single change-point test assumes that only one change point occurs and seeks to divide the time series into two sections separated by a single change point. Consequently, the effects of multiple changes can be ``canceled out''.
On the other hand, the unsupervised multiple change-point test exhibits the desired performance with increasing power as the magnitude of the change increases.
Hence, it is a promising candidate for detecting more complex patterns of changes in the tails of time series.

\section{Empirical Applications} \label{S5}

\subsection{Data and Expected Shortfall Estimation}

We study the changes in tail risk for two of the most important macroeconomic indicators that reflect the conditions in the U.S. stock market and for interest rates. First, we consider the daily log returns of the S\&P 500 Index, which is a proxy for a market risk factor, for the years 1928-2023.\footnote{We obtain the time-series from Yahoo Finance.} Second, we analyze the weekly log returns of U.S. 1-year and 10-year zero-coupon  bonds for the years 1961-2023. We obtain the data from the U.S. Treasury Discount Bond Database provided by \cite{filipovic2022shrinking}, which is shown to provide the most reliable estimates. These bonds are tradable portfolios of U.S. Treasuries and reflect information in interest rates.

%We study the changes in tail risk for two of the most important macroeconomic indicators that reflect the conditions in the U.S. stock market and for interest rates. First, we consider the daily log returns of the S\&P 500 Index from Yahoo Finance, which is a proxy for a market risk factor, for the years 1928-2023. Second, we analyze the weekly log returns of U.S. 1-year and 10-year zero-coupon bonds, derived from the U.S. Treasury Discount Bond Database provided by \cite{filipovic2022shrinking}, for the years 1961-2023. These bonds are tradable portfolios of U.S. Treasuries and reflect information in interest rates.

% We study the changes in tail risk for two of the most important macroeconomic indicators that reflect the conditions in the U.S. stock market and for interest rates. First, we consider the daily log returns of the SPY ETF, which tracks the S\&P 500 Index and hence is a proxy for a market risk factor, for the years 2004-2016. Second, we analyze the monthly log returns of US 30-Year Treasury bonds for the years 1942-2017. These returns reflect information in long term interest rates. Both time series are obtained from CSRP.

\begin{figure}[t!]
\tcapfig{Expected Shortfall of Market Returns over Time}
\centering
\minipage{\textwidth}
  \includegraphics[width=\linewidth]{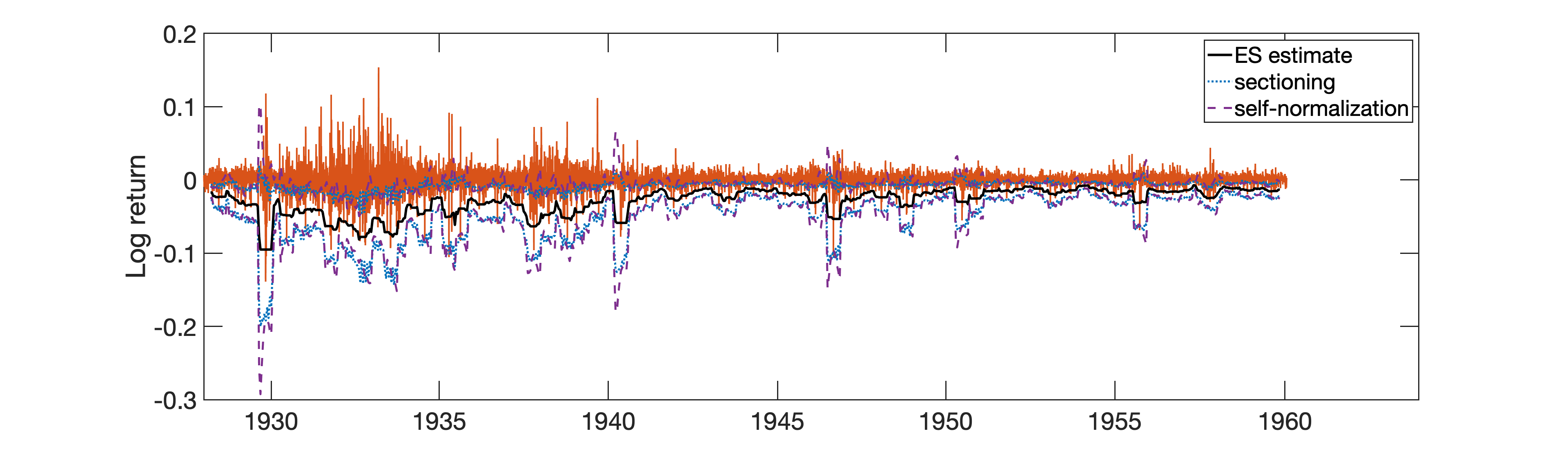}
\endminipage
\\
\minipage{\textwidth}
  \includegraphics[width=\linewidth]{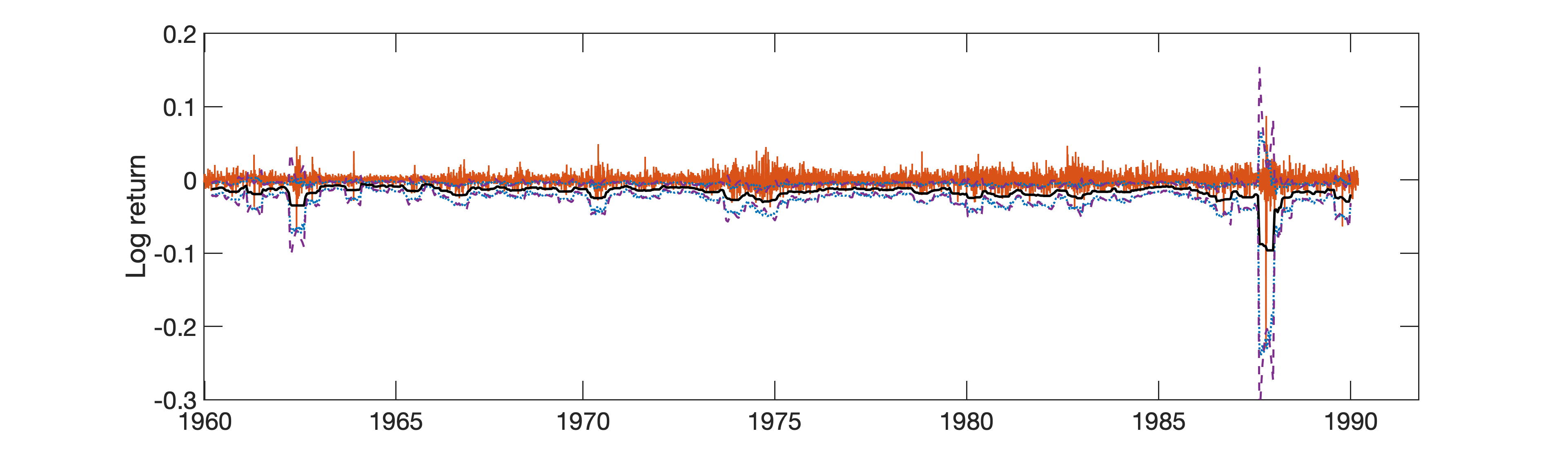}
\endminipage
\\
\minipage{\textwidth}
  \includegraphics[width=\linewidth]{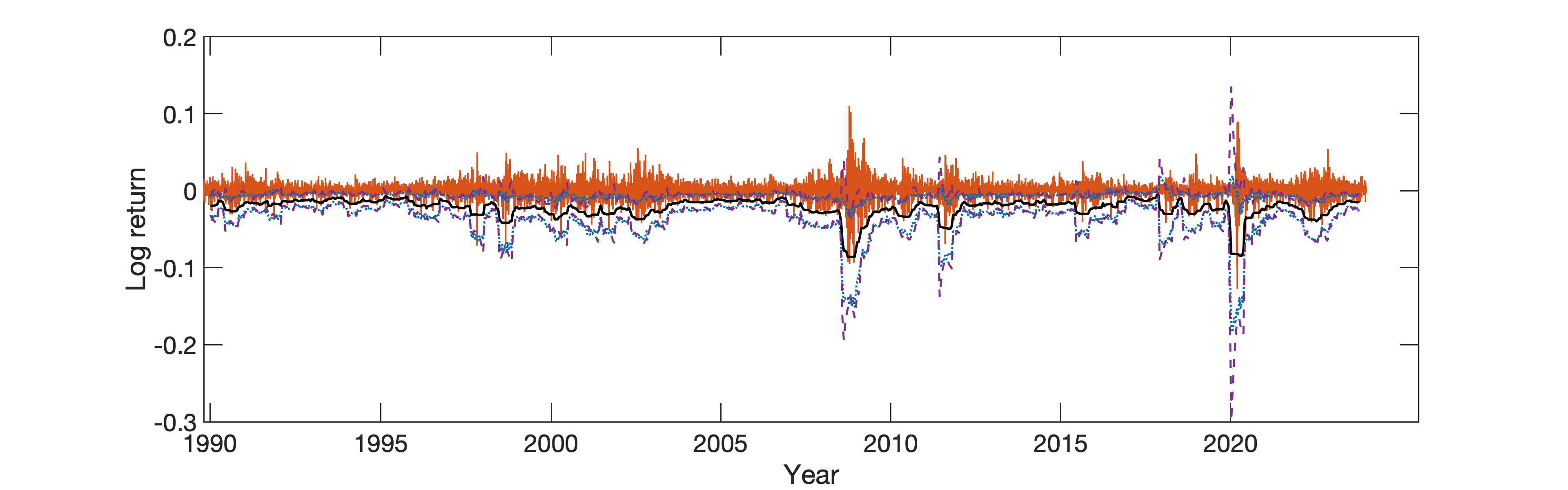}
\endminipage
\bnotefig{This figure shows the daily log returns and expected shortfall with confidence intervals for the S\&P 500 index that approximates a market return time-series. We plot the ES estimate and 95\% confidence bands for the lower 5th percentile. ES is computed using a rolling window of 100 days with 10 day shifts.}
\label{F6}
\end{figure}

\begin{figure}[t!]
\tcapfig{Expected Shortfall of Bond Returns over Time}
\centering
\begin{subfigure}[b]{\textwidth}
    \includegraphics[width=\linewidth]{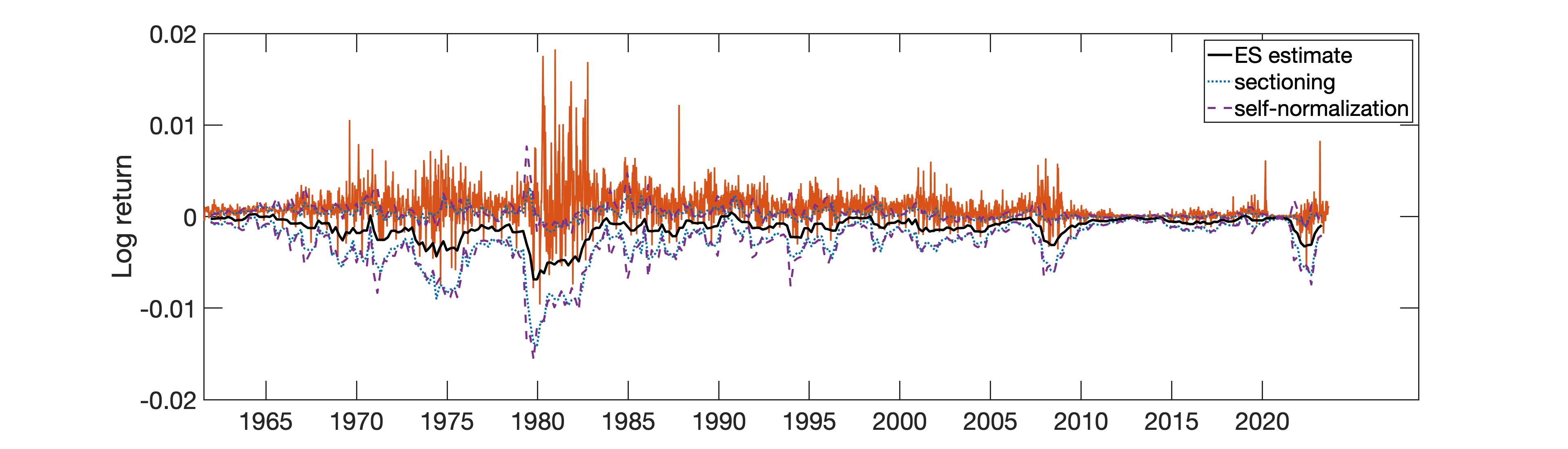}
    \caption{U.S. 1-year zero-coupon bond}
\end{subfigure}
\begin{subfigure}[b]{\textwidth}
    \includegraphics[width=\linewidth]{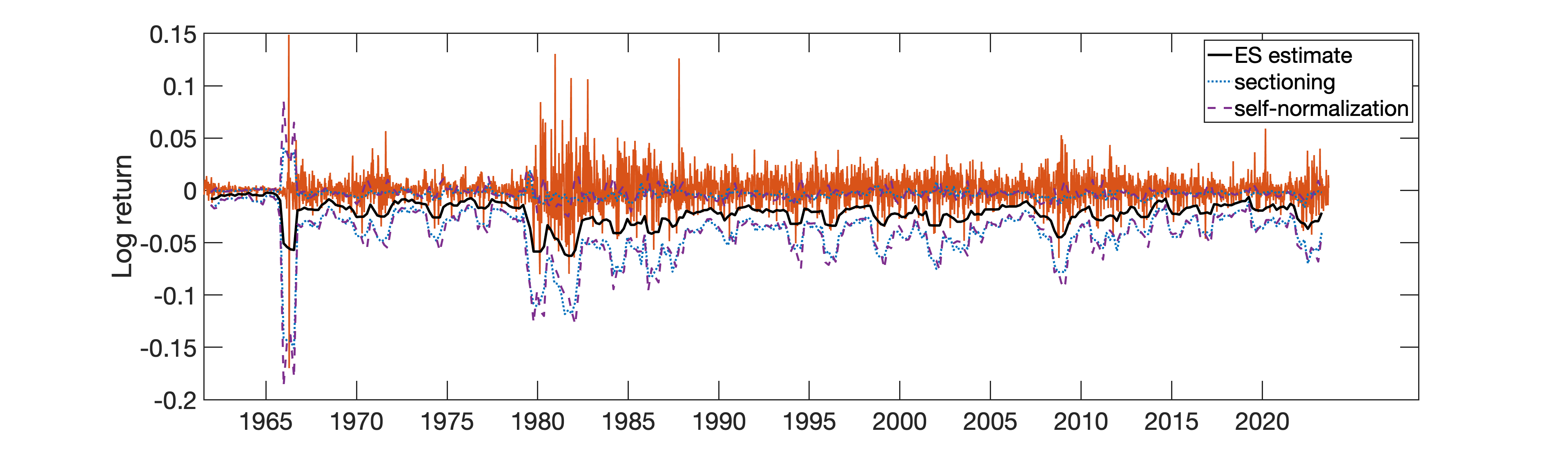}
    \caption{U.S. 10-year zero-coupon bond}
\end{subfigure}
\bnotefig{This figure shows the weekly log returns and expected shortfall with confidence intervals for long- and short-maturity zero-coupon bonds. We plot the ES estimate and 95\% confidence bands for the lower 10th percentile of log returns. ES is computed using a rolling window of 40 weeks with 10 week shifts.}
\label{F7}
\end{figure}

We begin our study with an analysis of the expected shortfall (ES) over time. We first apply the two methods of confidence interval construction from Section \ref{S3p1} to the daily log returns of S\&P 500 for the period from January 03, 1928 to December 29, 2023. In Figure \ref{F6}, we show ES estimates of the lower 5th percentile of log returns throughout this time period along with 95\% confidence bands computed using the sectioning and self-normalization methods. We use the sectioning method with $m=10$.
(Our simulation study demonstrates that the coverage results are robust to the choices of the tuning parameter $m$.)
We use a rolling window of 100 days with 90 days of overlap between successive windows. The self-normalization method appears to be more conservative and yields a wider confidence band compared to the sectioning method, which agrees with the results presented in Figure \ref{F1}. Overall, the ES estimates and confidence bands appear to capture well the increased volatility of returns during periods of financial instability, such as the 1929 Wall Street Crash, World War II in 1940, the 1987 Black Monday Crash, the 2008 Financial Crisis, and the COVID-19 Pandemic in 2020.

We next apply the two methods of confidence interval construction from Section \ref{S3p1} to weekly log returns of U.S. 1-year and 10-year zero-coupon bonds for the period from July 03, 1961 to August 31, 2023. In Figure \ref{F7}, we show ES estimates of the lower 10th percentile of log returns throughout this time period along with 95\% confidence bands computed using the sectioning and self-normalization methods. We use a rolling window of 40 weeks with 30 weeks of overlap between successive windows. Again, the self-normalization method appears to be more conservative and yields a wider confidence band compared to the sectioning method. Although time series for bond returns have a greater degree of autocorrelation compared to S\&P 500 returns, our methods are still applicable.\footnote{For comparison, we also estimate the ES with a parametric GARCH(1,1) model. Figures \ref{fig:stockGARCH} and \ref{fig:bondGARCH} in the Appendix collect the results. We show that for both the S\&P 500 log returns and U.S. 10-year bond log returns, the ES estimates are broadly similar using the parametric GARCH model and our method. However, the parametric GARCH model can lead to unstable and spurious results. For example, for the U.S. 1-year bond log returns, the ES estimates show some noticeable differences, indicating that our method and the parametric model can yield different results. The estimation of the GARCH model on a rolling window can also lead to issues, and on approximately 10\% of the rolling windows the optimizer failed to converge. For example, for the period from March 3, 1930 to July 23, 1930, the optimization method failed for both GARCH(1,1) models with either $t$-distributed innovations or normal innovations. We only plot the ES estimates from the GARCH model when the fitting process succeeds. Overall, our method appears to be more robust and reliable.}

These rolling window estimates of ES with confidence intervals provide a valuable tool to get insight into the variation of the tail behavior of these time series. However, they are not a formal test for change points. Such tests are based on our novel statistics, which we study next.

\subsection{Change-Point Detection} \label{empirical_change_point_detection}

%{\color{magenta} In this section, we illustrate the behavior of the single and multiple change-point tests on real data.
%We apply our single and multiple change-point testing methodology for ES to the lower 5th percentile of S\&P 500 daily log returns and to the lower 10th percentile of U.S. 1-year and 10-year zero-coupon bonds weekly log returns. 
%For a systematic illustration, we run the tests on rolling windows over the full return time series.
%We compare the behavior of the tests and also visualize the return time series on windows of interest.}

We now apply our change-point testing methodology to the equity and bond return time-series. We use our single and multiple change-point testing methodology for ES of the lower 5th percentile of S\&P 500 daily log returns and of the lower 10th percentile of U.S. 1-year and 10-year zero-coupon bonds weekly log returns. For a systematic study, we run the tests on rolling windows over the full return time series. We compare the behavior of the tests and also visualize the return time-series on windows of interest.\footnote{Recall that our change-point testing methodology is meant to test for change(s) in a \textit{given} time window, and help assess whether or not risk measures are stationary over the window. For the rolling window demonstrations in this section, if one is interested in performing statistical inference simultaneously for all windows, then one could run a multiple hypothesis testing procedure that controls the False Discovery Rate (FDR). A robust (albeit conservative) choice of procedure, which is valid for arbitrary dependence among the tests, is the Benjamini-Yekutieli (B-Y) procedure \citep{benjamini_yekutieli_2001}. A description of the B-Y procedure is provided at the end of Appendix C.}
%\textcolor{red}{Our goal is to illustrate and compare the behavior of the single and multiple change-point tests on local windows where changes could have potentially occurred.}
%In a second step, we compare the tests on the local windows where either the single or multiple change points test detects changes. 
%We also visualize the return time series on selected local windows.

%- We apply our analysis to the full time-series
%- then we zoom into particular subsets of the time-series namely, where the multiple change point test detects significant change points
%- For these particular times, we also provide examples of the time-series, to interpret the results
%- multiple testing adjustment? 

In more detail, we apply our single (based on (\ref{E11})-(\ref{E12})) and multiple (based on (\ref{E15})-(\ref{E16})) change-point tests to S\&P 500 returns in successive 6-month and 1-year time windows, with 1 month of shift between successive windows. In Figures \ref{fig:empir_sp_single_6m}-\ref{fig:empir_sp_multiple_12m} in Appendix C, we plot the test statistics for single and multiple change-point tests for ES in the lower 5th percentile of S\&P 500 log returns for different time windows. We also indicate the critical value for the 0.05 significance level.
%\footnote{As we are running multiple test, we can apply a multiple testing adjustment based on a Bonferroni correction. This would increase the critical value but not change the test statistic.} 
Then, we apply the two change-point tests to weekly log returns of the zero-coupon bonds. We use successive 5-year and 10-year time windows for both bonds, with 3 months of shift between successive windows. In Figures \ref{fig:empir_b1_5y}-\ref{fig:empir_b10_10y} in Appendix C, we plot the test statistics for single and multiple change-point tests for ES in the lower 10th percentile of U.S. 1-year and 10-year zero-coupon bond log returns for different time windows. 

In Tables \ref{tab:empirical_change_points} and \ref{tab:empirical_change_points_bond}, for S\&P 500 returns and bond returns, respectively, we report examples of the single and multiple change-point test results. For each window, we report the test statistics and the corresponding p-values.\footnote{ In Table \ref{tab:empirical_change_points_bond_benjamini_yekutieli}, we also report results from applying the B-Y multiple hypothesis testing adjustment for single and multiple change-point tests over rolling windows for the bond returns. Despite the conservativeness of the B-Y procedure, the bond returns generally exhibit less volatility relative to the S\&P 500 returns, and allow for discoveries with FDR control. Unfortunately, the B-Y procedure is too conservative to reliably make discoveries with FDR control for S\&P 500 returns.}

\begin{table}[t!]
\tcapfig{Examples of Change-Point Tests for S\&P 500 Returns}
\label{tab:empirical_change_points}
\centering
\begin{tabular}{ccccc}
\toprule
\multicolumn{1}{c}{} 
& \multicolumn{2}{c}{single CP test} & \multicolumn{2}{c}{multiple CP test} \\
\cmidrule(rl){2-5}
time window & statistic & p-value & statistic & p-value \\
\midrule
    01/02/1929 - 12/31/1929 &64.0 &0.018 &226.8 &0.004\\
    12/01/1939 - 05/31/1940 &157.3 &0.001 &387.6 & \hspace{-2.8mm}$<$0.001\\
    06/01/1939 - 05/31/1940 &43.2 &0.045 &328.2 & \hspace{-2.8mm}$<$0.001\\
    06/03/1946 - 11/29/1946 &58.6 &0.023 &155.0 &0.029\\
    02/01/1950 - 07/31/1950 &77.0 &0.008 &241.2 &0.003\\
    08/01/1957 - 07/31/1958 &66.8 &0.015 &199.6 &0.010\\
    08/01/1962 - 07/31/1963 &62.7 &0.020 &149.7 &0.035\\
    08/01/1969 - 07/31/1970 &69.7 &0.014 &168.6 &0.020\\
    05/01/1987 - 10/30/1987 &142.0 &0.001 &167.5 &0.021\\
    01/02/1987 - 12/31/1987 &41.7 &0.049 &216.8 &0.006\\
    05/01/2008 - 10/31/2008 &72.6 &0.011 &186.6 &0.012 \\
    09/01/2017 - 02/28/2018 &245.1 & \hspace{-2.8mm}$<$0.001 &306.1 &0.001 \\
    10/01/2019 - 03/31/2020 &130.4 &0.001 &276.4 &0.001 \\
    \cdashline{1-5}
    11/01/1963 - 04/30/1964 &58.5 &0.023 &119.6 &0.085\\
    10/02/2006 - 03/30/2007 &73.9 &0.010 &99.1 &0.156\\
    03/01/2011 - 08/31/2011 &80.8 &0.007 &132.8 &0.059\\
    \cdashline{1-5}
    04/01/1955 - 03/29/1956 &7.3 &0.514 &177.3 &0.016\\
    05/01/2008 - 04/30/2009 &8.0 &0.487 &171.7 &0.018 \\
    09/03/2019 - 08/31/2020 &4.6 &0.702 &220.4 &0.005 \\
    \bottomrule
\end{tabular}
\bnotefig{This table shows results for single and multiple change-point tests for expected shortfall (ES) in the lower 5th percentile for S\&P 500 daily log returns in different time windows. We report the change-point test statistics and the corresponding p-values. The top blocks show examples of where both the single and multiple change-point tests are significant. The middle blocks show examples of where the single change-point test is significant, but the multiple change-point test is not. The lower blocks show examples of where the multiple change-point test is significant, but the single change-point test is not.} 
\end{table}

\begin{table}[t!]
\tcapfig{Examples of Change-Point Tests for Bond Returns}
\label{tab:empirical_change_points_bond}
\centering
\begin{tabular}{ccccc}
\toprule
\multicolumn{1}{c}{} 
& \multicolumn{2}{c}{single CP test} & \multicolumn{2}{c}{multiple CP test} \\
\cmidrule(rl){2-5}
time window & statistic & p-value & statistic & p-value \\
\midrule
    \multicolumn{5}{c}{\textbf{US 1-year bond weekly log returns}}\\
    \midrule
    10/18/1961 - 09/19/1971 &104.5 &0.002 &173.5 &0.018 \\
    06/24/1973 - 06/11/1978 &107.9 &0.002 &172.3 &0.018 \\
    % 08/17/1975 - 03/16/1980 &242.1 &0.000 &202.5 &0.009\\
    06/20/1976 - 06/07/1981 &253.1 & \hspace{-2.8mm}$<$0.001 &350.4 & \hspace{-2.8mm}$<$0.001 \\ % significant
    06/15/1980 - 06/02/1985 &244.3 & \hspace{-2.8mm}$<$0.001 &293.3 &0.001 \\ % significant
    12/16/1979 - 11/26/1989 &418.1 & \hspace{-2.8mm}$<$0.001 &492.4 & \hspace{-2.8mm}$<$0.001 \\ % significant
    08/12/2007 - 07/29/2012 &98.2 &0.003 &180.1 &0.015 \\
    % 07/08/2007 - 09/11/2016 &157.3 &0.001 &296.1 &0.001\\
    % 06/03/2012 - 01/01/2017 &161.5 &0.001 &292.6 &0.001\\
    08/05/2012 - 07/23/2017 &155.1 &0.001 &290.0 &0.001 \\
    02/03/2013 - 01/15/2023 &93.3 &0.004 &339.0 & \hspace{-2.8mm}$<$0.001 \\ 
    01/28/2018 - 01/15/2023 &151.3 &0.001 &328.9 & \hspace{-2.8mm}$<$0.001  \\ 
    % 04/06/2014 - 06/11/2023 &81.6 &0.007 &184.4 &0.013\\ 
    \cdashline{1-5} 
    10/06/1963 - 09/22/1968 &65.3 &0.016 &87.4 &0.222 \\
    % 12/22/1963 - 02/25/1973 &77.0 &0.008 &113.5 &0.102\\
    % 05/26/1974 - 12/24/1978 &54.7 &0.028 &78.6 &0.289\\
    % 09/24/1989 - 04/24/1994 &51.7 &0.032 &97.3 &0.165\\
    11/22/1998 - 11/09/2003 &84.9 &0.005 &84.1 &0.246 \\
    04/30/2017 - 04/17/2022 &94.1 &0.003 &124.4 &0.075 \\  
    \cdashline{1-5}    
    % 08/22/1971 - 10/26/1980 &25.4 &0.129 &321.7 &0.000\\
    08/14/2005 - 08/01/2010 &7.1 &0.524 &188.1 &0.012 \\
    09/19/1976 - 08/31/1986 &5.0 &0.671 &477.8 & \hspace{-2.8mm}$<$0.001 \\
    11/23/1997 - 11/04/2007 &2.8 &0.924 &162.4 &0.024 \\
    11/14/2004 - 10/26/2014 &9.1 &0.433 &189.1 &0.012 \\
    % 09/28/2008 - 12/03/2017 &15.1 &0.266 &156.8 &0.028\\
    11/06/2011 - 10/17/2021 &2.4 &0.978 &228.0 &0.004 \\
    \midrule
    \multicolumn{5}{c}{\textbf{US 10-year bond weekly log returns}} \\
    \midrule
    % 06/24/1973 - 01/22/1978 &155.7 &0.002 &161.6 &0.024\\
    12/26/1971 - 12/06/1981 &471.5 & \hspace{-2.8mm}$<$0.001 &542.3 & \hspace{-2.8mm}$<$0.001\\
    06/20/1976 - 06/07/1981 &281.0 & \hspace{-2.8mm}$<$0.001 &362.0 & \hspace{-2.8mm}$<$0.001\\
    % 02/22/1981 - 09/22/1985 &160.2 &0.001 &207.0 &0.007\\
    06/17/1979 - 05/28/1989 &47.1 &0.039 &151.4 &0.033\\
    06/12/1983 - 05/23/1993 &222.5 &0.001 &354.1 & \hspace{-2.8mm}$<$0.001\\
    08/15/2004 - 08/02/2009 &105.7 &0.002 &159.7 &0.025\\
    % 02/17/2008 - 09/16/2012 &132.9 &0.001 &139.9 &0.047\\
    % 06/03/2012 - 09/11/2016 &141.3 &0.001 &220.2 &0.005\\
    01/28/2018 - 01/15/2023 &106.3 & 0.002 &203.9 &0.008\\
    % 12/15/2013 - 02/19/2023 &42.3 &0.047 &161.7 &0.024\\
    \cdashline{1-5}
    01/06/1963 - 12/24/1967 &144.3 &0.001 &96.7 &0.168\\
    09/14/1980 - 09/01/1985 &58.6 &0.023 &127.3 &0.068\\
    % 09/24/1989 - 04/24/1994 &51.7 &0.032 &97.3 &0.165\\
    % 12/06/1998 - 07/06/2003 &115.1 &0.002 &91.8 &0.195\\
    11/17/2002 - 11/04/2007 &77.4 &0.008 &106.6 &0.128\\
    07/29/2018 - 07/16/2023 &68.6 &0.014 &127.2 &0.069\\
    \cdashline{1-5}
    12/18/1977 - 11/29/1987 &3.2 &0.874 &233.0 &0.004\\
    12/14/1980 - 11/25/1990 &16.8 &0.237 &171.7 &0.018\\
    11/19/2000 - 10/31/2010 &2.9 &0.916 &162.6 &0.024\\
\bottomrule
\end{tabular}
\bnotefig{This table shows results for single and multiple change-point tests for expected shortfall (ES) in the lower 10th percentile for U.S. 1-year and 10-year zero-coupon bond weekly log returns in different time windows. We report the change-point test statistics and the corresponding p-values. The top blocks show examples of where both the single and multiple change-point tests are significant. The middle blocks show examples of where the single change-point test is significant, but the multiple change-point test is not. The lower blocks show examples of where the multiple change-point test is significant, but the single change-point test is not.} 
\end{table}

\begin{figure}[t!]
\tcapfig{Examples of S\&P 500 Return Time-Series Around Detected Change Points}
\centering
\includegraphics[width=\linewidth]{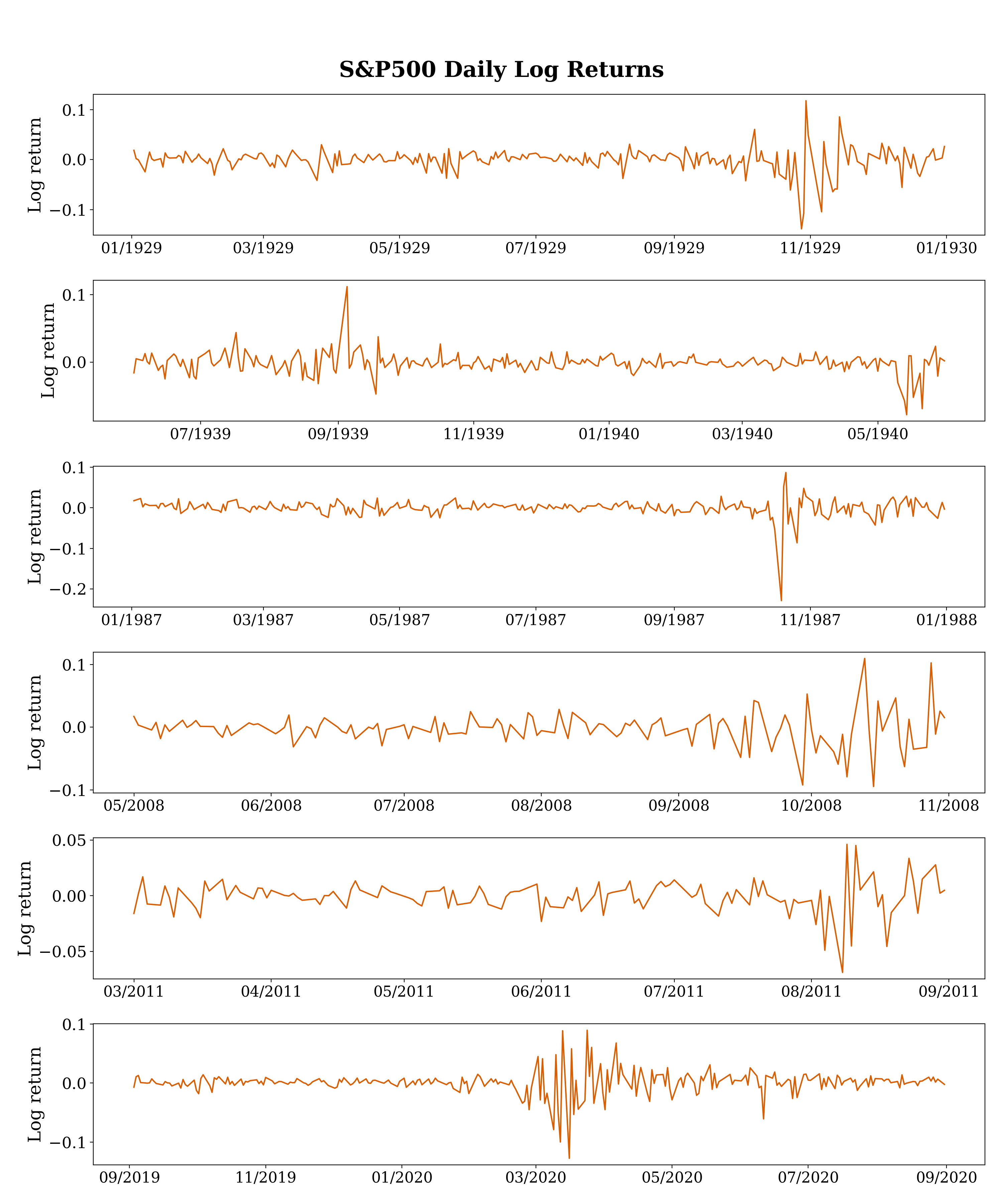}
\bnotefig{This figure shows the S\&P 500 log return time-series on local windows around suspected change points.}
\label{F8}
\end{figure}

\begin{figure}[t!]
\tcapfig{Examples of Bond Return Time-Series Around Detected Change Points}
\centering
\includegraphics[width=\linewidth]{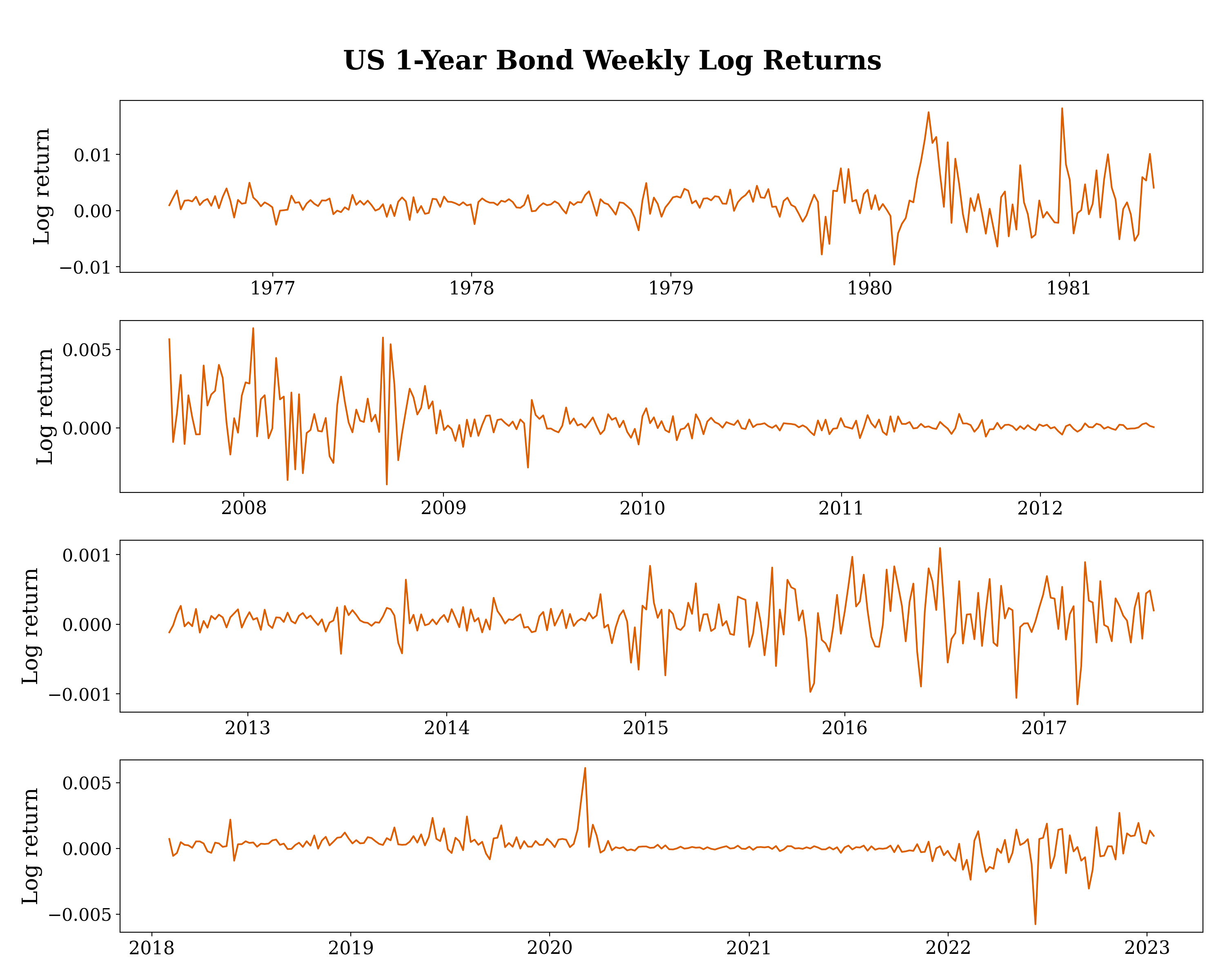}
\includegraphics[width=\linewidth]{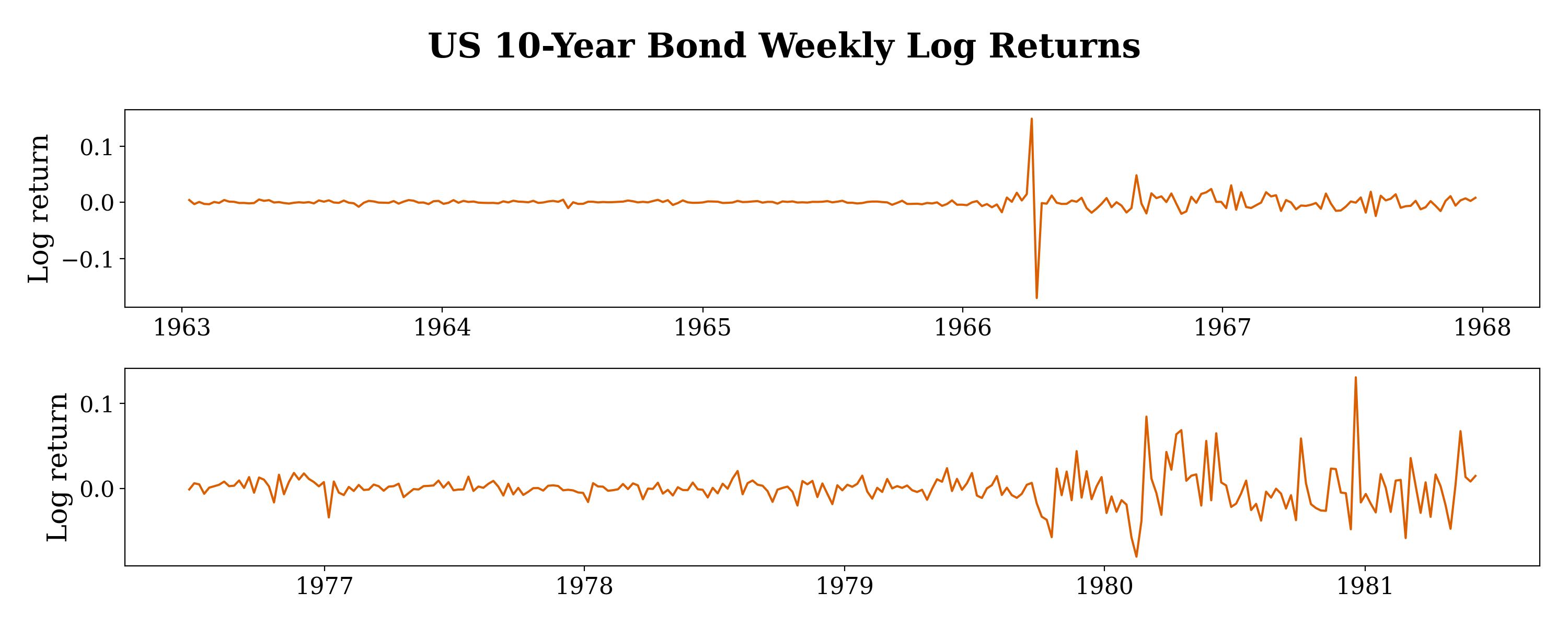}
\bnotefig{This figure shows the U.S. 1-year and 10-year bond log return time-series on local windows around suspected change points.}
\label{F8_2}
\end{figure}

There are examples of time windows where both the single and multiple change-point tests detect change at the 0.05 significance level. 
We collect some of these examples in the top block of Tables \ref{tab:empirical_change_points} and \ref{tab:empirical_change_points_bond}. To gain better intuition for the results, we also plot the return time series for some of these examples. In Figure \ref{F8}, we plot the return time series for the S\&P 500 log returns for periods including major events such as the 1929 Wall Street Crash, World War II in 1940, the 1987 Black Monday Crash, the 2008 Global Financial Crisis, the 2011 August Market Decline, and the 2020 Covid-19 Pandemic. Similarly, for the 1-year and 10-year bond returns, the top block of Table \ref{tab:empirical_change_points_bond} shows examples of time windows where both the single and multiple change-point tests detect change at the 0.05 significance level. In Figure \ref{F8_2}, we plot the return time series for some of these time windows, including the 1980-1982 US recession (due in part to government restrictive monetary policy, and to a lesser degree the Iranian Revolution of 1979, which resulted in significant oil price increases) and the 2008 Financial Crisis. These findings suggest that structural breaks in ES for stock returns closely align with major financial crises. This highlights the sensitivity of tail risk to systemic market disruptions. For bond returns, the detected breaks often coincide with significant shifts in monetary policy, which implies that changes in risk may reflect expectations about interest rate and policy uncertainty.

Interestingly, the change points detected for U.S. 1-year bond returns can differ from those for the U.S. 10-year bond returns. For example, during the period from August 05, 2012 to July 23, 2017, the p-values of both single and multiple-change tests for U.S. 1-year bond returns are less than 0.001, indicating the presence of change points. This is illustrated in the third plot in Figure \ref{F8_2}. However, the p-values of both tests for U.S. 10-year bond returns exceed 0.7, suggesting no significant change points during the same period. This discrepancy might be due to the Federal Reserve's decision in 2015 to raise the federal funds rate, initiating a monetary tightening cycle after the Great Recession. Short-term yields, such as those on 1-year bonds, are directly tied to the federal funds rate and quickly reflect such policy changes. In contrast, long-term yields, such as those on 10-year bonds, are less sensitive to short-term policy changes and more influenced by long-term expectations of future economic growth and inflation. 

%Furthermore, the change points detected with the single- and multiple-change point tests can differ. The middle block in Tables \ref{tab:empirical_change_points} and \ref{tab:empirical_change_points_bond} shows examples where single-change point test are significant, but the multiple change point tests are not, whereas the lower block shows examples where the multiple-change point tests but not the single-change point tests are significant. 

On the other hand, different window sizes and change-point testing methods can yield different results. There are time windows where the single change-point test fails, but the multiple change-point test succeeds; see the bottom block of Table \ref{tab:empirical_change_points} for the S\&P 500 daily returns. One example is the period between May 01, 2008 and April 30, 2009. Here, the p-value is 0.018 for the unsupervised multiple change-point test, with the test statistic (in (\ref{E15})) having value 171.7, which indicates that there are one or more change points during this time window. However, the single change-point test performs poorly on this time window, with a p-value of 0.487 and the test statistic (in (\ref{E11})) having a value of 8.0. As seen in Figure \ref{F8}, changes in the tail of the distribution of the S\&P 500 log returns are clearly present at the end of 2008, corresponding to the 2008 Financial Crisis. Because the single change-point test assumes that only one change point occurs and seeks to divide the time series into two sections separated by a single change point, the effects of multiple changes can be ``canceled out''. The same issue arises for the U.S. 1-year bond log returns in the period from September 19, 1976 to August 31, 1986, and for U.S. 10-year bond log returns in the period from December 18, 1977 to November 29, 1987. This problem is addressed by the multiple change-point test.

However, there is also a trade-off associated with using the multiple change-point test, as it can suffer from low power when there is a single change point and the time window is short. This is supported by the examples in the middle block in Table \ref{tab:empirical_change_points} for the S\&P 500 log returns. In these instances, the single change-point test has p-values that are all less than 0.025, but the multiple change-point test has p-values that are between 0.059 and 0.156. This suggests that the multiple change-point test detects some evidence of a change point, but not enough to achieve statistical significance at the 0.05 level.

In light of the above discussion, we recommend the following use of single and multiple change-point tests. In general, the multiple change-point test can be run first. If the test fails to reach statistical significance, and it is plausible that there is only one change point, then to increase power, the single change-point test can be run (perhaps within a local window). Alternatively, the multiple change-point test can be used to check the result of the single change-point test, and help rule out the possibility that the effects of multiple changes are canceled out. In summary, the single and multiple change-point tests can be deployed in complementary ways.

\section{Conclusion} \label{C}

We propose a novel methodology to perform confidence interval construction and change-point testing for fundamental nonparametric estimators of risk such as ES. This allows for evaluation of the homogeneity of ES and related measures such as the conditional tail moments, and in particular allows the researcher to detect general tail structural changes in time series observations. While current approaches to tail structural change testing typically involve quantities such as the tail index and thus require parametric modeling of the tail, our approach does not require such assumptions. Moreover, we are able to detect more general structural changes in the tail using ES, for example, location and scale changes, which are undetectable using tail index. Hence, we advocate for the use of ES for general purpose monitoring for tail structural change.

We view our methods as being more robust compared to extant approaches which involve consistent estimation of standard errors or blockwise versions of the bootstrap or empirical likelihood, which can be more sensitive to tuning parameters. We note that our proposed sectioning and self-normalization methods for confidence interval construction and change-point testing still require some user choice, for example, the number of sections to use in sectioning or which particular functional to use for self-normalization. Our simulations demonstrate that our methods are robust to these user choices. 

Our simulations show the promising finite-sample performance of our procedures. Our empirical study of S\&P 500 and US Treasury bond returns illustrates the practical use of our methods in detecting and quantifying market instability via the tails of financial time series.

%Furthermore, we are able to construct pointwise ES confidence bands for SPY ETF log returns in the period 2004-2016 and for US 30-Year Treasury log returns in the period 1942-2017 that well capture periods of market distress such as the 2008 Financial Crisis. In similar spirit, our change-point tests are able to detect tail structural changes through the ES for both the SPY ETF and the US 30-Year Treasury log returns during key times of financial instability in the recent past.

\bibliographystyle{plainnat} % or try abbrvnat or unsrtnat
\bibliography{example} % refers to example.bib

\setcounter{table}{0}
\setcounter{figure}{0}
\renewcommand{\thetable}{A.\arabic{table}}
\renewcommand{\thefigure}{A.\arabic{figure}}
 
\section*{Appendix}
\subsection*{Appendix A: Proofs}
In what follows, let $p \in (0,1)$ be a fixed probability level of interest for VaR, ES and their estimators.
To keep notation simple in our proofs, we use the shorthand $q := VaR(p)$, and based on samples $X_1,\dots,X_n$, we set $\widehat{Q}_n := \widehat{VaR}_n$ and $\widehat{F}_n(\cdot)$ denote the empirical distribution function.
We let $\widehat{Q}_{l:m}$ denote the VaR estimator and $\widehat{F}_{l:m}(\cdot)$ the empirical distribution function based on samples $X_l, \dots, X_m$.
For a sample $X_1, X_2, \dots, X_n$, denote the order statistics by $X_{n,1} \le X_{n,2} \le \dots \le X_{n,n}$.

\subsection*{Proof of Theorem \ref{thm1}}

First, for VaR, we show that 
\begin{align}
\sup_{t \in [0,1]} \sqrt{n} t\abs{\widehat{Q}_{[nt]} - q - f(q)^{-1}(\widehat{F}_{[nt]}(q)-p)} = o_P(1). \label{X1}
\end{align}
As in the proof of Theorem 6.2 of \cite{sen1972}, let $\epsilon > 0$, and for every positive integer $k$, define
$$k^* = \abs{q(k^{-1-\epsilon})} + \abs{q(1-k^{-1-\epsilon})} + 1.$$
Consider a sequence $k_n$ of positive integers such that $k_n \to \infty$, but $n^{-1/2}k_n k^*_n \to 0$.
We then have
\begin{align*}
& \sup_{t \in [0,1]} \sqrt{n}t\abs{\widehat{Q}_{[nt]}-q - f(q)^{-1}(\widehat{F}_{[nt]}(q)-p)} \\
& \le \sup_{t \in [0,n^{-1}k_n]} \sqrt{n}t\abs{\widehat{Q}_{[nt]}-q} + \sup_{t \in [0,n^{-1}k_n]} \sqrt{n}t\abs{f(q)^{-1}(\widehat{F}_{[nt]}(q)-p)} \\
& \quad + \sup_{t \in [n^{-1}k_n, 1]} \sqrt{n}t\abs{\widehat{Q}_{[nt]}-q - f(q)^{-1}(\widehat{F}_{[nt]}(q)-p)}.
\end{align*}
For the first term on the right hand side, notice that
$$\sup_{t \in [0,n^{-1}k_n]} \sqrt{n}t\abs{\widehat{Q}_{[nt]}-q} = \max_{1 \le k \le k_n} \frac{k}{\sqrt{n}\sigma}\abs{\widehat{Q}_k - q} \le \frac{k_n}{\sqrt{n}\sigma}(\abs{X_{k_n,k_n} - q} + \abs{X_{k_n,1} - q}).$$
Then, with $n(1-F(X_{n,n}))$ converging in distribution to a standard exponential,
$$\Parg{q \le X_{k_n,k_n} \le q(1-k_n^{-1-\epsilon})} = \Parg{k_n(1-p) \ge k_n(1-F(X_{k_n,k_n})) \ge k_n^{-\epsilon}} \to 1.$$
Similarly, with $nF(X_{n,1})$ converging in distribution to a standard exponential,
$$\Parg{q(k_n^{-1-\epsilon}) \le X_{k_n,1} \le q} = \Parg{k_n^{-\epsilon} \le k_nF(X_{k_n,1}) \le k_n p} \to 1.$$
So asymptotically with probability one, 
$$\abs{q(1-k_n^{-1-\epsilon}) - q} \ge \abs{X_{k_n,k_n} - q}$$
$$\abs{q(k_n^{-1-\epsilon}) - q} \ge \abs{X_{k_n,1} - q}.$$
Thus, asymptotically with probability one,
$$\sup_{t \in [0,n^{-1}k_n]} \sqrt{n}t\abs{\widehat{Q}_{[nt]}-q} \le \frac{k_n}{\sqrt{n}\sigma}(k^*_n + 2\abs{q}) \to 0.$$
Under Assumption \ref{A3} and the functional convergence $\sqrt{n}tf(q)^{-1}(\widehat{F}_{[nt]}(q)-p) \overset{d}{\to} \sigma W(t)$ for some $\sigma \ge 0$ (see, for example, Theorem 1 of \cite{douckhan_etal1994}), we have
$$\sup_{t \in [0,n^{-1}k_n]} \sqrt{n}t\abs{f(q)^{-1}(\widehat{F}_{[nt]}(q)-p)} \overset{P}{\to} 0.$$
Lastly, by the Bahadur representation for $\widehat{Q}_n$ in Theorem 1 of \cite{wendler2011}, which holds under Assumption \ref{A3},
\begin{align*}
& \sup_{t \in [n^{-1}k_n, 1]} \sqrt{n}t\abs{\widehat{Q}_{[nt]}-q - f(q)^{-1}(\widehat{F}_{[nt]}(q)-p)} \\
& = \max_{k_n \le k \le n} \frac{1}{\sqrt{n}}k \abs{\widehat{Q}_k - q - f(q)^{-1}(\widehat{F}_k(q)-p)} \\
& = \max_{k_n \le k \le n} \frac{1}{\sqrt{n}}k o_{a.s.}(k^{-5/8} (\log k)^{3/4} (\log \log k)^{1/2}) \\
& = o_{a.s.}(1),
\end{align*}
and we have established (\ref{X1}).

Next, for ES, we show that
\begin{align}\sup_{t \in [0,1]} \frac{1}{\sqrt{n}} \abs{\frac{1}{1-p} \sum_{k=1}^{[nt]} X_k \indic{\widehat{Q}_{[nt]} \le X_k} - \biggl([nt]q + \frac{1}{1-p} \sum_{k=1}^{[nt]} [X_k - q]_+ \biggr)} = o_P(1). \label{X2}
\end{align}
Using (\ref{res3}) from the proof of Proposition \ref{prop1}, we have
\begin{align*}
& \sup_{t \in [0,1]} \frac{1}{\sqrt{n}} \abs{\frac{1}{1-p} \sum_{k=1}^{[nt]} X_k \indic{\widehat{Q}_{[nt]} \le X_k} - \biggl([nt]q + \frac{1}{1-p} \sum_{k=1}^{[nt]} [X_k - q]_+ \biggr)} \\
& \qquad \le \sup_{t \in [0,1]} \frac{1}{1-p}\frac{[nt]}{\sqrt{n}}\left(\abs{q}\abs{\widehat{F}_{[nt]}(\widehat{Q}_{[nt]}) - p} + \abs{\widehat{Q}_{[nt]} - q} \left(3\abs{\widehat{F}_{[nt]}(\widehat{Q}_{[nt]}) - p} + \abs{\widehat{F}_{[nt]}(q) - p}\right)\right) \\
& \qquad = \sup_{1 \le k \le n} \frac{1}{1-p}\frac{k}{\sqrt{n}}\left(\abs{q}\abs{\widehat{F}_{k}(\widehat{Q}_{k}) - p} + \abs{\widehat{Q}_{k} - q} \left(3\abs{\widehat{F}_{k}(\widehat{Q}_{k}) - p} + \abs{\widehat{F}_{k}(q) - p}\right)\right).
\end{align*}
Observe that for some $C > 0$, we have
$$\sup_{1 \le k \le n} \frac{1}{1-p}\frac{k}{\sqrt{n}} \abs{q}\abs{\widehat{F}_{k}(\widehat{Q}_{k}) - p} \overset{a.s.}{\le} C\frac{1}{1-p}\frac{1}{\sqrt{n}}\abs{q} = O(n^{-1/2}).$$
Let $k_n \to \infty$ be the sequence as used above in the proof of (\ref{X1}) for VaR.
Then observe that
\begin{align*}
& \sup_{1 \le k \le n} \frac{1}{1-p}\frac{k}{\sqrt{n}} \abs{\widehat{Q}_{k} - q} \left(3\abs{\widehat{F}_{k}(\widehat{Q}_{k}) - p} + \abs{\widehat{F}_{k}(q) - p}\right) \\
& \le \sup_{1 \le k \le k_n} \frac{4}{1-p}\frac{k}{\sqrt{n}} \abs{\widehat{Q}_{k} - q} + \sup_{k_n \le k \le n} \frac{1}{1-p}\frac{k}{\sqrt{n}} \abs{\widehat{Q}_{k} - q} \left(3\abs{\widehat{F}_{k}(\widehat{Q}_{k}) - p} + \abs{\widehat{F}_{k}(q) - p}\right).
\end{align*}
As was shown in the proof of (\ref{X1}) for VaR,
$$\sup_{1 \le k \le k_n} \frac{4}{1-p}\frac{k}{\sqrt{n}} \abs{\widehat{Q}_{k} - q} = o_{a.s.}(1).$$
Finally, using the Bahadur representation from Proposition \ref{prop1}, we have
\begin{align*}
& \sup_{k_n \le k \le n} \frac{1}{1-p}\frac{k}{\sqrt{n}} \abs{\widehat{Q}_{k} - q} \left(3\abs{\widehat{F}_{k}(\widehat{Q}_{k}) - p} + \abs{\widehat{F}_{k}(q) - p}\right) \\
& \qquad \le \sup_{k_n \le k \le n} \frac{1}{1-p}\frac{k}{\sqrt{n}} o_{a.s.}(k^{-1 + 1/(2a) + \gamma'} \log k) \\
& \qquad \le \sup_{k_n \le k \le n} \frac{1}{1-p} o_{a.s.}(k^{-1/2 + 1/(2a) + \gamma'} \log k) \\
& \qquad = o_{a.s.}(1).
\end{align*}
The last equality is due to $-1/2 + 1/(2a) + \gamma' < 0$, as in the proof of Proposition \ref{prop1}.
We have thus shown the stronger version of (\ref{X2}):
\begin{align}
\sup_{t \in [0,1]} \frac{1}{\sqrt{n}} \abs{\frac{1}{1-p} \sum_{k=1}^{[nt]} X_k \indic{\widehat{Q}_{[nt]} \le X_k} - \biggl([nt]q + \frac{1}{1-p} \sum_{k=1}^{[nt]} [X_k - q]_+ \biggr)} = o_{a.s.}(1). \label{es_bahadur}
\end{align}

By Theorem 1 of \cite{douckhan_etal1994} (and the remark following the theorem), under Assumption \ref{A3}, the bivariate process
\begin{align}
    \left\{ \frac{1}{\sqrt{n}} \sum_{k=1}^{[nt]}\begin{bmatrix} f(q)^{-1}(\indic{X_k \le q}-p) \\ (1-p)^{-1}([X_k - q]_+ - \Earg{[X_k - q]_+}) \end{bmatrix} : t \in [0,1]\right\} \label{bivariate1}    
\end{align}
converges in distribution in $D[0,1] \times D[0,1]$ to $\Sigma (W_1,W_2)$, where $W_1$ and $W_2$ are independent standard Brownian motions and $\Sigma \in \mathbb{R}^{2\times 2}$ is a positive semidefinite matrix with components (\ref{sigma_11})-(\ref{sigma_22}).
In light of (\ref{X1}) and (\ref{X2}), the proof is complete.

\subsection*{Proof of Proposition \ref{prop1}}
First, we have
\begin{align}
& \abs{\frac{1}{1-p} \frac{1}{n} \sum_{k=1}^n X_k \indic{\widehat{Q}_n \le X_k} - \biggl( \widehat{Q}_n + \frac{1}{1-p} \frac{1}{n} \sum_{k=1}^n [X_k - \widehat{Q}_n]_+ \biggr)} \nonumber \\
& \qquad \le \frac{1}{1-p} \abs{\widehat{Q}_n} \abs{\widehat{F}_n(\widehat{Q}_n) - p}. \label{res4}
\end{align}
Next, observe the following.
\begin{align}
& \abs{\biggl(\widehat{Q}_n + \frac{1}{1-p} \frac{1}{n} \sum_{k=1}^n [X_k - \widehat{Q}_n]_+ \biggr) - \biggl(q + \frac{1}{1-p} \frac{1}{n} \sum_{k=1}^n [X_k - q]_+ \biggr)} \nonumber \\
& = \abs{(\widehat{Q}_n - q) + \frac{1}{1-p} \frac{1}{n} \sum_{k=1}^n (q - \widehat{Q}_n) \indic{\widehat{Q}_n \le X_k} + \frac{1}{1-p} \frac{1}{n} \sum_{k=1}^n (X_k - q) \left(\indic{\widehat{Q}_n \le X_k} - \indic{q \le X_k} \right)} \nonumber \\
& \overset{a.s.}{=} \abs{\frac{1}{1-p}(\widehat{Q}_n - q)(\widehat{F}_n(\widehat{Q}_n) - p) + \frac{1}{1-p} \frac{1}{n} \sum_{k=1}^n (X_k - q)\left(\indic{\widehat{Q}_n \le X_k} - \indic{q \le X_k}\right)} \nonumber \\
& \le \frac{1}{1-p}\abs{\widehat{Q}_n - q}\abs{\widehat{F}_n(\widehat{Q}_n) - p} + \abs{\frac{1}{1-p} \frac{1}{n} \sum_{k=1}^n (X_k - q)\left(\indic{\widehat{Q}_n \le X_k} - \indic{q \le X_k}\right)} \nonumber \\
& \le \frac{1}{1-p}\abs{\widehat{Q}_n - q}\abs{\widehat{F}_n(\widehat{Q}_n) - p} + \frac{1}{1-p} \abs{\widehat{Q}_n -q} \abs{\frac{1}{n} \sum_{k=1}^n \indic{\widehat{Q}_n \le X_k} - \frac{1}{n} \sum_{k=1}^n \indic{q \le X_k}} \nonumber \\
& \overset{a.s.}{=} \frac{1}{1-p}\abs{\widehat{Q}_n - q}\abs{\widehat{F}_n(\widehat{Q}_n) - p} + \frac{1}{1-p} \abs{\widehat{Q}_n -q} \abs{\widehat{F}_n(\widehat{Q}_n) - \widehat{F}_n(q)} \nonumber \\
& \le \frac{1}{1-p} \abs{\widehat{Q}_n - q} \left(2\abs{\widehat{F}_n(\widehat{Q}_n) - p} + \abs{\widehat{F}_n(q) - p}\right) \label{res5}
\end{align}
Under the assumptions of Proposition \ref{prop1}, $\abs{\widehat{F}_n(\widehat{Q}_n) - p} = O_{a.s.}(n^{-1})$, which can be seen via the following arguments.
Let $\mathcal{O}$ denote the neighborhood of $q$ in Assumption \ref{A3} (iii).
\begin{align}
\abs{\widehat{F}_n(\widehat{Q}_n) - p} & \le n^{-1} \sum_{k=1}^n \indic{\widehat{Q}_n = X_k} \nonumber \\
& \le n^{-1} + \indic{\widehat{Q}_n = X_i = X_j, i \ne j} \nonumber \\
& = n^{-1} + \indic{\widehat{Q}_n = X_i = X_j, i \ne j, \widehat{Q}_n  \in \mathcal{O}} + \indic{\widehat{Q}_n = X_i = X_j, i \ne j, \widehat{Q}_n  \notin \mathcal{O}} \nonumber \\
& \le n^{-1} + \indic{X_i = X_j, i \ne j, X_i \in \mathcal{O}} + \indic{\widehat{Q}_n \notin \mathcal{O}} \nonumber \\
& \overset{a.s.}{=} n^{-1} + \indic{\widehat{Q}_n \notin \mathcal{O}} \label{res1} \\
& \overset{a.s.}{=} n^{-1} \label{res2}
\end{align}
In the above, (\ref{res1}) is due to Assumption \ref{A3} (iii), and (\ref{res2}) holds for sufficiently large $n$ due to strong consistency of $\widehat{Q}_n$ under the assumptions of Proposition \ref{prop1}, as established in Theorem 2.1 of \cite{xing_etal2012}.
Also, by Theorem 2.1 of \cite{xing_etal2012}, $\abs{\widehat{Q}_n - q} = o_{a.s.}(n^{-1/2} \log n)$.
\\ \\
Now we examine the term $\abs{\widehat{F}_n(q) - p}$.
Let $\gamma' > 0$ such that $-1/2 + 1/(2a) + \gamma' < 0$.
Note that
\begin{align*}
\Parg{\abs{\widehat{F}_n(q) - p} > n^{-1/2 + 1/(2a) + \gamma'}} & = \Parg{\abs{\sum_{k=1}^n (\indic{X_k \le q} - p)} > n^{1/2 + 1/(2a) + \gamma'}} \\
& \le \frac{\Earg{\abs{\sum_{k=1}^n (\indic{X_k \le q} - p)}^{2a}}}{n^{a + 1 + 2a\gamma'}} \\
& \le \frac{K n^{a}}{n^{a + 1 + 2a\gamma'}} \\
& = K n^{-1-2a\gamma'}.
\end{align*}
The second inequality above is due to Theorem 4.1 of \cite{shao_etal1996}, where $K$ is a global constant depending only on $C > 0$ and $a > 1$ ensuring that $\alpha(n) \le C n^{-a}$ for all $n$.
Since $\sum_{n=1}^\infty K n^{-1-2a\gamma'} < \infty$,
the Borel-Cantelli Lemma yields:
$$\abs{\widehat{F}_n(q) - p} = O_{a.s.}(n^{-1/2 + 1/(2a) + \gamma'}).$$
Finally, putting together (\ref{res4}) and (\ref{res5}) yields the following.
\begin{align}
& \abs{\frac{1}{1-p} \frac{1}{n} \sum_{k=1}^n X_k \indic{\widehat{Q}_n \le X_k} - \biggl(q + \frac{1}{1-p} \frac{1}{n} \sum_{k=1}^n [X_k - q]_+ \biggr)} \nonumber \\
& \qquad \le \frac{1}{1-p}\left(\abs{\widehat{Q}_n}\abs{\widehat{F}_n(\widehat{Q}_n) - p} + \abs{\widehat{Q}_n - q} \left(2\abs{\widehat{F}_n(\widehat{Q}_n) - p} + \abs{\widehat{F}_n(q) - p}\right)\right) \nonumber \\
& \qquad \le \frac{1}{1-p}\left(\abs{q}\abs{\widehat{F}_n(\widehat{Q}_n) - p} + \abs{\widehat{Q}_n - q}\abs{\widehat{F}_n(\widehat{Q}_n) - p} + \abs{\widehat{Q}_n - q} \left(2\abs{\widehat{F}_n(\widehat{Q}_n) - p} + \abs{\widehat{F}_n(q) - p}\right)\right) \nonumber \\
& \qquad = \frac{1}{1-p}\left(\abs{q}\abs{\widehat{F}_n(\widehat{Q}_n) - p} + \abs{\widehat{Q}_n - q} \left(3\abs{\widehat{F}_n(\widehat{Q}_n) - p} + \abs{\widehat{F}_n(q) - p}\right)\right) \label{res3} \\
& \qquad = o_{a.s.}(n^{-1 + 1/(2a) + \gamma'} \log n) \nonumber
\end{align}

\subsection*{Proof of Theorem \ref{thm2}} 
Recall the definition of the index set $\Delta = \{ s,t \in [0,1], t-s \ge \delta \}$, with $\delta > 0$.
First, for VaR, we show that
\begin{align}
\sup_{(s,t) \in \Delta} \sqrt{n}\abs{\widehat{Q}_{[ns]+1:[nt]} - q - f(q)^{-1}(\widehat{F}_{[ns]+1:[nt]}(q)-p)} = o_P(1). \label{var_st}
\end{align}
%We apply Theorem 2.5 of \cite{volgushev_etal2014}.
%To check that the required conditions of their theorem hold, we note the following.
By Theorem 1 of \cite{bucher2015}, the sequential empirical process
$$\left\{ \frac{1}{\sqrt{n}} \sum_{i=1}^{[nt]} (\indic{X_i \le x} - F(x)) : (t,x) \in [0,1] \times \mathbb{R} \right\}$$
converges in distribution in $\ell^\infty([0,1] \times \mathbb{R})$ to a Gaussian process $\{Y(t,x) : (t,x) \in [0,1] \times \mathbb{R}\}$.
%By Addendum 1.5.8 of \cite{vandervaart_etal1996}, almost surely, $Y$ has uniformly continuous paths, with respect to the Euclidean metric on $[0,1] \times \mathbb{R}$.
Consider the mapping from $\ell^\infty([0,1] \times \mathbb{R}) \to \ell^\infty(\Delta \times \mathbb{R})$ given by $\{Y(t,x) : (t,x) \in [0,1] \times \mathbb{R} \} \mapsto \{ Y(t,x) - Y(s,x) : (s,t,x) \in \Delta \times \mathbb{R} \}$. 
This is a continuous mapping when both spaces are endowed with their uniform topologies.
Then, the Continuous Mapping Theorem yields convergence in distribution of the process
\begin{align*}
    \left\{ \frac{1}{\sqrt{n}} \sum_{k=[ns]+1}^{[nt]} (\indic{X_i \le x} - F(x)) : ((s,t),x) \in \Delta \times \mathbb{R} \right\}
\end{align*}
in $\ell^\infty(\Delta \times \mathbb{R})$ to a Gaussian process $\{Z(s,t,x) : (s,t,x) \in \Delta \times \mathbb{R} \}$.
As discussed in Example 3.9.21 of \cite{vandervaart_etal1996}, for fixed $p \in (0,1)$, the mapping from a distribution function to its VaR at level $p$: $F \mapsto F^{-1}(p)$ is Hadamard-differentiable at every distribution function $F$ that is differentiable at $F^{-1}(p)$ with strictly positive derivative $f(F^{-1}(p))$, tangentially to the set of functions that are continuous at $F^{-1}(p)$.
By the Functional Delta Method (see, for example, Theorem 3.9.4 of \cite{vandervaart_etal1996}), (\ref{var_st}) is established.

Next, for ES, we show that
\begin{align}
    \sup_{(s,t) \in \Delta} \frac{1}{\sqrt{n}} \abs{ \frac{1}{1-p} \sum_{k=[ns]+1}^{[nt]} X_k \indic{X_k \ge \widehat{Q}_{[ns]+1:[nt]}} - \Biggl(([nt]-[ns])q + \frac{1}{1-p} \sum_{k=[ns]+1}^{[nt]} [X_k - q]_+ \Biggr)} = o_P(1). \label{es_st}
\end{align}
Using (\ref{res3}) from the proof of Proposition \ref{prop1}, we have
\begin{align*}
& \sup_{(s,t) \in \Delta} \frac{1}{\sqrt{n}} \abs{ \frac{1}{1-p} \sum_{k=[ns]+1}^{[nt]} X_k \indic{X_k \ge \widehat{Q}_{[ns]+1:[nt]}} - \Biggl(([nt]-[ns])q + \frac{1}{1-p} \sum_{k=[ns]+1}^{[nt]} [X_k - q]_+ \Biggr)} \\
 \le & \sup_{(s,t) \in \Delta} \frac{1}{1-p} \frac{[nt]-[ns]}{\sqrt{n}} \Biggl( \abs{q} \abs{\widehat{F}_{[ns]+1:[nt]}(\widehat{Q}_{[ns]+1:[nt]}) - p}  \\
&+ \abs{\widehat{Q}_{[ns]+1:[nt]} - q}\left(3\abs{\widehat{F}_{[ns]+1:[nt]}(\widehat{Q}_{[ns]+1:[nt]}) - p} + \abs{\widehat{F}_{[ns]+1:[nt]}(q) - p} \right) \Biggr).
\end{align*}
By the assumption that $(X_1,X_k)$ has joint density for all $k \ge 2$, there can be no ties among $X_1, X_2, \dots$.
So with $\delta > 0$ from the definition of the index set $\Delta$,
\begin{align*}
    \sup_{(s,t) \in \Delta} \abs{\widehat{F}_{[ns]+1:[nt]}(\widehat{Q}_{[ns]+1:[nt]}) - p} \le \frac{1}{[nt]-[ns]}.
\end{align*}
Hence, it suffices to show, as $n \to \infty$,
\begin{align*}
    \sup_{(s,t) \in \Delta} \frac{[nt]-[ns]}{\sqrt{n}} \abs{\widehat{Q}_{[ns]+1:[nt]} - q} \abs{\widehat{F}_{[ns]+1:[nt]}(q) - p} = o_P(1).
\end{align*}
In light of our above arguments and the result for VaR established above, as $n \to \infty$,
\begin{align*}
    & \sup_{(s,t) \in \Delta} \sqrt{n}(t-s)\abs{\widehat{F}_{[ns]+1:[nt]}(q) - p} = O_P(1) \\
    & \sup_{(s,t) \in \Delta} \sqrt{n}(t-s)\abs{\widehat{Q}_{[ns]+1:[nt]} - q} = O_P(1),
\end{align*}
from which (\ref{es_st}) follows.

Consider the mapping from $\ell^\infty([0,1]) \to \ell^\infty(\Delta)$ given by $\{Y(t) : (t) \in [0,1] \} \mapsto \{ Y(t) - Y(s) : (s,t) \in \Delta \}$.
By the Continuous Mapping Theorem and the convergence in distribution of the process in (\ref{bivariate1}) (from the proof of Theorem \ref{thm1}), we have that
\begin{align}
    \left\{ \frac{1}{\sqrt{n}} \sum_{k=[ns]+1}^{[nt]}\begin{bmatrix} f(q)^{-1}(\indic{X_k \le q}-p) \\ (1-p)^{-1}([X_k - q]_+ - \Earg{[X_k - q]_+}) \end{bmatrix} : (s,t) \in \Delta \right\} \label{bivariate2}    
\end{align}
converges in distribution in $\ell^\infty(\Delta) \times \ell^\infty(\Delta)$ to the process in (\ref{bivariate_limit}).
In light of (\ref{var_st}) and (\ref{es_st}), the proof is complete.

%Recall, under Assumption \ref{A2}, by Theorem 1.7 of \cite{ibragimov1962} and Theorem 0 of \cite{herrndorf1985}, 
%we have the convergence in distribution of the process
%\begin{align*}
%    \left\{n^{-1/2}\biggl([nt]q + \frac{1}{1-p} \sum_{k=1}^{[nt]} [X_k - q]_+ - [nt]\Earg{X \mid X \ge q} \biggr): t \in [0,1] \right\}
%\end{align*}
%to $\sigma_2 W$ in $D[0,1]$ (with $\sigma_2 \ge 0$), with respect to the uniform topology.
%Then, as in the case of VaR, the continuous mapping theorem yields convergence in distribution of the process
%\begin{align*}
%    \left\{n^{-1/2}\biggl(([nt] - [ns])q + \frac{1}{1-p} \sum_{k=[ns]+1}^{[nt]} [X_k - q]_+ - ([nt] - [ns])\Earg{X \mid X \ge q} \biggr): (s,t) \in \Delta \right\}
%\end{align*}
%to $\{\sigma_2 (W(t) - W(s)), \: (s,t) \in \Delta \}$ in $\ell^\infty(\Delta)$, with respect to the uniform topology.

\subsection*{Proof of Proposition \ref{prop2}}
We are concerned with the following variant of the CUSUM process:
$$\left\{\frac{1}{\sqrt{n}}\left(\frac{n-[nt]}{n} \sum_{k=1}^{[nt]} \frac{X_k \indic{\widehat{Q}_{1:[nt]} \le X_k}}{1 - p} - \frac{[nt]}{n} \sum_{k=[nt]+1}^n \frac{X_k \indic{\widehat{Q}_{[nt]+1:n} \le X_k}}{1-p}\right) : t \in [0,1] \right\}.$$
Under Assumption \ref{A3}, using Theorem \ref{thm1}, it is straightforward to obtain convergence in distribution in $D[0,1]$ of the above process to $\sigma' B$, where $B(t) := W(t) - tW(1)$ (for $t \in [0,1]$) is a standard Brownian bridge on $[0,1]$ and $\sigma' \ge 0$.
Note that
$$\left\{\frac{1}{\sqrt{n}}\left(\frac{n-[nt]}{n} \sum_{k=1}^{[nt]} \frac{\max(X_k,q)}{1 - p} - \frac{[nt]}{n} \sum_{k=[nt]+1}^n \frac{\max(X_k,q)}{1 - p}\right) : t \in [0,1] \right\}$$
converges in distribution in $D[0,1]$ to $\sigma' B$.
Then, we get the desired conclusion by observing the following.
\begin{align*}
& \sup_{t \in [0,1]} \frac{1}{\sqrt{n}} \Biggl\vert \frac{n-[nt]}{n} \sum_{k=1}^{[nt]} \frac{X_k \indic{\widehat{Q}_{1:[nt]} \le X_k}}{1 - p} - \frac{[nt]}{n} \sum_{k=[nt]+1}^n \frac{X_k \indic{\widehat{Q}_{[nt]+1:n} \le X_k}}{1-p} \\
& \qquad \qquad \qquad - \frac{n-[nt]}{n} \sum_{k=1}^{[nt]} \frac{\max(X_k,q)}{1 - p} + \frac{[nt]}{n} \sum_{k=[nt]+1}^n \frac{\max(X_k,q)}{1 - p} \Biggr\vert \\
& \qquad \le \sup_{t \in [0,1]} \frac{1}{\sqrt{n}} \abs{\sum_{k=1}^{[nt]} \frac{X_k \indic{\widehat{Q}_{1:[nt]} \le X_k}}{1 - p} - \sum_{k=1}^{[nt]} \frac{\max(X_k,q)-qp}{1 - p}} \\
& \qquad \quad + \sup_{t \in [0,1]} \frac{1}{\sqrt{n}} \abs{\sum_{k=[nt]+1}^n \frac{X_k \indic{\widehat{Q}_{[nt]+1:n} \le X_k}}{1-p} - \sum_{k=[nt]+1}^n \frac{\max(X_k,q) - qp}{1 - p}} \\
& \qquad \le \sup_{t \in [0,1]} \frac{1}{\sqrt{n}} \abs{\sum_{k=1}^{[nt]} \frac{X_k \indic{\widehat{Q}_{1:[nt]} \le X_k}}{1 - p} - \left([nt]q + \sum_{k=1}^{[nt]} \frac{[X_k - q]_+}{1-p} \right)} \\
& \qquad \quad + \sup_{t \in [0,1]} \frac{1}{\sqrt{n}} \abs{\sum_{k=[nt]+1}^n \frac{X_k \indic{\widehat{Q}_{[nt]+1:n} \le X_k}}{1-p} - \left((n-[nt])q + \sum_{k=[nt]+1}^n \frac{[X_k - q]_+}{1-p} \right)} \\
& \qquad = o_p(1)
\end{align*}
The convergence in probability to zero follows by stationarity of the sequence $X_1, X_2, \dots$ and (\ref{es_bahadur}) from the proof of Theorem \ref{thm1}.

\newpage
\subsection*{Appendix B: Simulations}

% {\color{red} This page and next show new results}

% By fitting a GARCH(1,1) model to the daily log returns of the S\&P 500 Index for the years 1928-2023, we obtain estimates of $\lambda_1\approx 0.09$ and $\lambda_2\approx 0.88$.
% \begin{figure}[htbp]
% \tcapfig{Coverage Probability and Confidence Intervals for GARCH(1,1) with i.i.d. Normal Innovations and Parameters $\lambda_1=0.09$ and $\lambda_2=0.88$}
% \centering
% \minipage{0.45\textwidth}
%   \includegraphics[trim={0 2.5cm 0 3.5cm},clip,width=\linewidth]{new_plots/normal_lambda1_0.09_lambda2_0.88.pdf}
% \endminipage
% \minipage{0.45\textwidth}
%   \includegraphics[trim={0 2.5cm 0 3.5cm},clip,width=\linewidth]{new_plots/normal_width_lambda1_0.09_lambda2_0.88.pdf}
% \endminipage
% \bnotefig{This figure shows the coverage probability and confidence interval width for different sample size. We use the GARCH(1,1) process with i.i.d. standard normal innovations, and set the model parameters as $\omega=0.01, \lambda_1=0.09$, and $\lambda_2=0.88$. Left subfigure: relationship between sample size and empirical coverage probability of 95\% confidence intervals for ES in the upper 5th percentile computed for stationary GARCH(1,1) process. Right subfigure: relationship between time series sample size and width of 95\% confidence intervals for ES in the upper 5th percentile computed for stationary GARCH(1,1) process. Each plotted point is the averaged result over 10,000 replications.}
% \end{figure}

\begin{figure}[htbp]
\tcapfig{Coverage Probability and Confidence Intervals for GARCH(1,1) with i.i.d. Normal Innovations and Parameters $\lambda_1=0.05$ and $\lambda_2=0.9$}
\centering
\minipage{0.45\textwidth}
  \includegraphics[trim={0 2.5cm 0 3.5cm},clip,width=\linewidth]{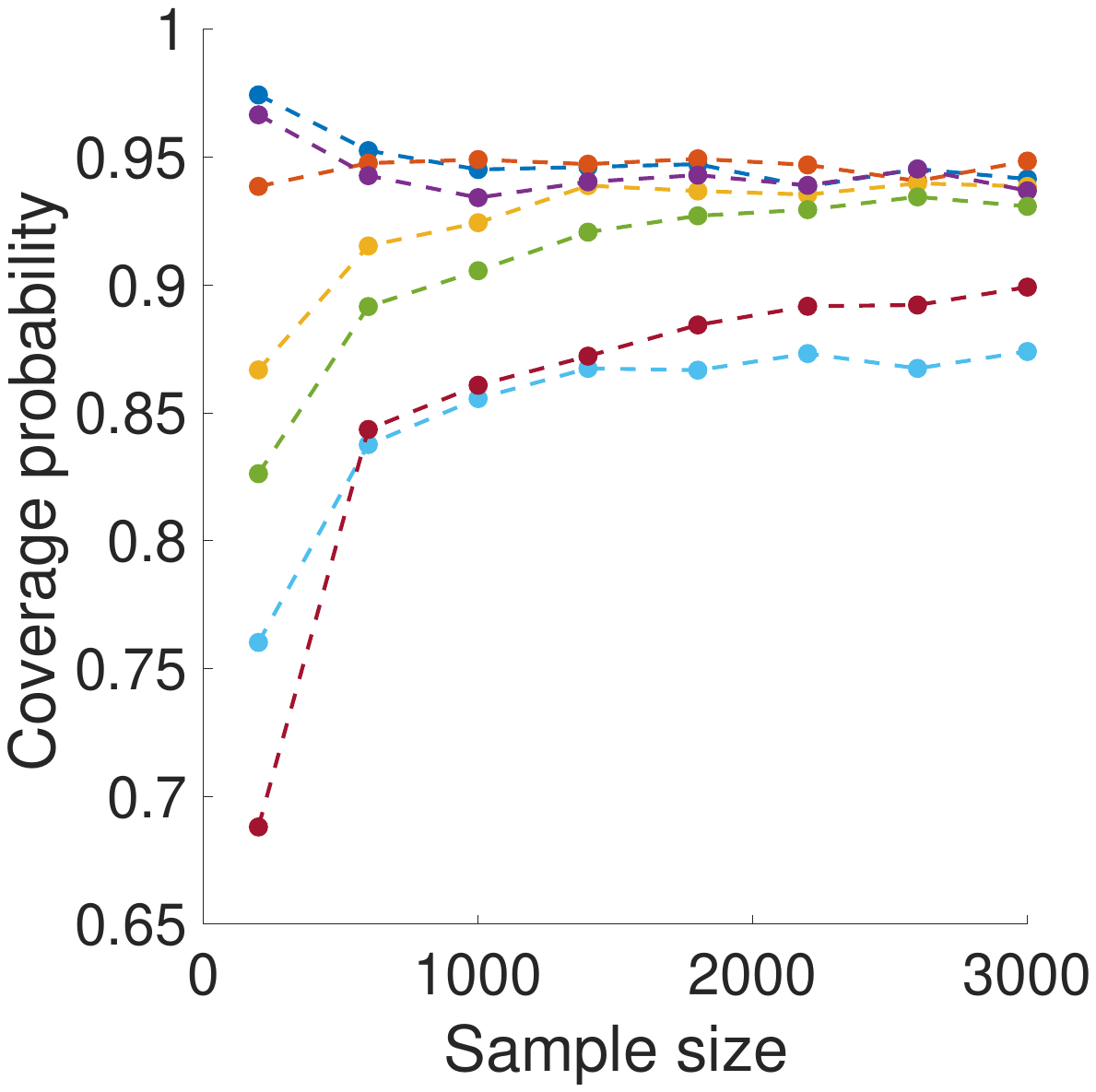}
\endminipage
\minipage{0.45\textwidth}
  \includegraphics[trim={0 2.5cm 0 3.5cm},clip,width=\linewidth]{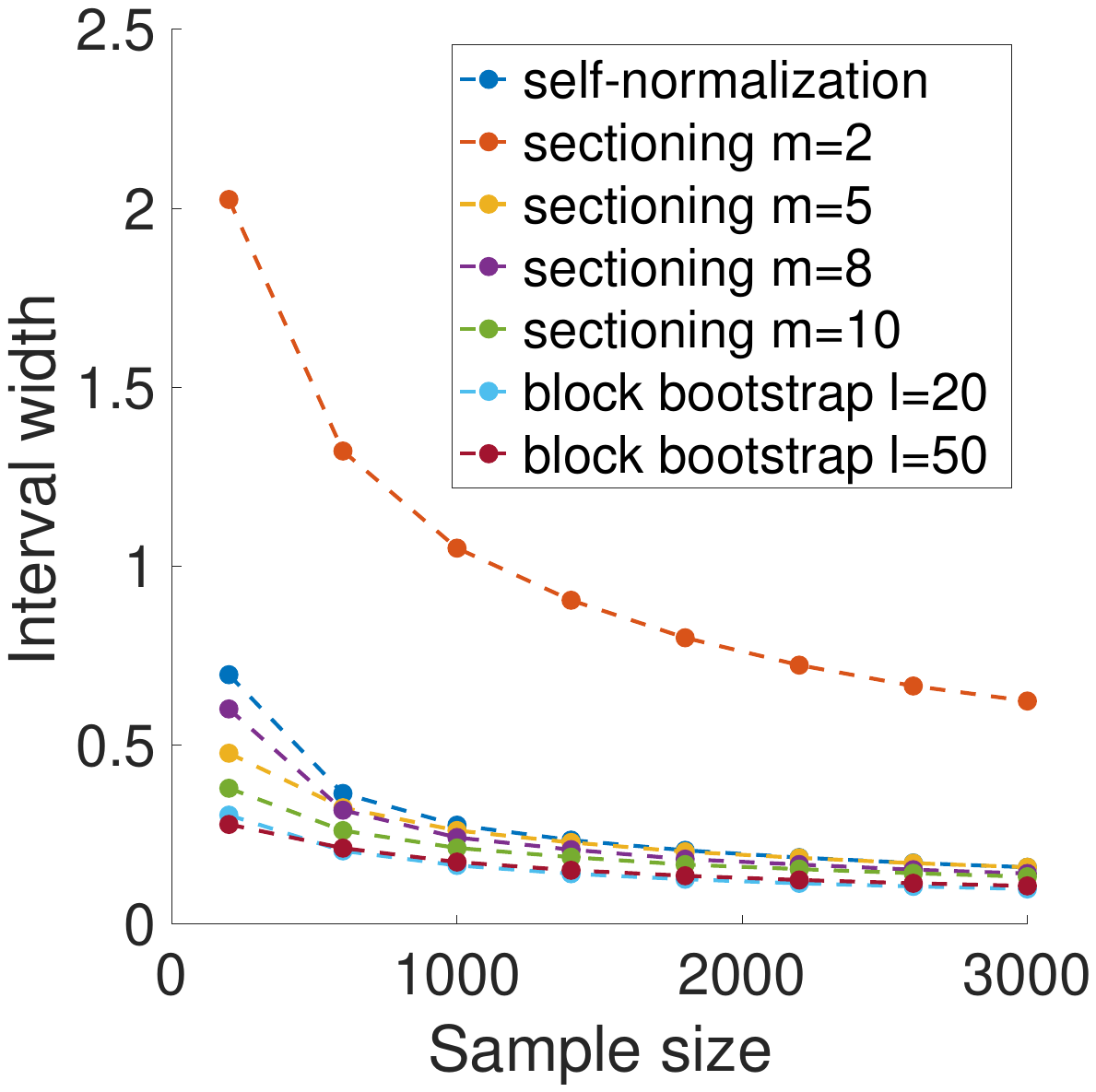}
\endminipage
\bnotefig{This figure shows the coverage probability and confidence interval width for different sample size. We use the GARCH(1,1) process with i.i.d. standard normal innovations, and set the model parameters as $\omega=0.01, \lambda_1=0.05$, and $\lambda_2=0.9$. Left subfigure: relationship between sample size and empirical coverage probability of 95\% confidence intervals for ES in the upper 5th percentile computed for stationary GARCH(1,1) process. Right subfigure: relationship between time series sample size and width of 95\% confidence intervals for ES in the upper 5th percentile computed for stationary GARCH(1,1) process. Each plotted point is the averaged result over 10,000 replications.}
\label{fig:A1}
\end{figure}

\begin{figure}[htbp]
\tcapfig{Coverage Probability and Confidence Intervals for GARCH(1,1) with i.i.d. Normal Innovations and Parameters $\lambda_1=0.1$ and $\lambda_2=0.7$}
\centering
\minipage{0.45\textwidth}
  \includegraphics[trim={0 2.5cm 0 3.5cm},clip,width=\linewidth]{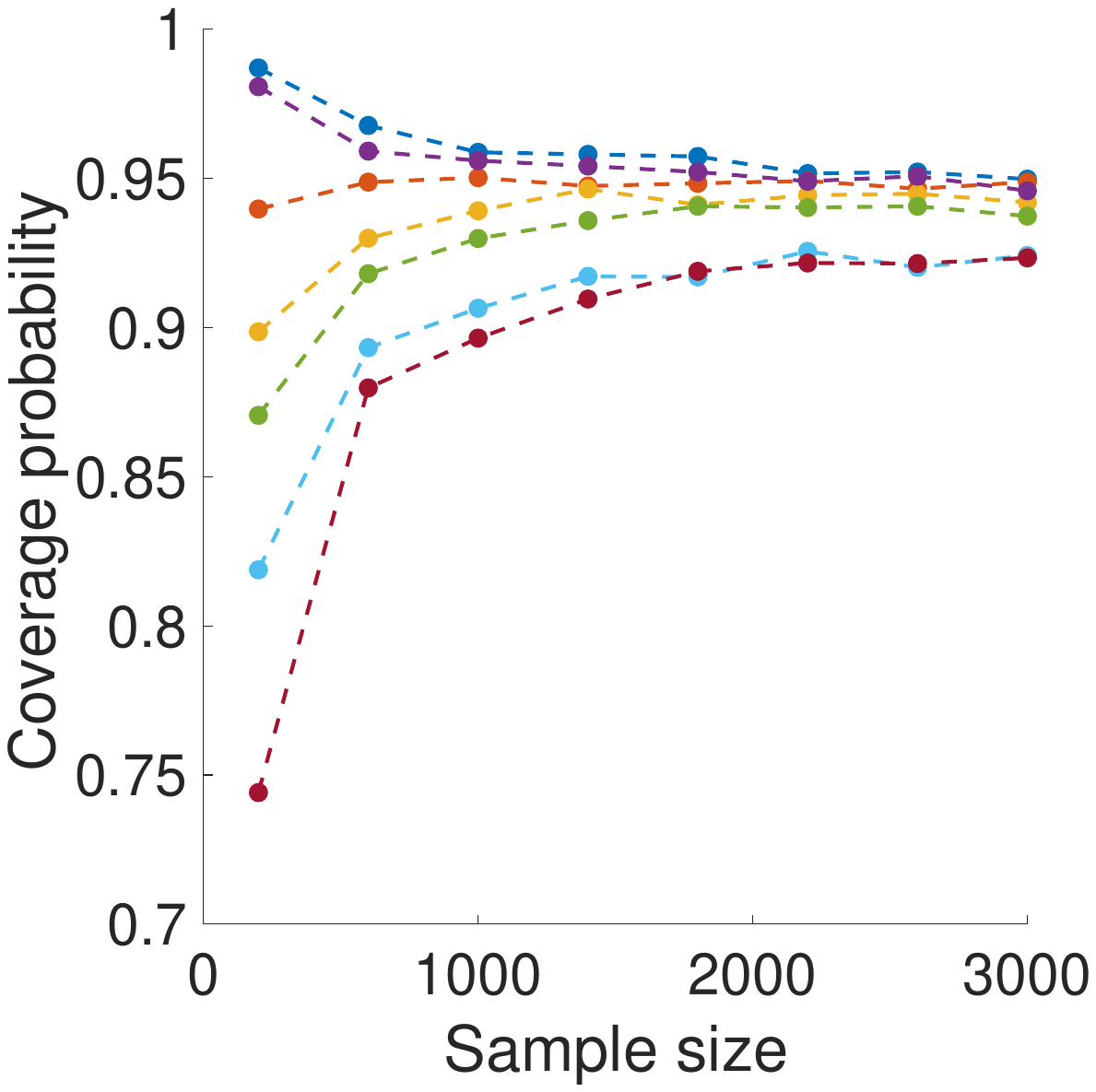}
\endminipage
\minipage{0.45\textwidth}
  \includegraphics[trim={0 2.5cm 0 3.5cm},clip,width=\linewidth]{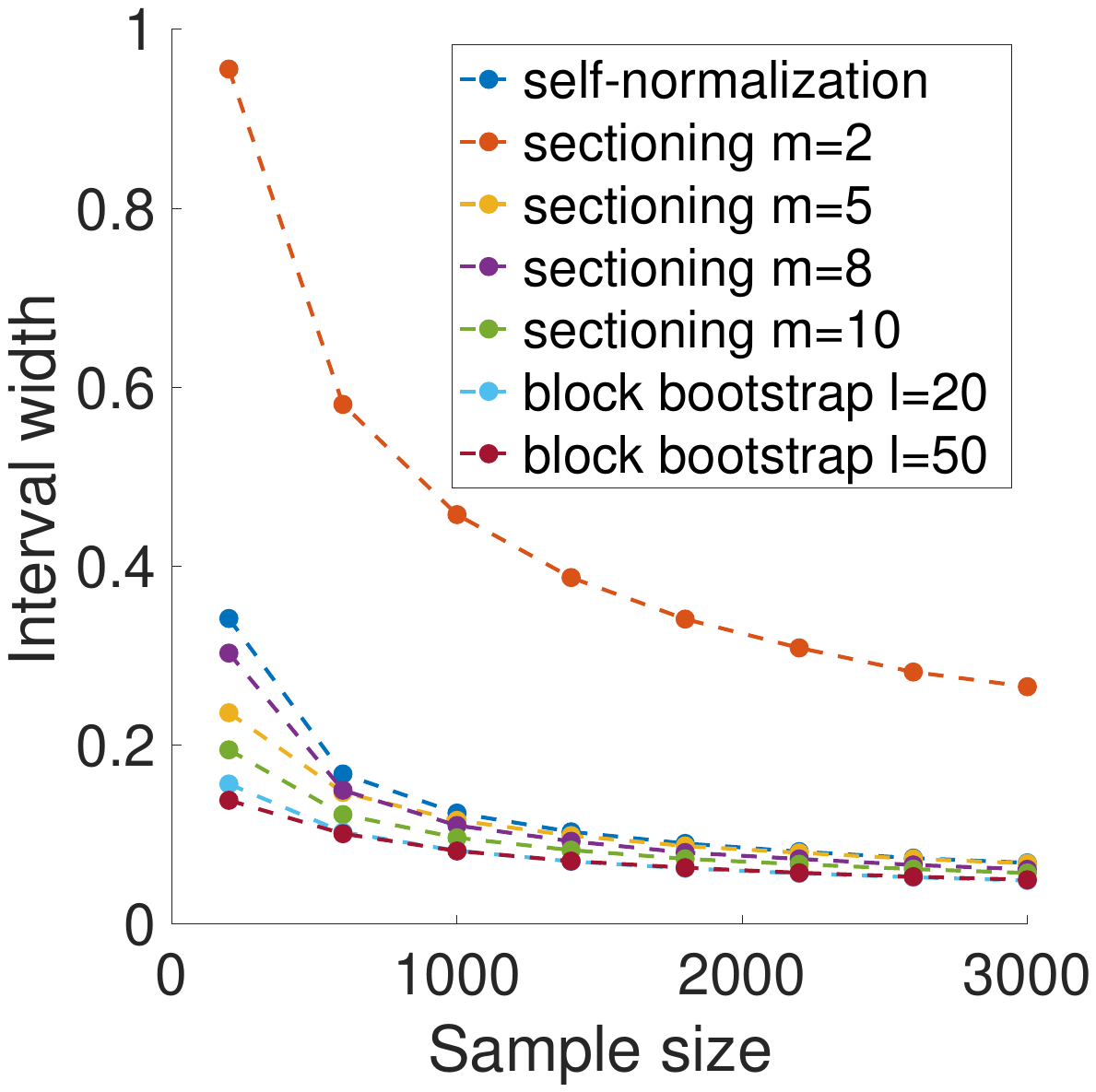}
\endminipage
\bnotefig{This figure shows the coverage probability and confidence interval width for different sample size. We use the GARCH(1,1) process with i.i.d. standard normal innovations, and set the model parameters as $\omega=0.01, \lambda_1=0.1$, and $\lambda_2=0.7$. Left subfigure: relationship between sample size and empirical coverage probability of 95\% confidence intervals for ES in the upper 5th percentile computed for stationary GARCH(1,1) process. Right subfigure: relationship between time series sample size and width of 95\% confidence intervals for ES in the upper 5th percentile computed for stationary GARCH(1,1) process. Each plotted point is the averaged result over 10,000 replications.}
\end{figure}

\begin{figure}[htbp]
\tcapfig{Coverage Probability and Confidence Intervals for GARCH(1,1) with i.i.d. Normal Innovations and Parameters $\lambda_1=0.1$ and $\lambda_2=0.6$}
\centering
\minipage{0.45\textwidth}
  \includegraphics[trim={0 2.5cm 0 3.5cm},clip,width=\linewidth]{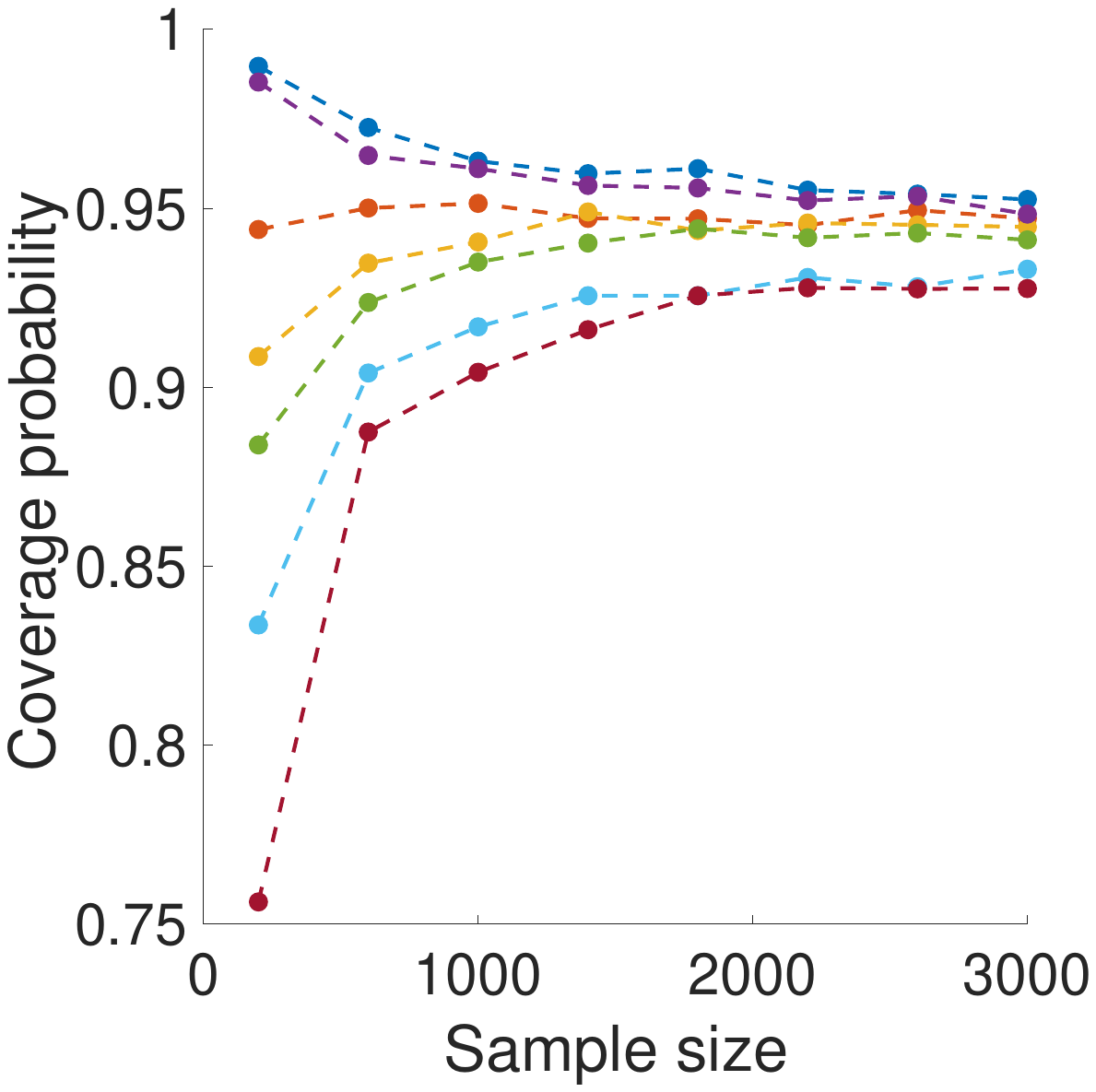}
\endminipage
\minipage{0.45\textwidth}
  \includegraphics[trim={0 2.5cm 0 3.5cm},clip,width=\linewidth]{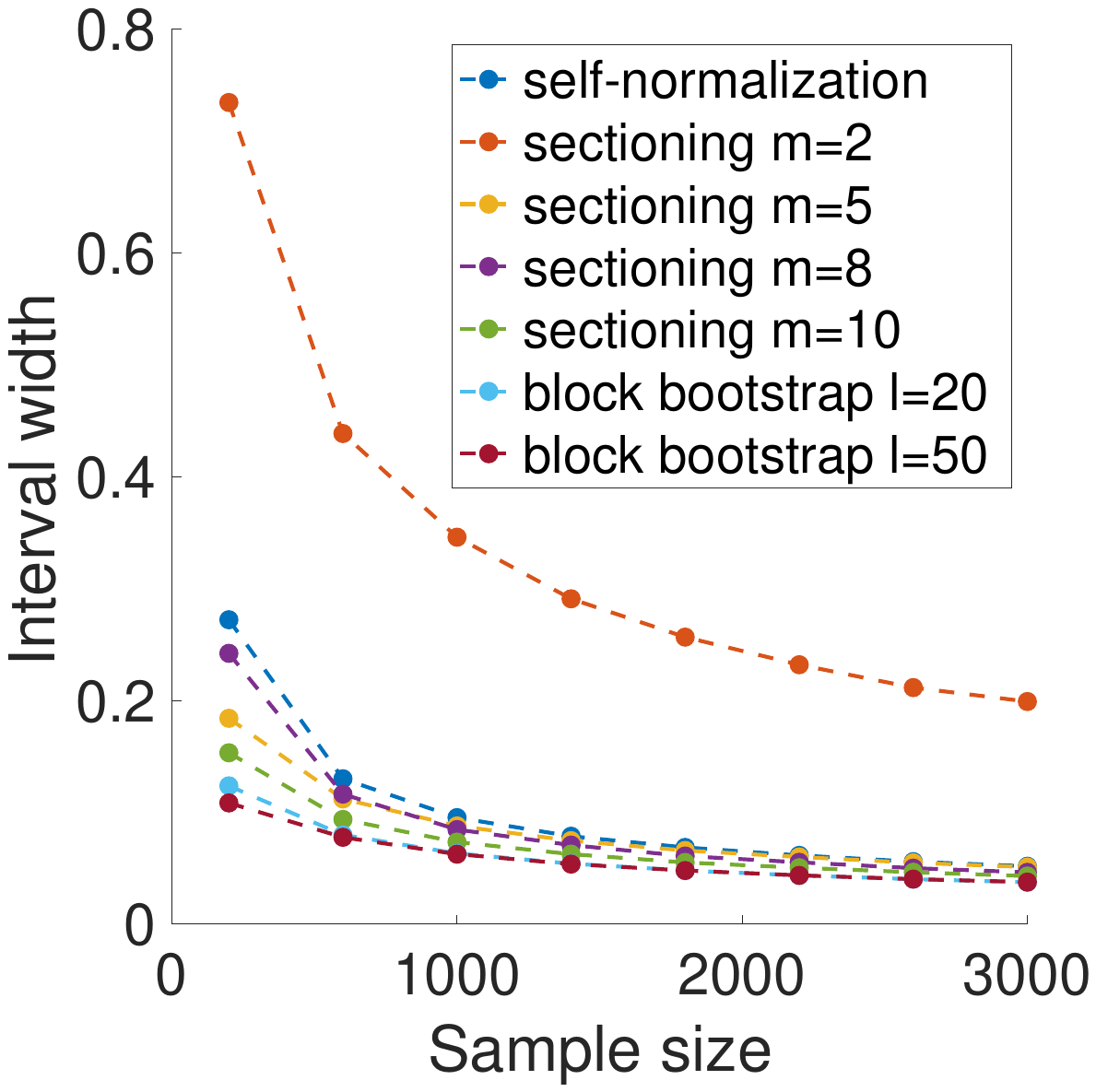}
\endminipage
\bnotefig{This figure shows the coverage probability and confidence interval width for different sample size. We use the GARCH(1,1) process with i.i.d. standard normal innovations, and set the model parameters as $\omega=0.01, \lambda_1=0.1$, and $\lambda_2=0.6$. Left subfigure: relationship between sample size and empirical coverage probability of 95\% confidence intervals for ES in the upper 5th percentile computed for stationary GARCH(1,1) process. Right subfigure: relationship between time series sample size and width of 95\% confidence intervals for ES in the upper 5th percentile computed for stationary GARCH(1,1) process. Each plotted point is the averaged result over 10,000 replications.}
\end{figure}

\begin{figure}[htbp]
\tcapfig{Coverage Probability and Confidence Intervals for GARCH(1,1) with i.i.d. Normal Innovations and Parameters $\lambda_1=0.1$ and $\lambda_2=0.5$}
\centering
\minipage{0.45\textwidth}
  \includegraphics[trim={0 2.5cm 0 3.5cm},clip,width=\linewidth]{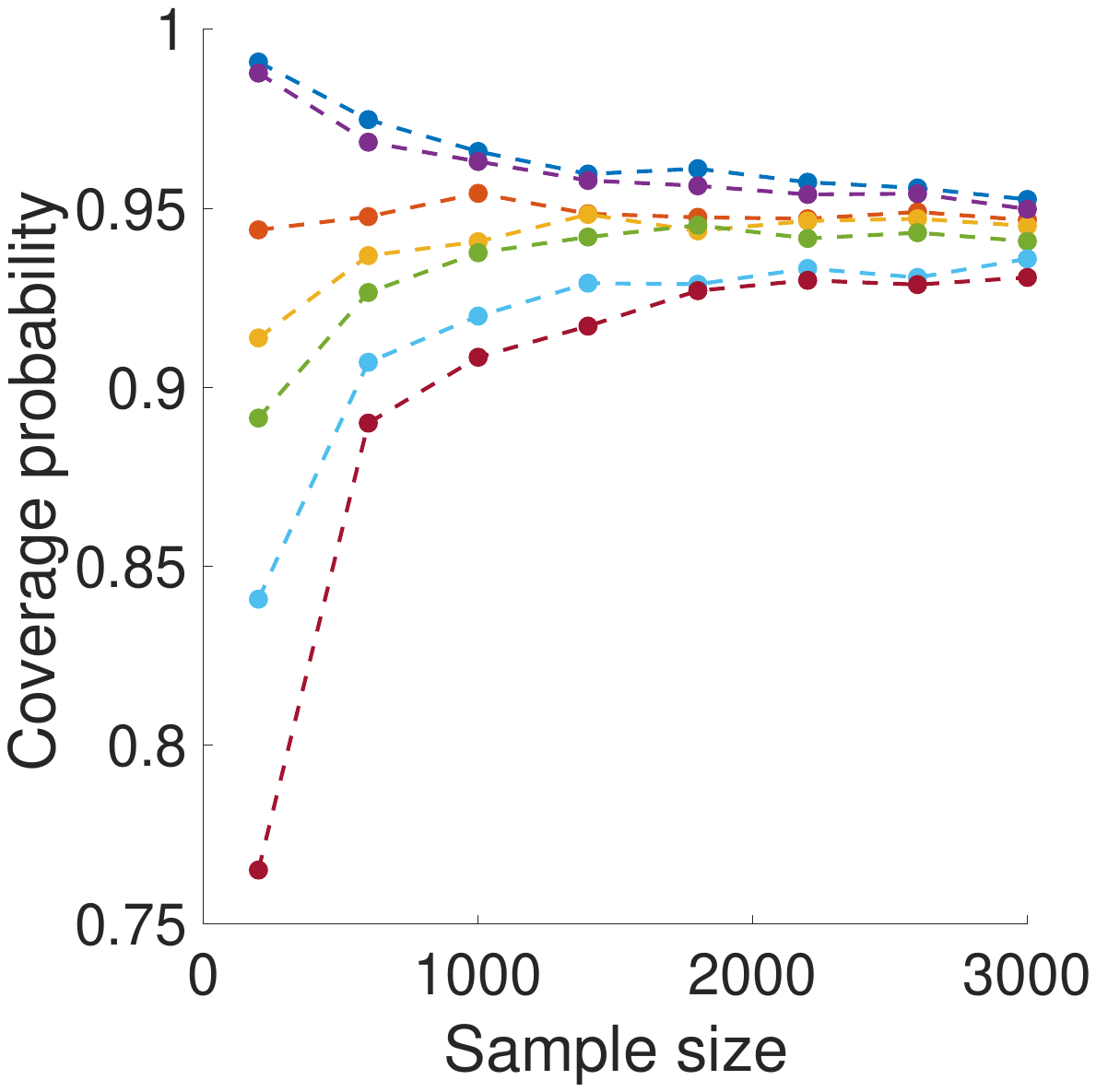}
\endminipage
\minipage{0.45\textwidth}
  \includegraphics[trim={0 2.5cm 0 3.5cm},clip,width=\linewidth]{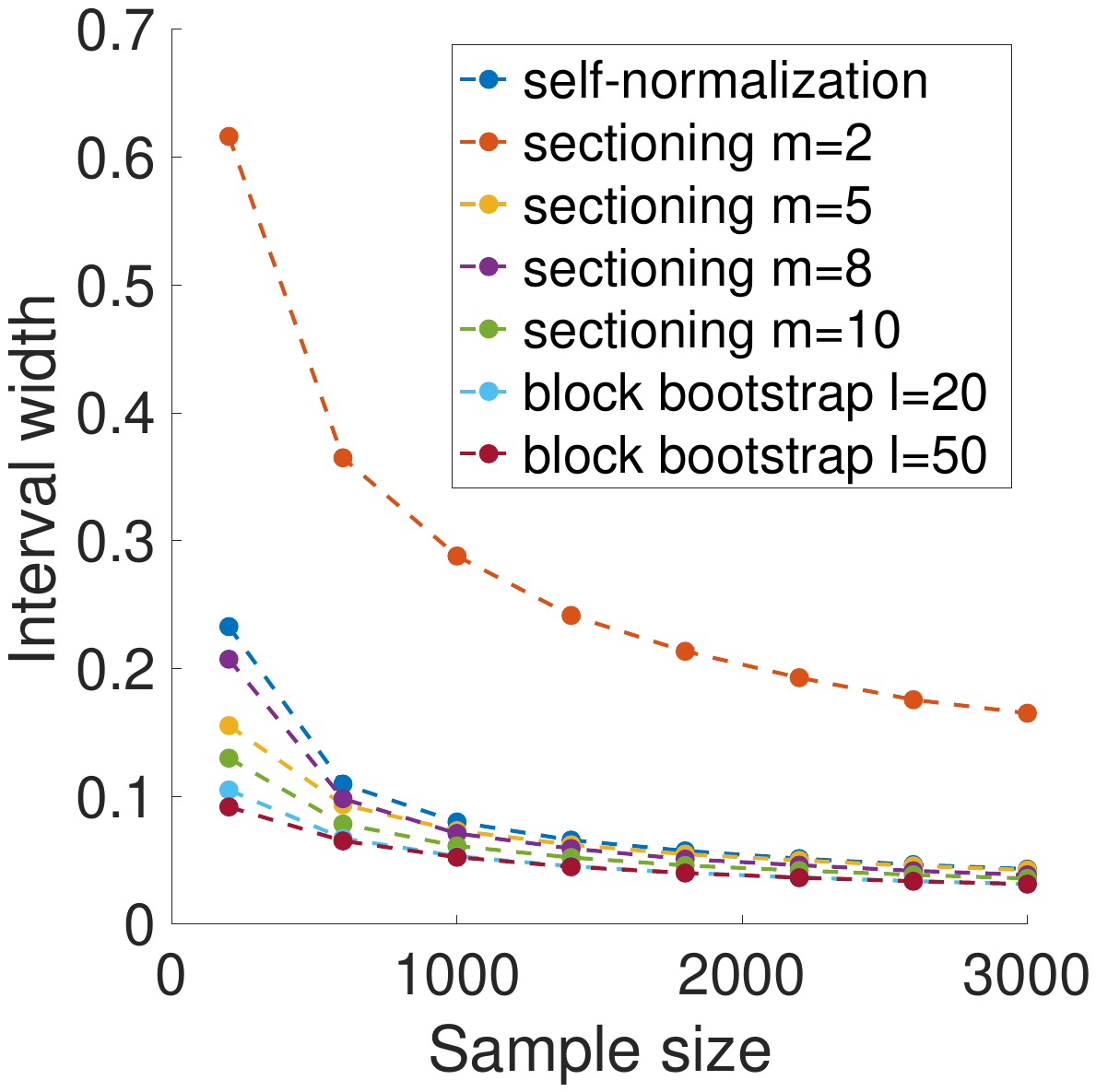}
\endminipage
\bnotefig{This figure shows the coverage probability and confidence interval width for different sample size. We use the GARCH(1,1) process with i.i.d. standard normal innovations, and set the model parameters as $\omega=0.01, \lambda_1=0.1$, and $\lambda_2=0.5$. Left subfigure: relationship between sample size and empirical coverage probability of 95\% confidence intervals for ES in the upper 5th percentile computed for stationary GARCH(1,1) process. Right subfigure: relationship between time series sample size and width of 95\% confidence intervals for ES in the upper 5th percentile computed for stationary GARCH(1,1) process. Each plotted point is the averaged result over 10,000 replications.}
\end{figure}

\begin{figure}[htbp]
\tcapfig{Coverage Probability and Confidence Intervals for GARCH(1,1) with i.i.d. $t(15)$-Distributed Innovations and Parameters $\lambda_1=0.1$ and $\lambda_2=0.8$}
\centering
\minipage{0.45\textwidth}
  \includegraphics[trim={0 2.5cm 0 3.5cm},clip,width=\linewidth]{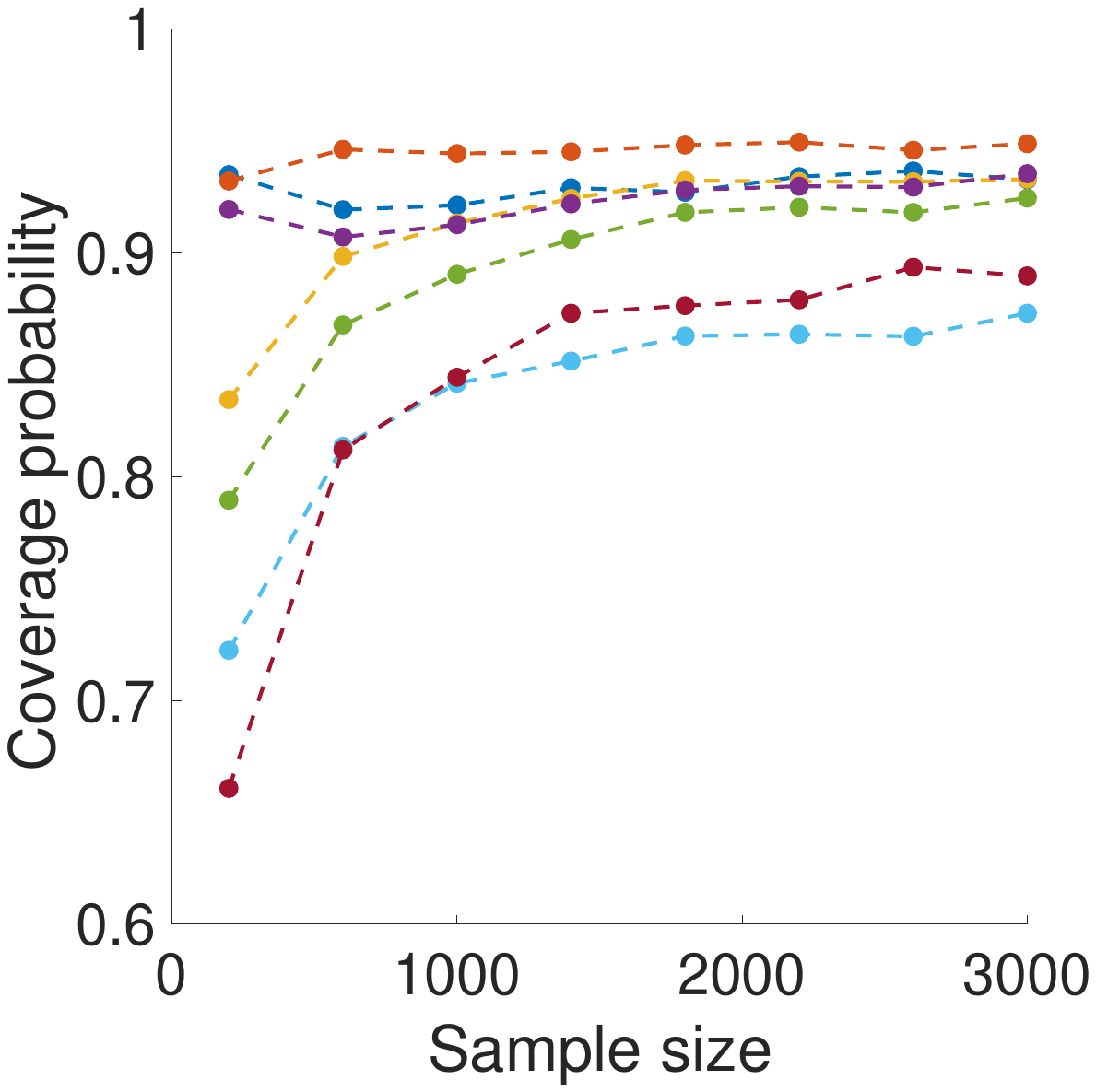}
\endminipage
\minipage{0.45\textwidth}
  \includegraphics[trim={0 2.5cm 0 3.5cm},clip,width=\linewidth]{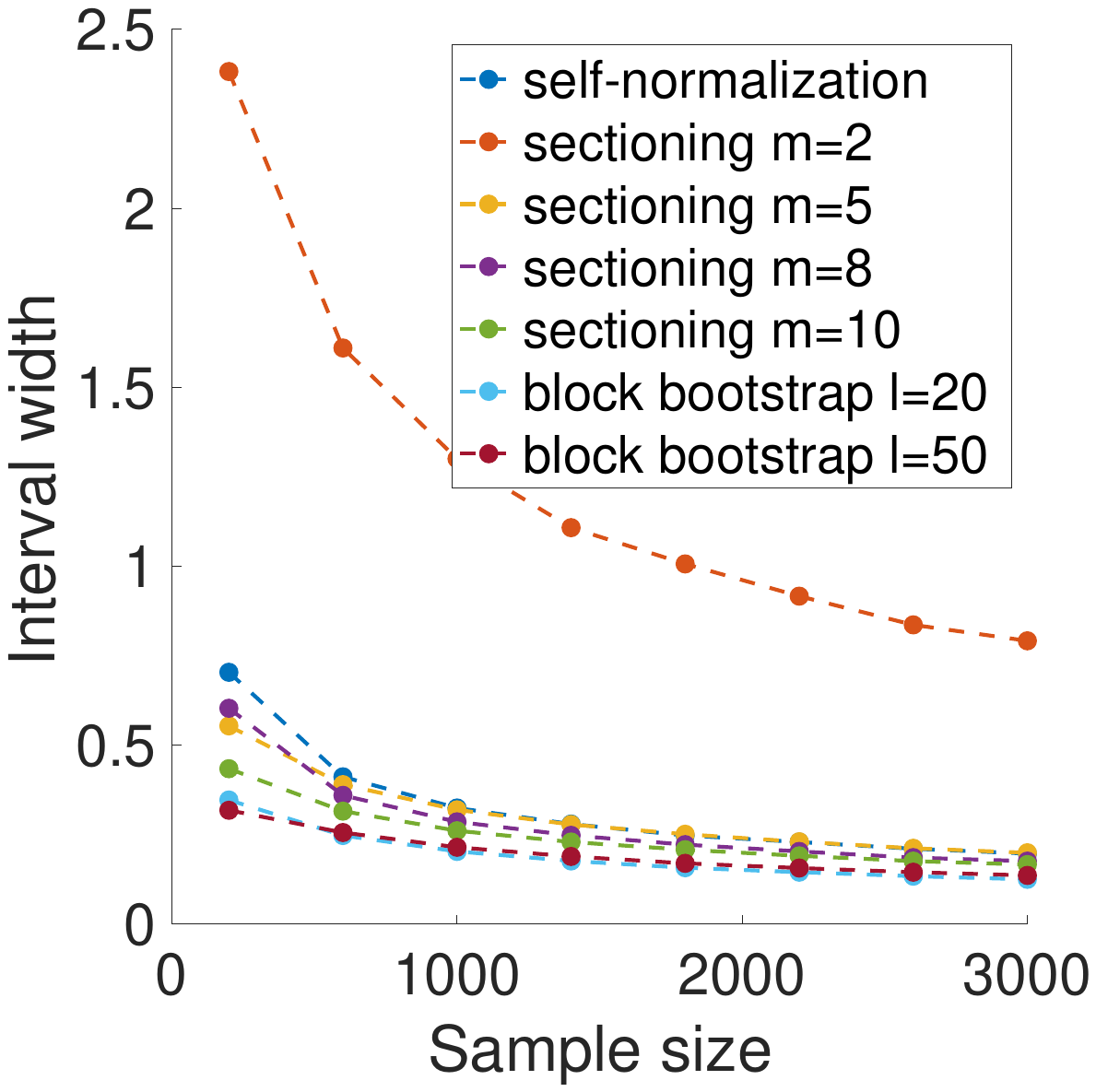}
\endminipage
\bnotefig{This figure shows the coverage probability and confidence interval width for different sample size. We use the GARCH(1,1) process with i.i.d. $t(15)$-distributed innovations, and set the model parameters as $\omega=0.01, \lambda_1=0.1$, and $\lambda_2=0.8$. Left subfigure: relationship between sample size and empirical coverage probability of 95\% confidence intervals for ES in the upper 5th percentile computed for stationary GARCH(1,1) process. Right subfigure: relationship between time series sample size and width of 95\% confidence intervals for ES in the upper 5th percentile computed for stationary GARCH(1,1) process. Each plotted point is the averaged result over 10,000 replications.}
\end{figure}

\begin{figure}[htbp]
\tcapfig{Coverage Probability and Confidence Intervals for GARCH(1,1) with i.i.d. $t(15)$-Distributed Innovations and Parameters $\lambda_1=0.1$ and $\lambda_2=0.7$}
\centering
\minipage{0.45\textwidth}
  \includegraphics[trim={0 2.5cm 0 3.5cm},clip,width=\linewidth]{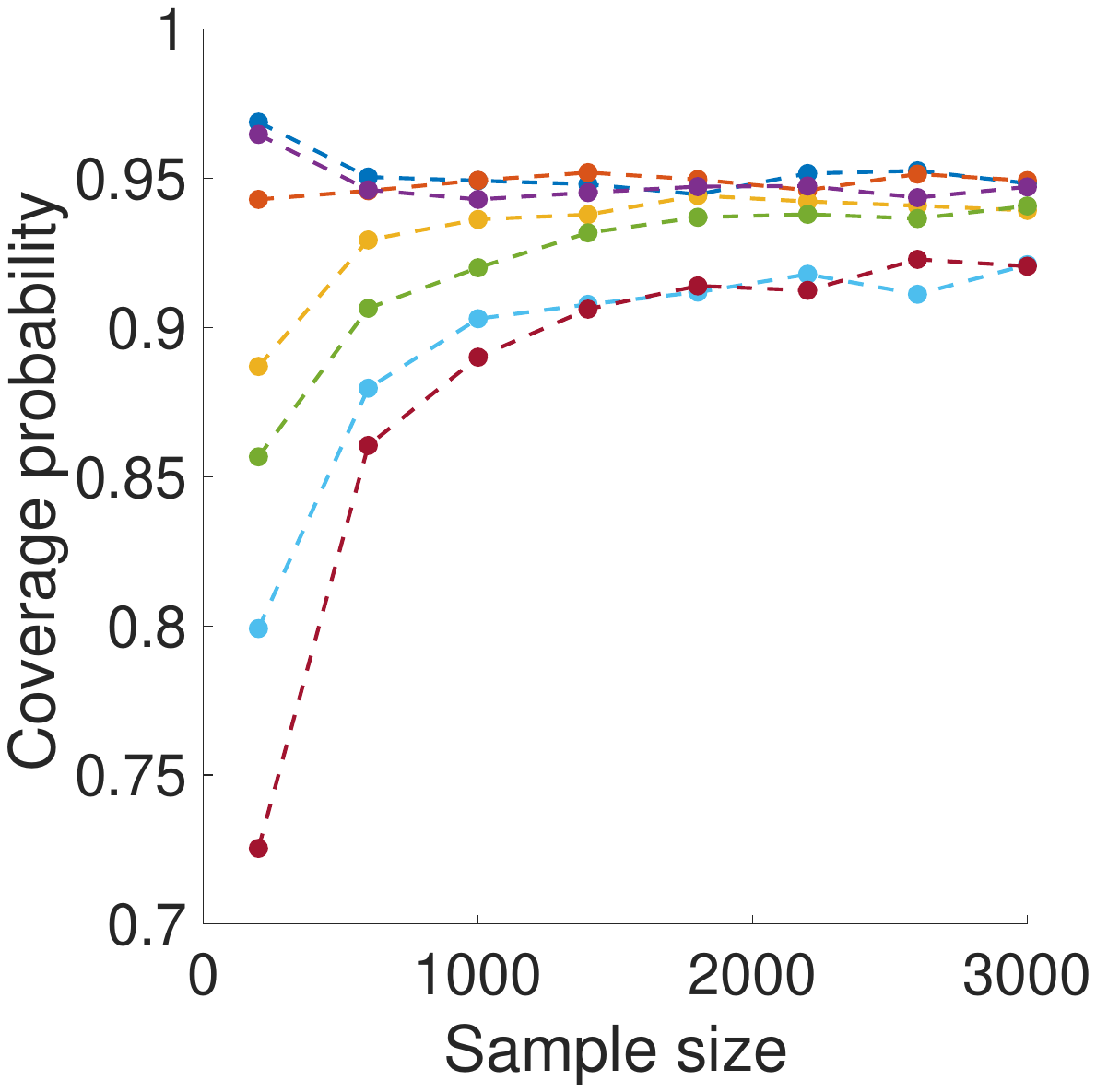}
\endminipage
\minipage{0.45\textwidth}
  \includegraphics[trim={0 2.5cm 0 3.5cm},clip,width=\linewidth]{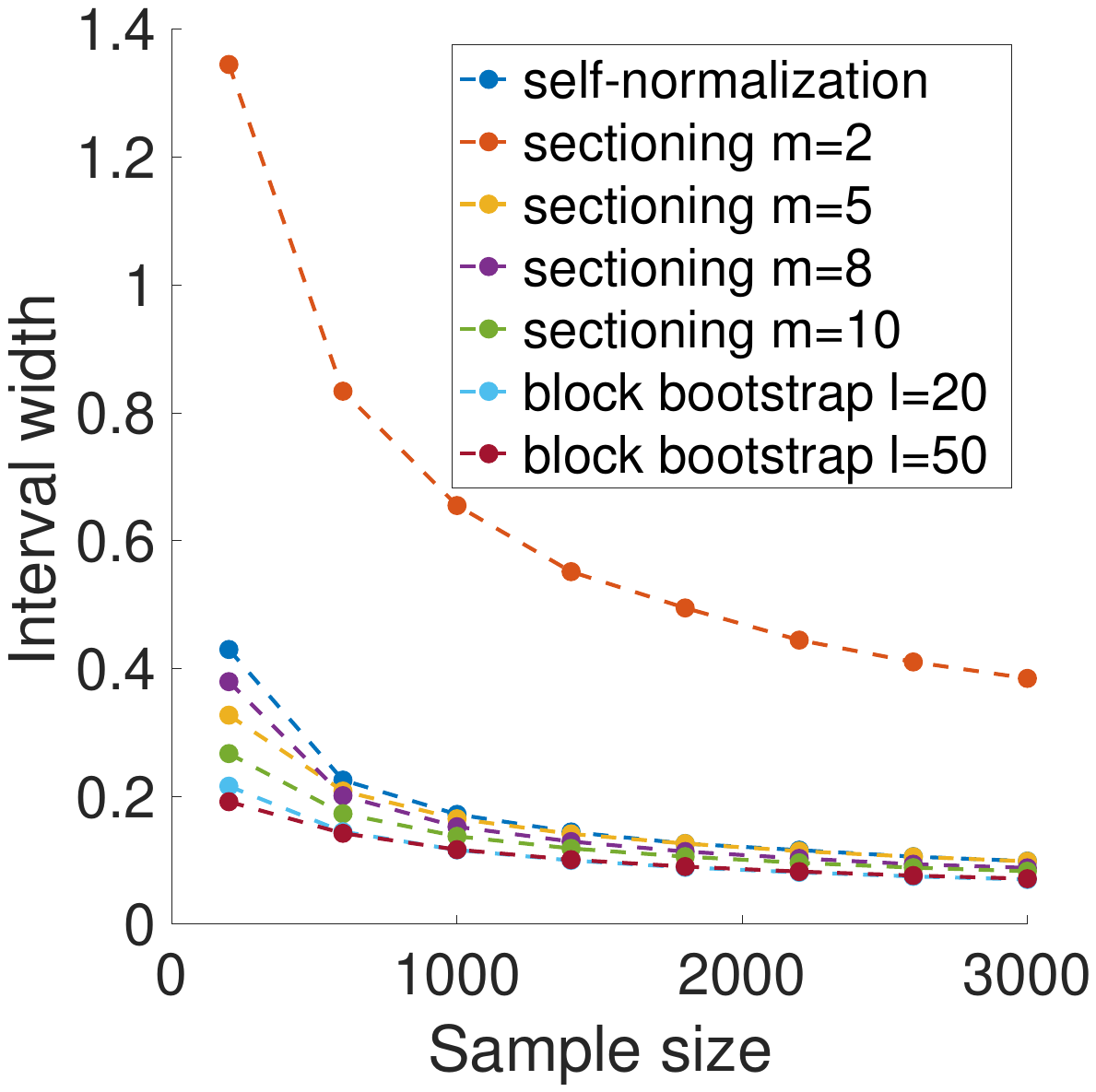}
\endminipage
\bnotefig{This figure shows the coverage probability and confidence interval width for different sample size. We use the GARCH(1,1) process with i.i.d. $t(15)$-distributed innovations, and set the model parameters as $\omega=0.01, \lambda_1=0.1$, and $\lambda_2=0.7$. Left subfigure: relationship between sample size and empirical coverage probability of 95\% confidence intervals for ES in the upper 5th percentile computed for stationary GARCH(1,1) process. Right subfigure: relationship between time series sample size and width of 95\% confidence intervals for ES in the upper 5th percentile computed for stationary GARCH(1,1) process. Each plotted point is the averaged result over 10,000 replications.}
\end{figure}

\begin{figure}[htbp]
\tcapfig{Coverage Probability and Confidence Intervals for GARCH(1,1) with i.i.d. $t(15)$-Distributed Innovations and Parameters $\lambda_1=0.1$ and $\lambda_2=0.6$}
\centering
\minipage{0.45\textwidth}
  \includegraphics[trim={0 2.5cm 0 3.5cm},clip,width=\linewidth]{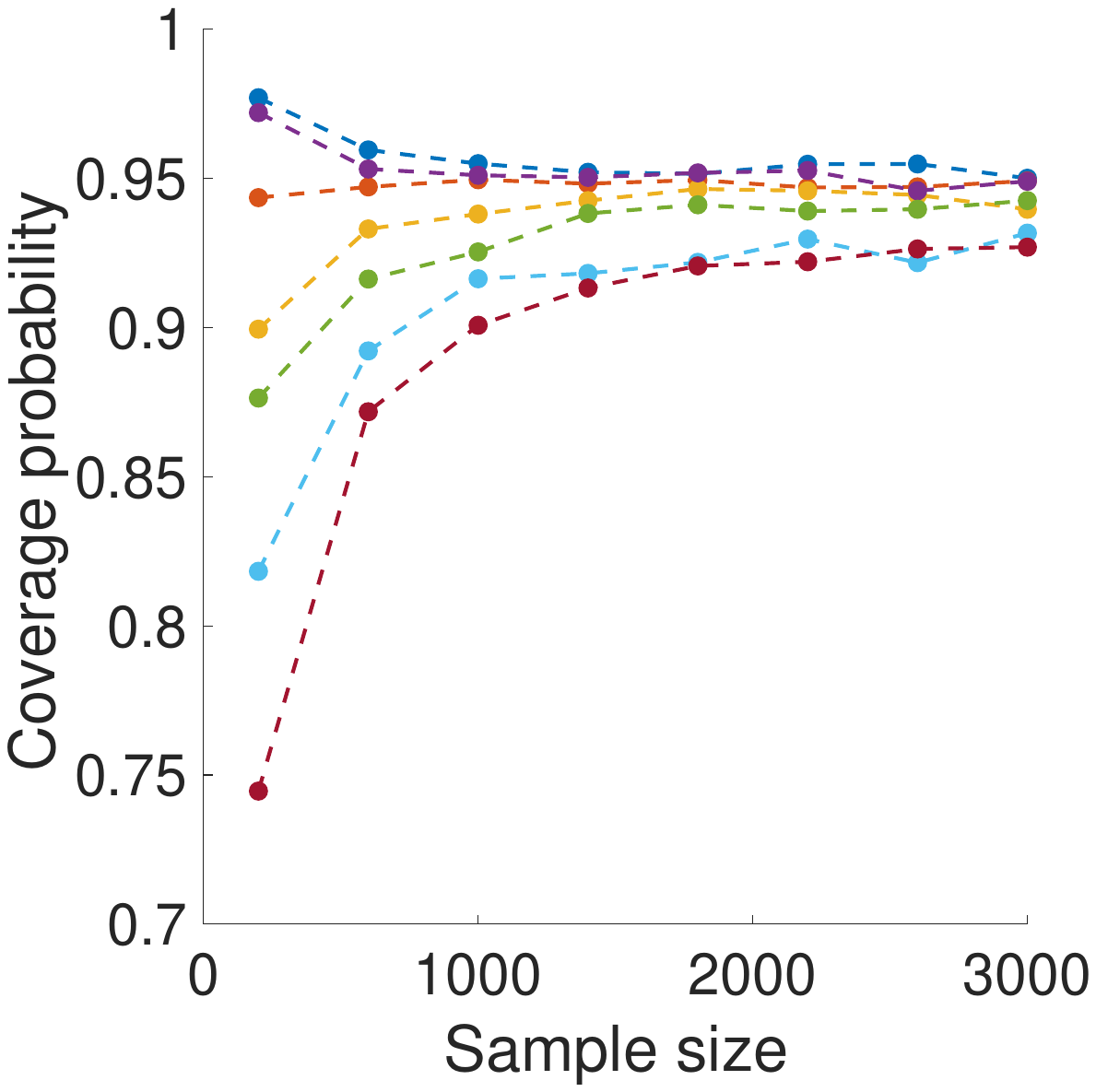}
\endminipage
\minipage{0.45\textwidth}
  \includegraphics[trim={0 2.5cm 0 3.5cm},clip,width=\linewidth]{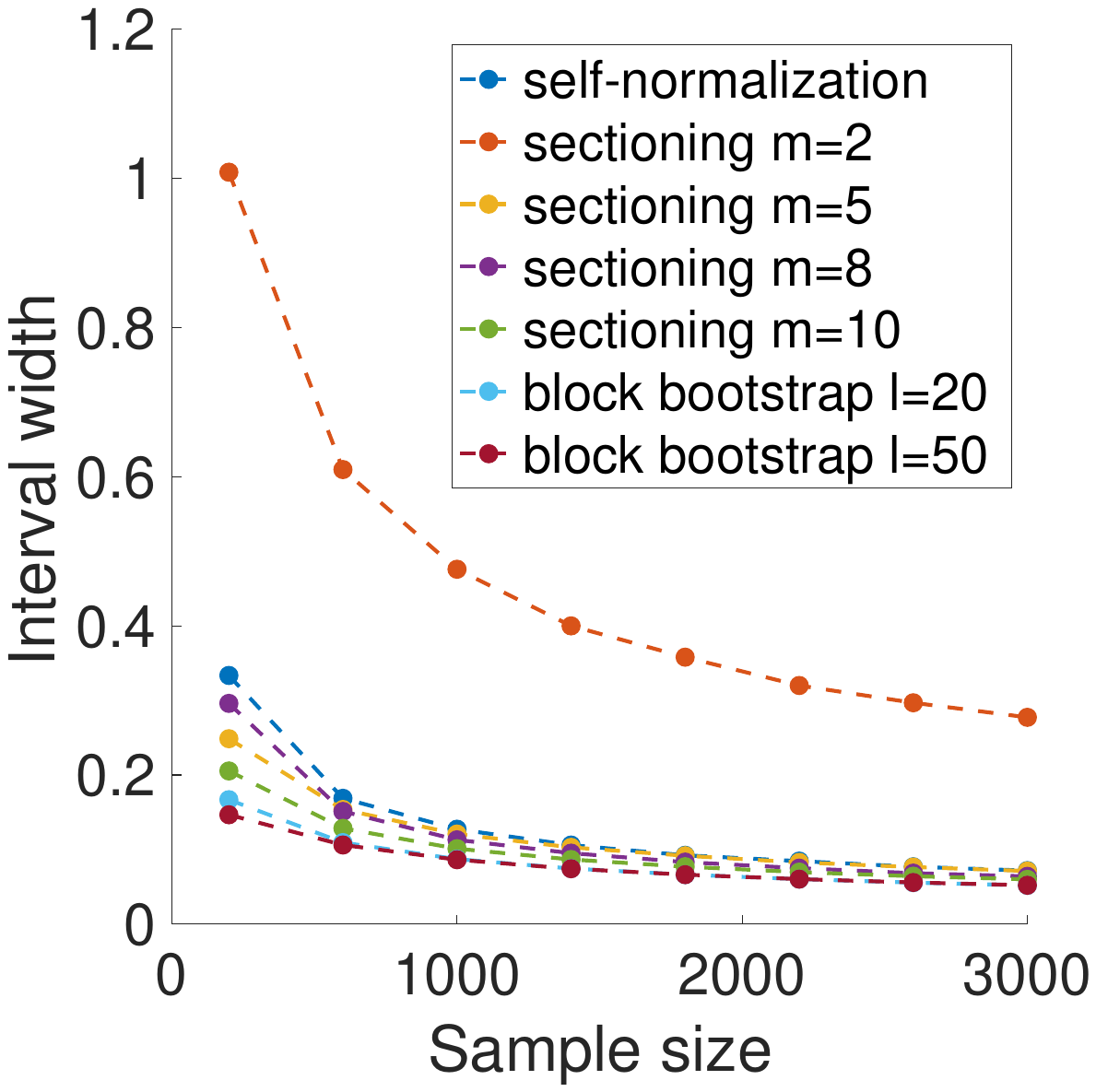}
\endminipage
\bnotefig{This figure shows the coverage probability and confidence interval width for different sample size. We use the GARCH(1,1) process with i.i.d. $t(15)$-distributed innovations, and set the model parameters as $\omega=0.01, \lambda_1=0.1$, and $\lambda_2=0.6$. Left subfigure: relationship between sample size and empirical coverage probability of 95\% confidence intervals for ES in the upper 5th percentile computed for stationary GARCH(1,1) process. Right subfigure: relationship between time series sample size and width of 95\% confidence intervals for ES in the upper 5th percentile computed for stationary GARCH(1,1) process. Each plotted point is the averaged result over 10,000 replications.}
\end{figure}

\begin{figure}[H]
\tcapfig{Coverage Probability and Confidence Intervals for GARCH(1,1) with i.i.d. $t(15)$-Distributed Innovations and Parameters $\lambda_1=0.1$ and $\lambda_2=0.5$}
\centering
\minipage{0.45\textwidth}
  \includegraphics[trim={0 2.5cm 0 3.5cm},clip,width=\linewidth]{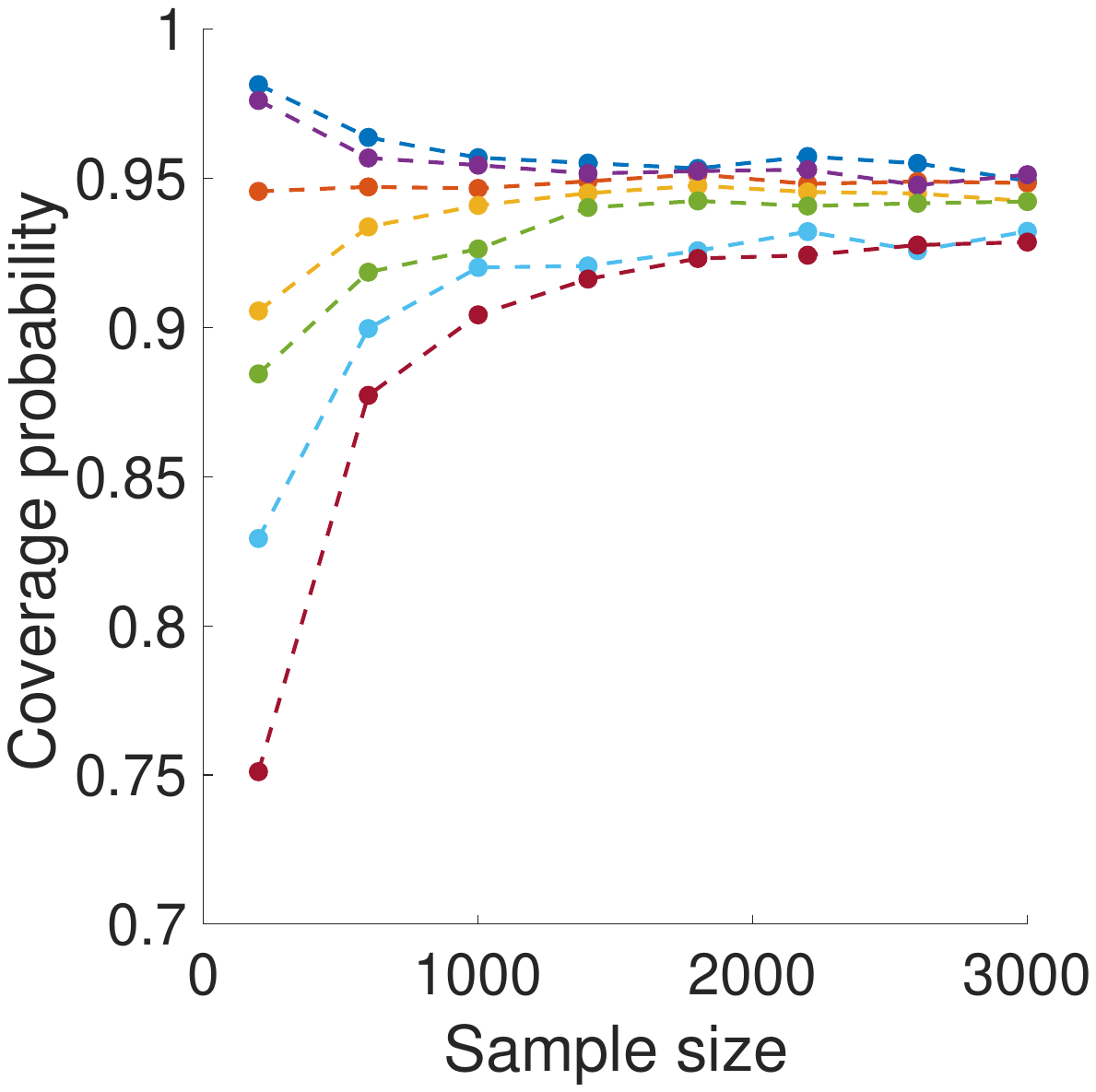}
\endminipage
\minipage{0.45\textwidth}
  \includegraphics[trim={0 2.5cm 0 3.5cm},clip,width=\linewidth]{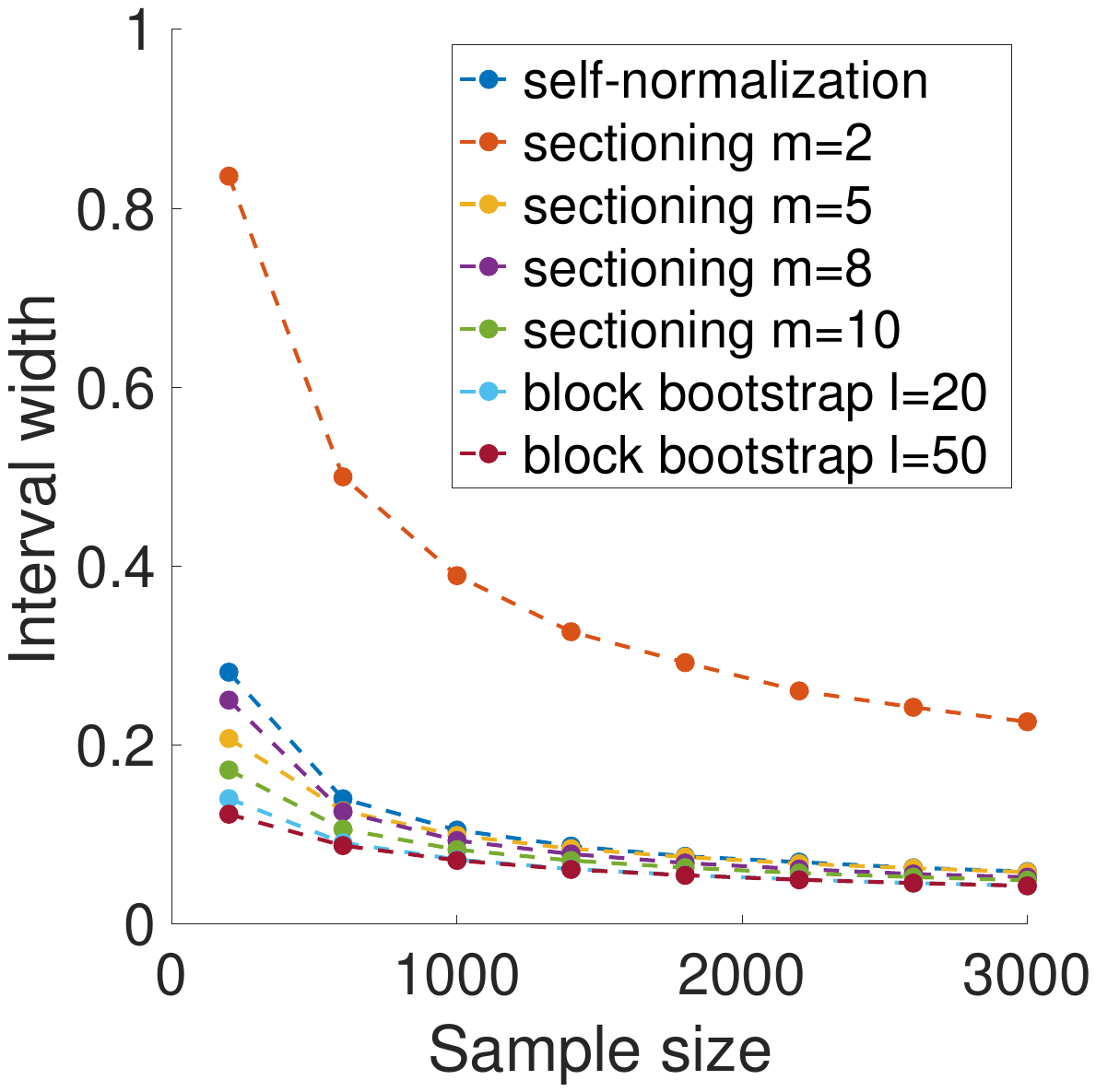}
\endminipage
\bnotefig{This figure shows the coverage probability and confidence interval width for different sample size. We use the GARCH(1,1) process with i.i.d. $t(15)$-distributed innovations, and set the model parameters as $\omega=0.01, \lambda_1=0.1$, and $\lambda_2=0.5$. Left subfigure: relationship between sample size and empirical coverage probability of 95\% confidence intervals for ES in the upper 5th percentile computed for stationary GARCH(1,1) process. Right subfigure: relationship between time series sample size and width of 95\% confidence intervals for ES in the upper 5th percentile computed for stationary GARCH(1,1) process. Each plotted point is the averaged result over 10,000 replications.}
\end{figure}

\newpage
\subsection*{Appendix C: Empirical Results}\label{sec:appendixC}

\begin{figure}[htbp]
\tcapfig{Single Change-Point Tests for S\&P 500 Daily Returns with 6-Month Windows}
\centering
\includegraphics[width=\linewidth]{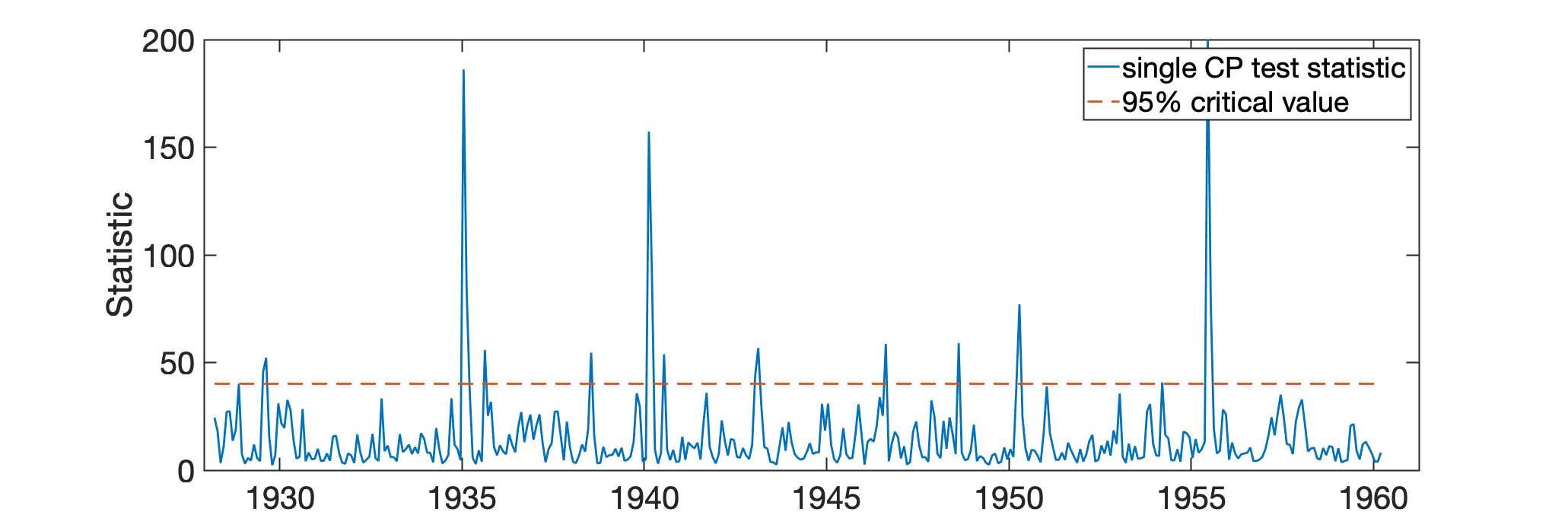}
\includegraphics[width=\linewidth]{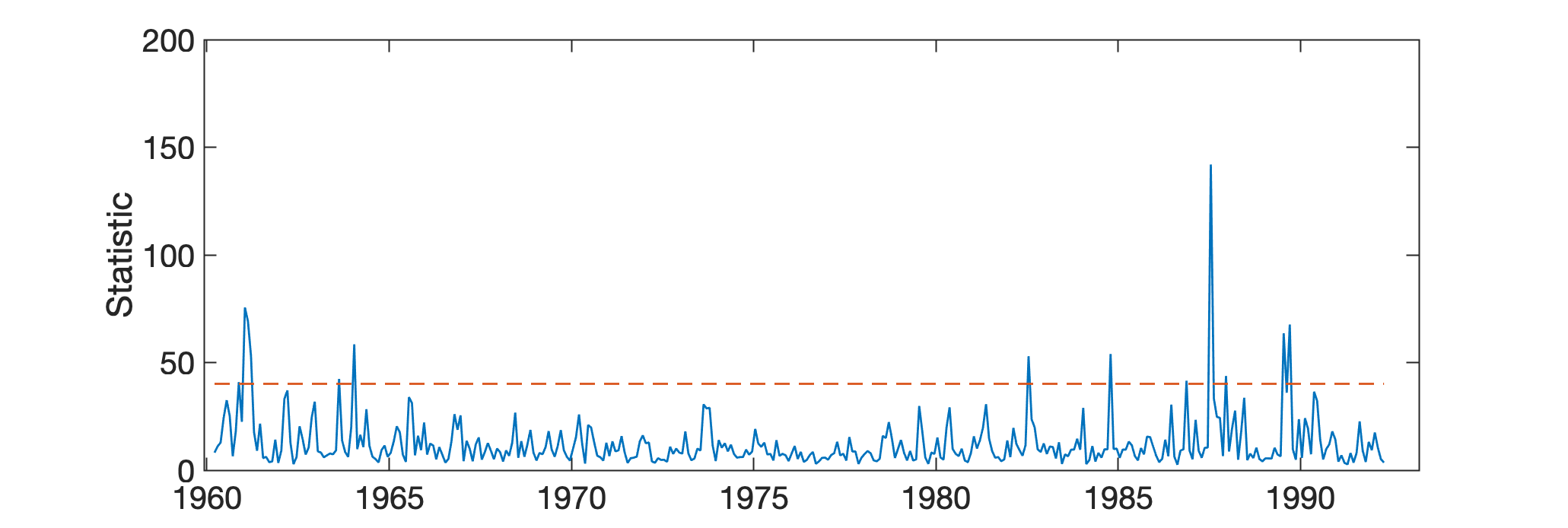}
\includegraphics[width=\linewidth]{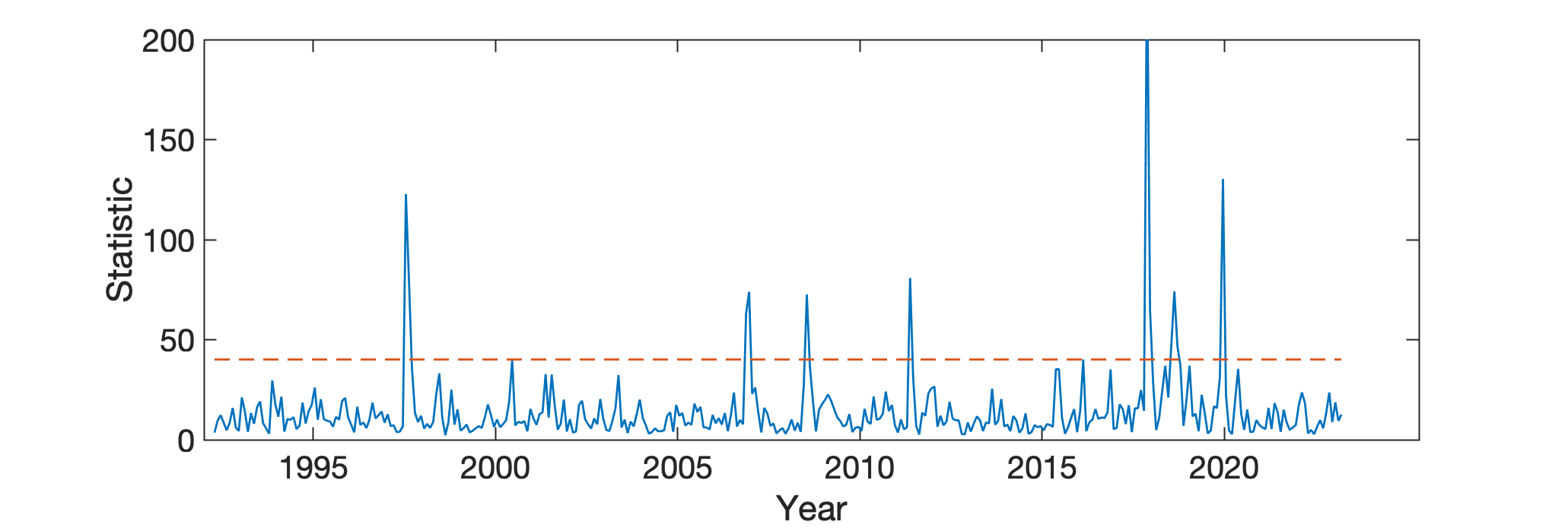}
\bnotefig{This figure shows the test statistics for single change-point tests for expected shortfall (ES) in the lower 5th percentile for S\&P 500 log returns in different 6-month time windows.}
\label{fig:empir_sp_single_6m}    
\end{figure}

\begin{figure}[htbp]
\tcapfig{Single Change-Point Tests for S\&P 500 Daily Returns with 12-Month Windows}
\centering
\includegraphics[width=\linewidth]{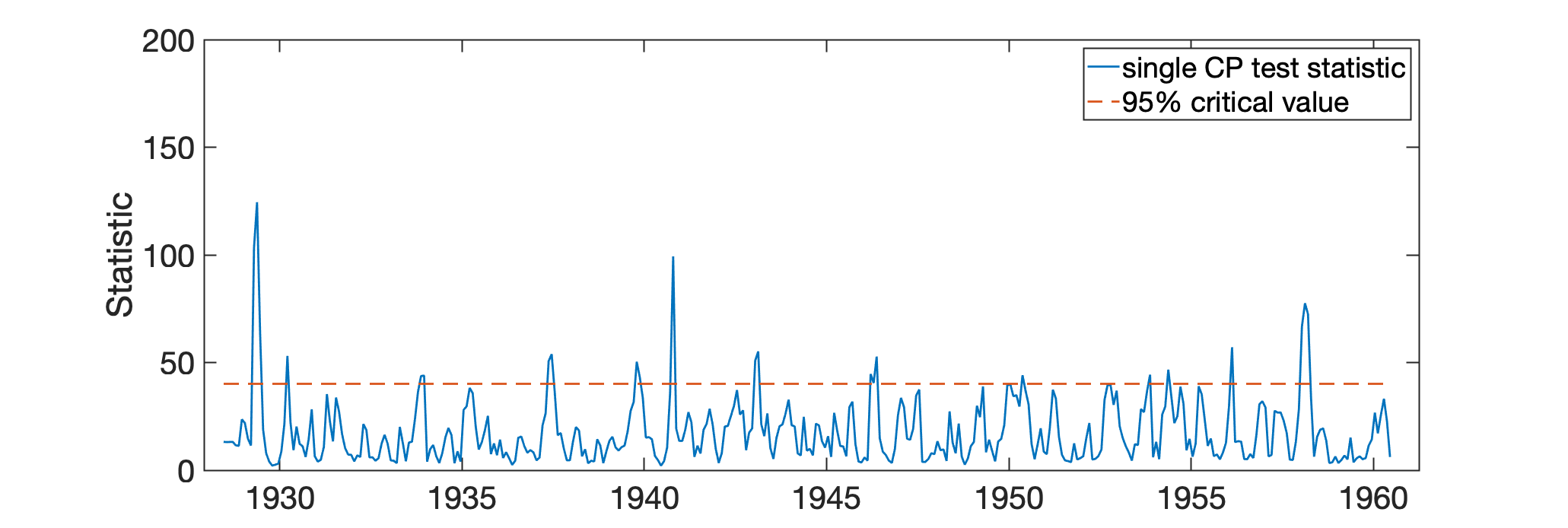}
\includegraphics[width=\linewidth]{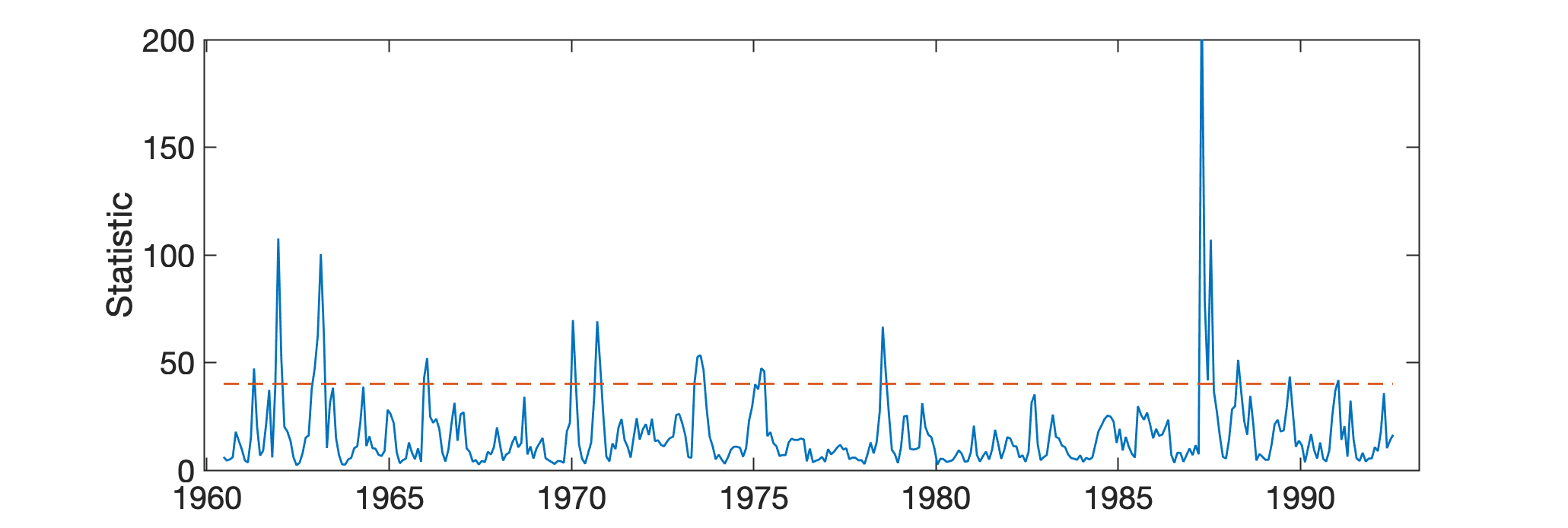}
\includegraphics[width=\linewidth]{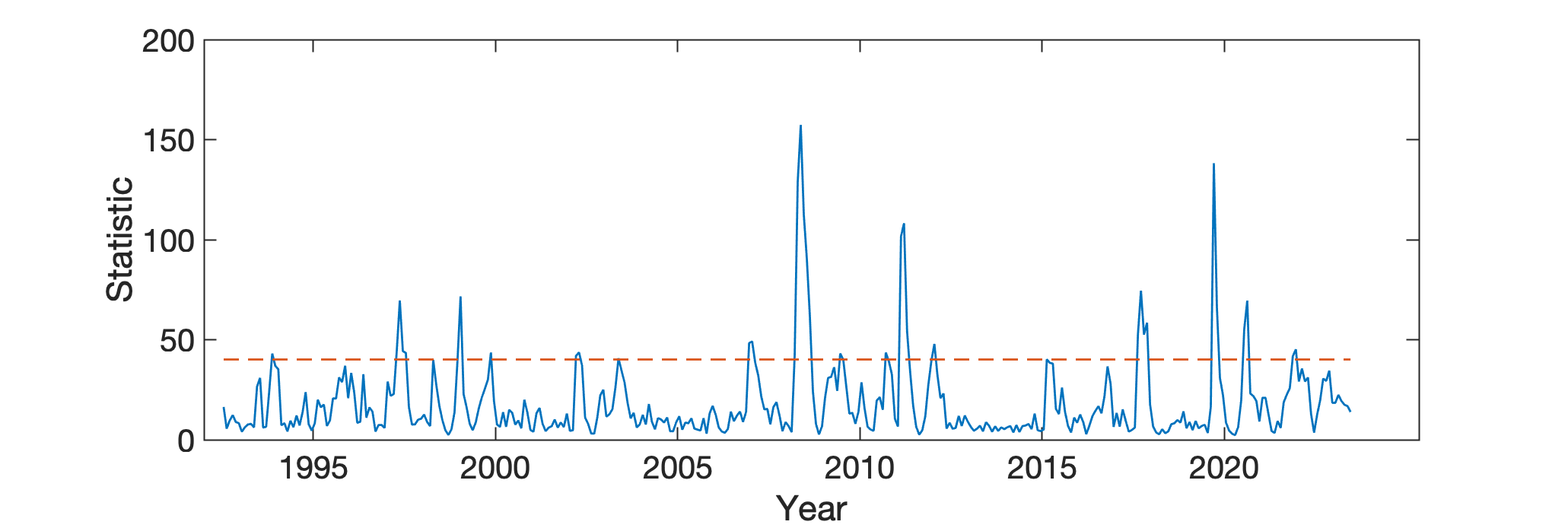}
\bnotefig{This figure shows the test statistics for single change-point tests for expected shortfall (ES) in the lower 5th percentile for S\&P 500 log returns in different 12-month time windows.}
\label{fig:empir_sp_single_12m}    
\end{figure}

\begin{figure}[htbp]
\tcapfig{Multiple Change-Point Tests for S\&P 500 Daily Returns with 6-Month Windows}
\centering
\includegraphics[width=\linewidth]{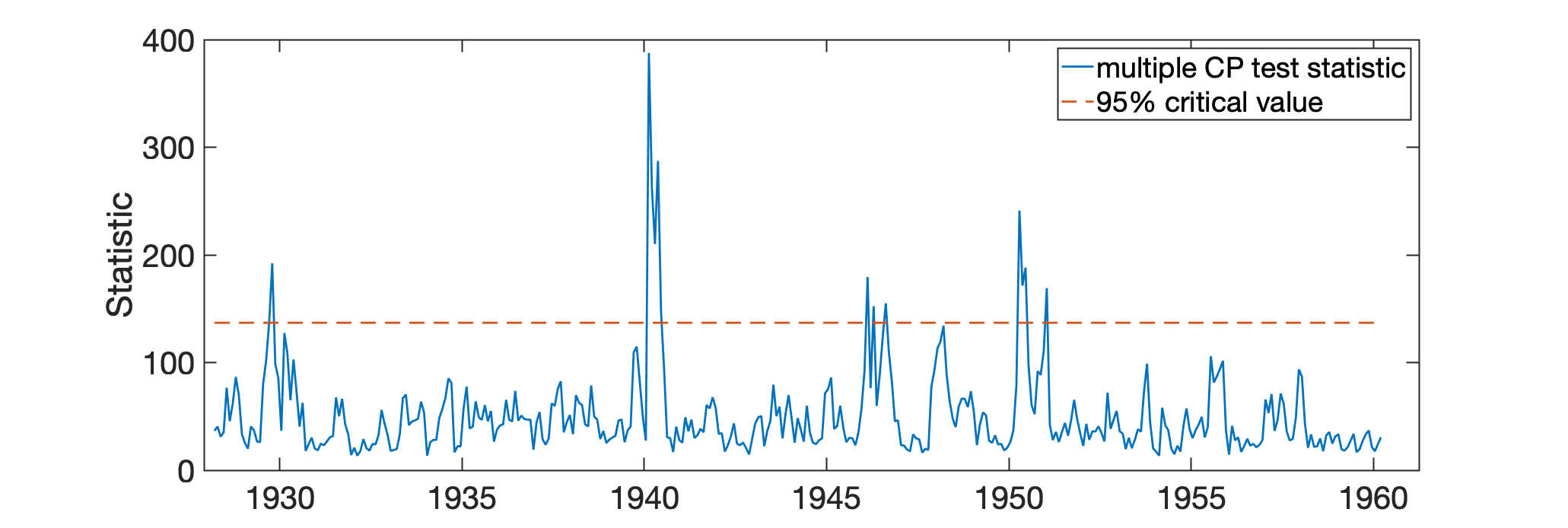}
\includegraphics[width=\linewidth]{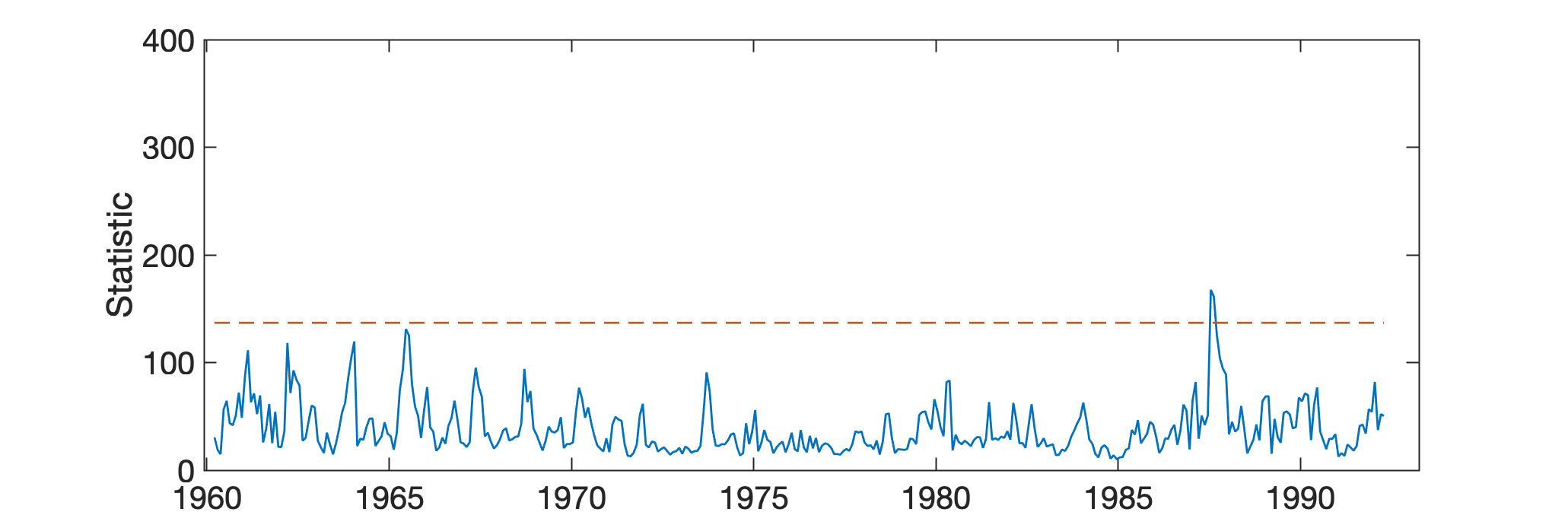}
\includegraphics[width=\linewidth]{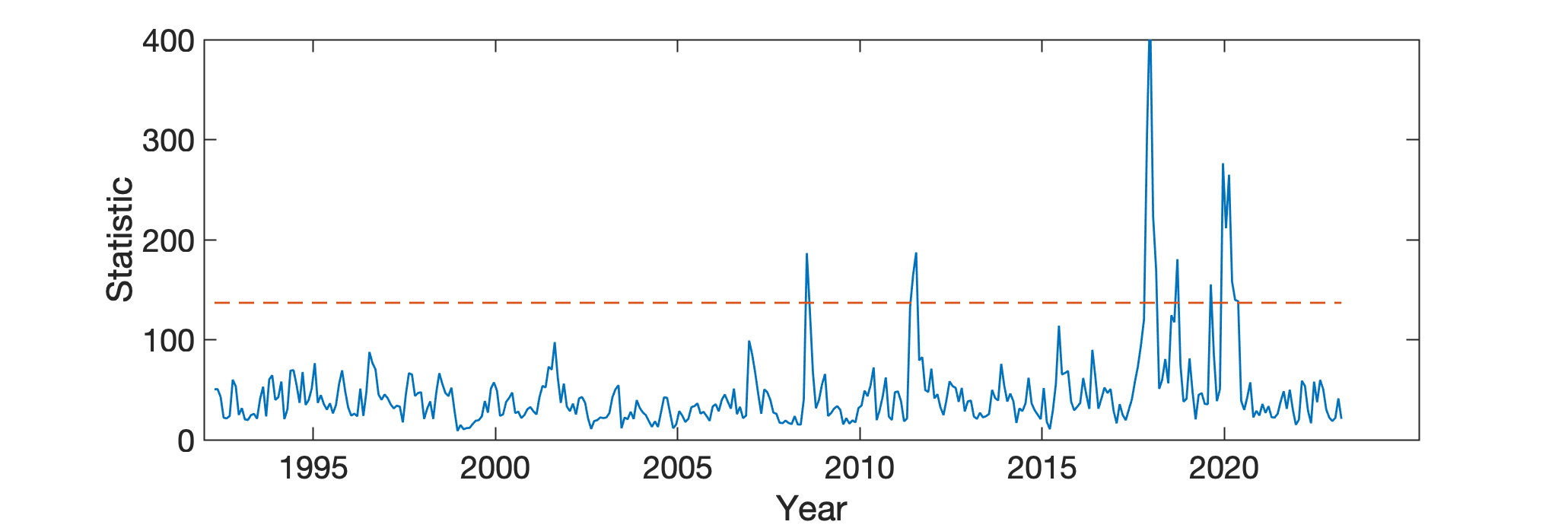}
\bnotefig{This figure shows the test statistics for multiple change-point tests for expected shortfall (ES) in the lower 5th percentile for S\&P 500 log returns in different 6-month time windows.}
\label{fig:empir_sp_multiple_6m}    
\end{figure}

\begin{figure}[htbp]
\tcapfig{Multiple Change-Point Tests for S\&P 500 Daily Returns with 12-Month Windows}
\centering
\includegraphics[width=\linewidth]{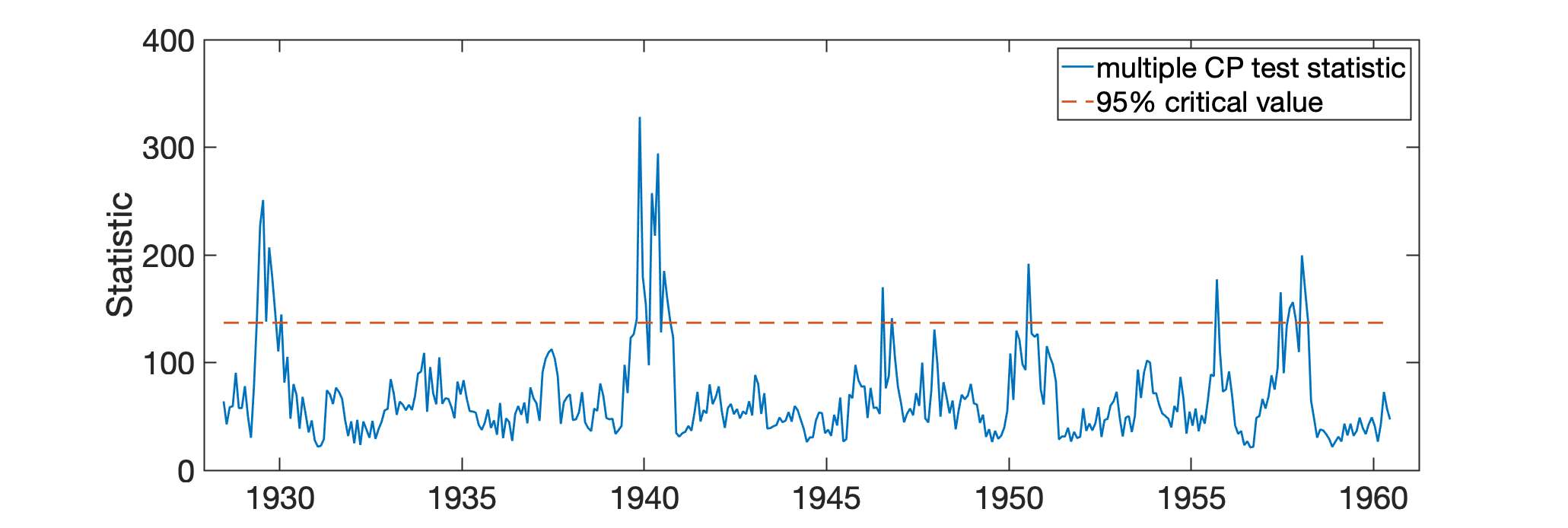}
\includegraphics[width=\linewidth]{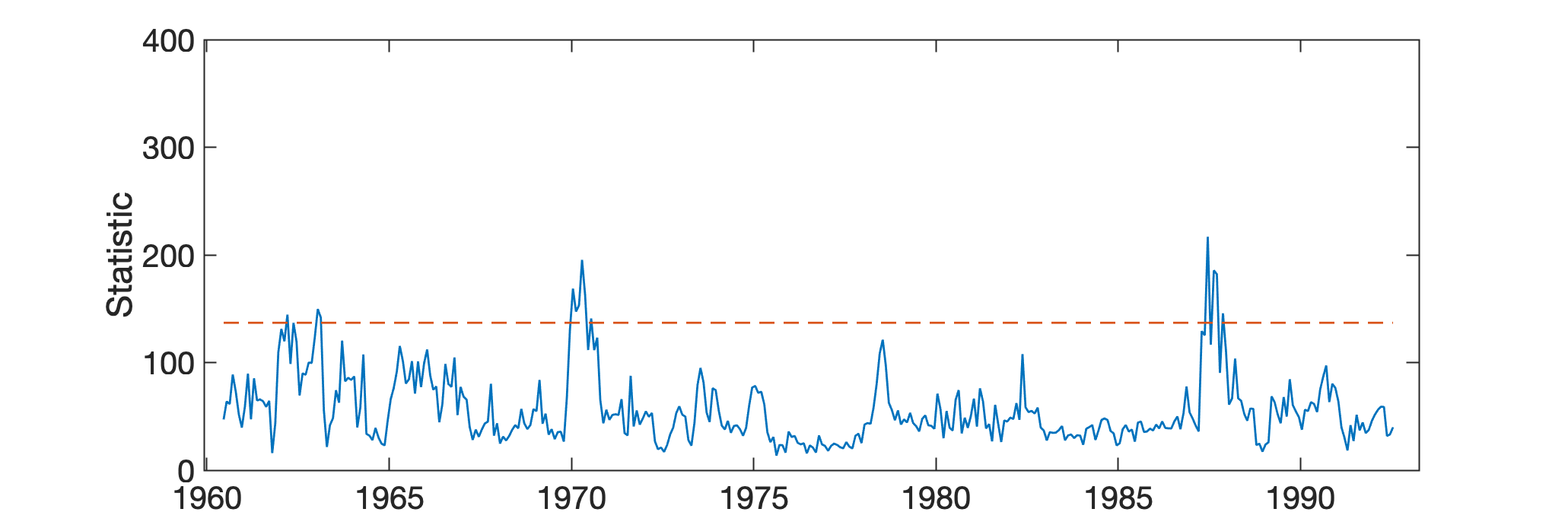}
\includegraphics[width=\linewidth]{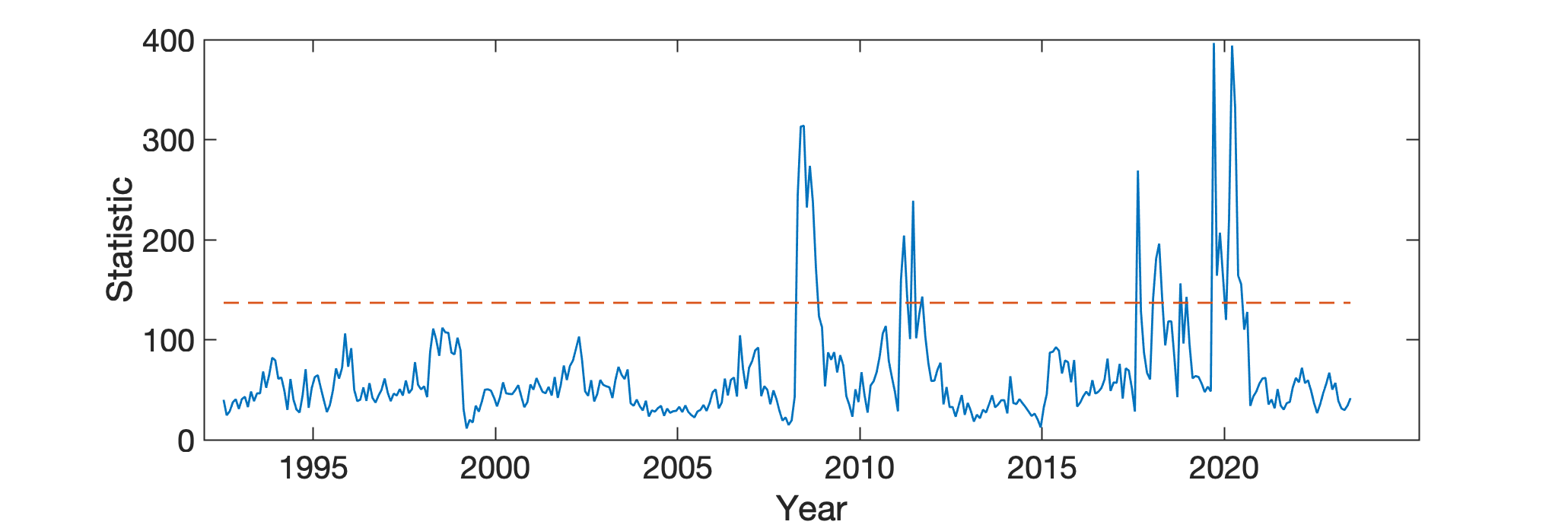}
\bnotefig{This figure shows the test statistics for multiple change-point tests for expected shortfall (ES) in the lower 5th percentile for S\&P 500 log returns in different 12-month time windows.}
\label{fig:empir_sp_multiple_12m}    
\end{figure}

\begin{figure}[htbp]
\tcapfig{Change-Point Tests for U.S. 1-Year Bond Weekly Returns with 5-Year Windows}
\centering
\begin{subfigure}[b]{\textwidth}
    \includegraphics[width=\linewidth]{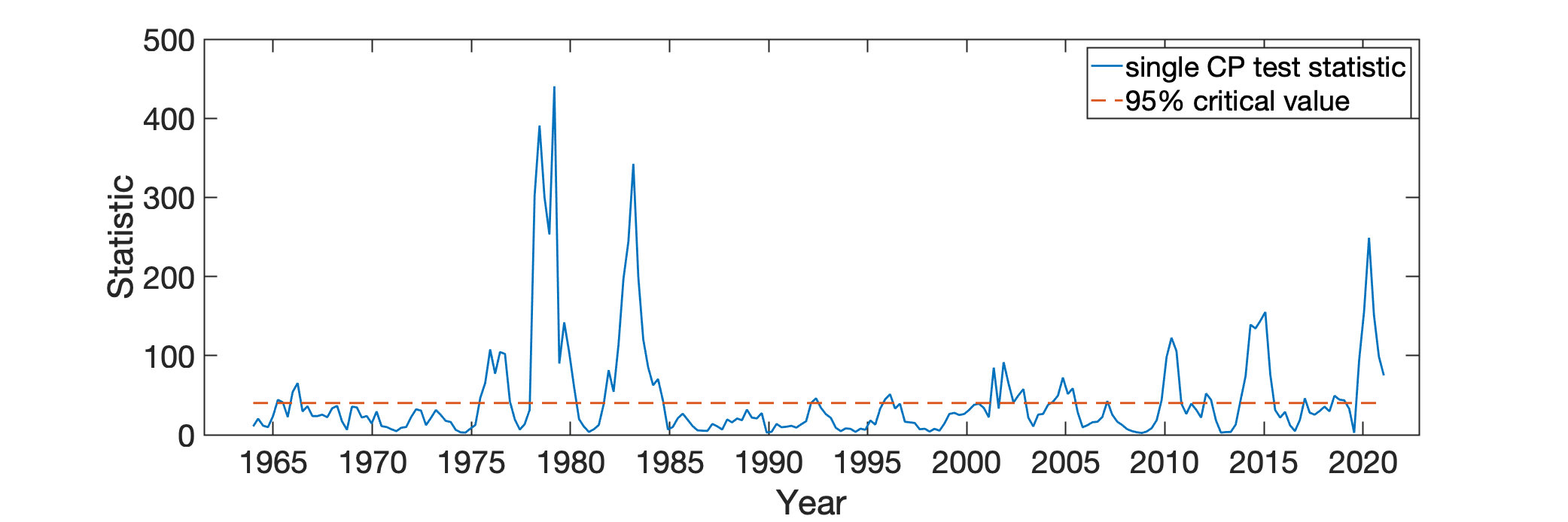}
    \caption{Single change-point test statistics}
\end{subfigure}
\begin{subfigure}[b]{\textwidth}
    \includegraphics[width=\linewidth]{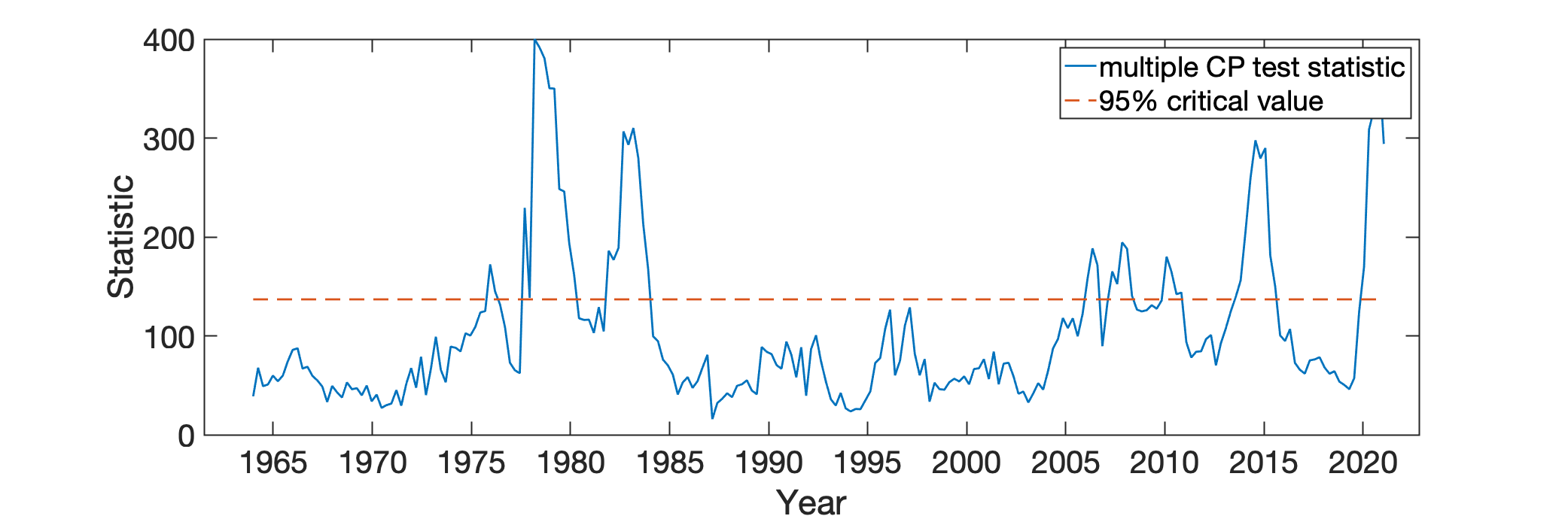}
    \caption{Multiple change-point test statistics}
\end{subfigure}
\bnotefig{This figure shows the test statistics for single and multiple change-point tests for expected shortfall (ES) in the lower 10th percentile for U.S. 1-year zero-coupon bond log returns in different 5-year time windows.}
\label{fig:empir_b1_5y}
\end{figure}

\begin{figure}[htbp]
\tcapfig{Change-Point Tests for U.S. 1-Year Bond Weekly Returns with 10-Year Windows}
\centering
\begin{subfigure}[b]{\textwidth}
    \includegraphics[width=\linewidth]{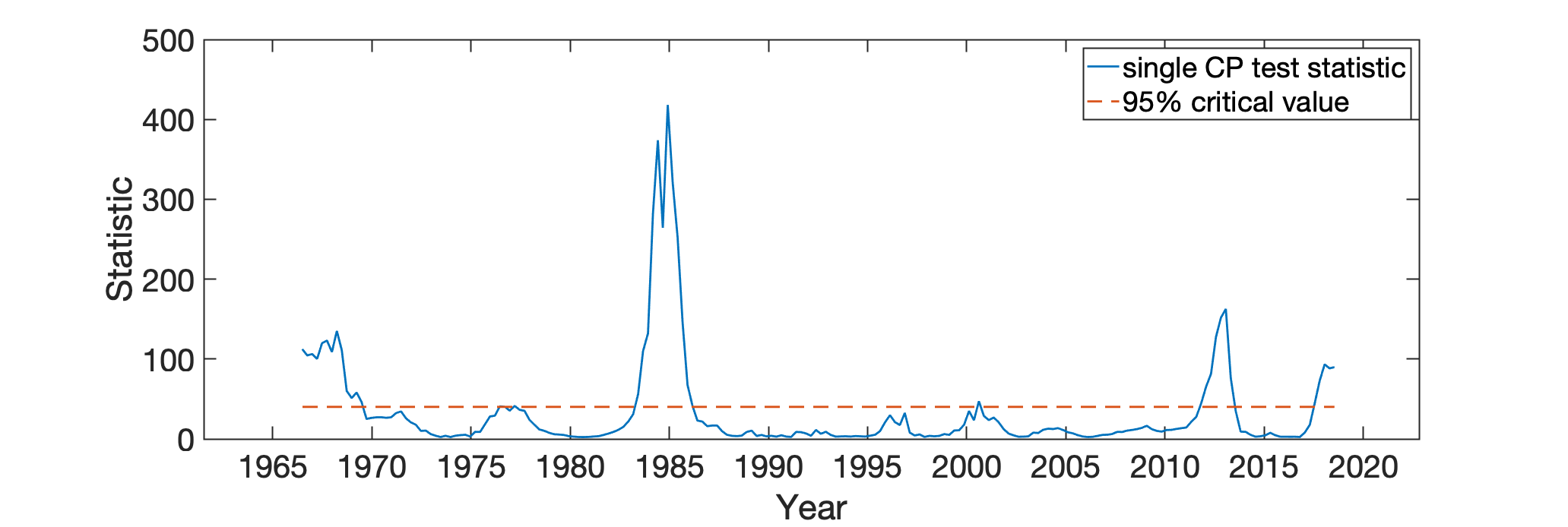}
    \caption{Single change-point test statistics}
\end{subfigure}
\begin{subfigure}[b]{\textwidth}
    \includegraphics[width=\linewidth]{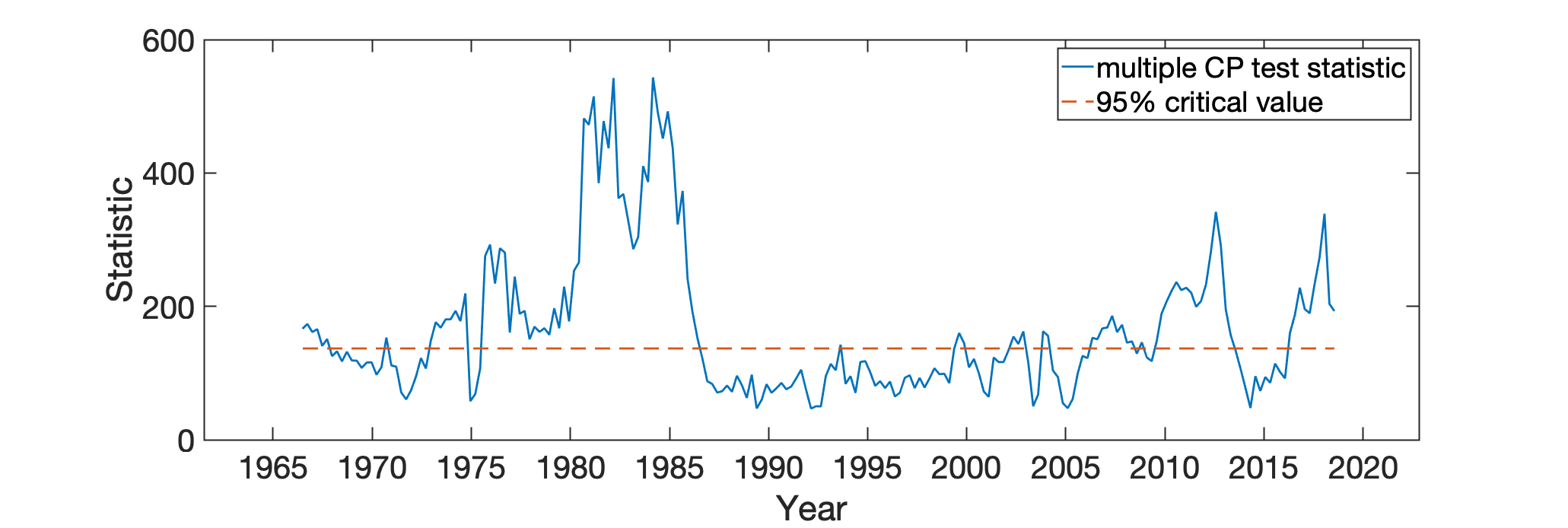}
    \caption{Multiple change-point test statistics}
\end{subfigure}
\bnotefig{This figure shows the test statistics for single and multiple change-point tests for expected shortfall (ES) in the lower 10th percentile for U.S. 1-year zero-coupon bond log returns in different 10-year time windows.}
\label{fig:empir_b1_10y}
\end{figure}

\begin{figure}[htbp]
\tcapfig{Change-Point Tests for U.S. 10-Year Bond Weekly Returns with 5-Year Windows}
\centering
\begin{subfigure}[b]{\textwidth}
    \includegraphics[width=\linewidth]{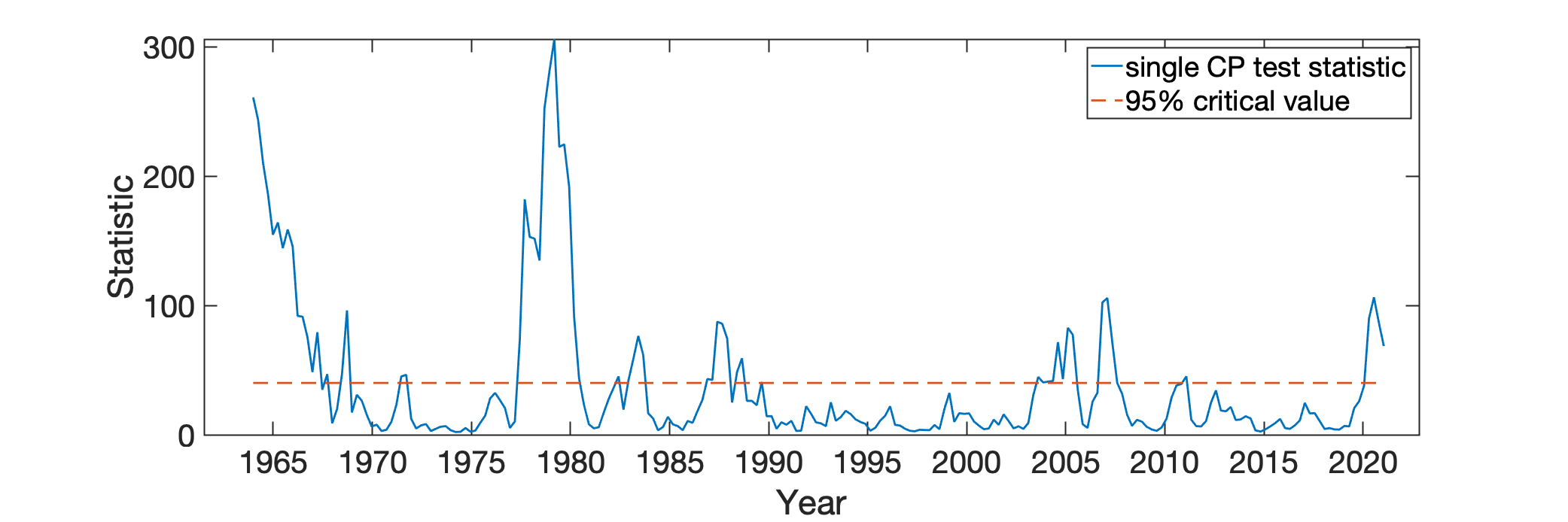}
    \caption{Single change-point test statistics}
\end{subfigure}
\begin{subfigure}[b]{\textwidth}
    \includegraphics[width=\linewidth]{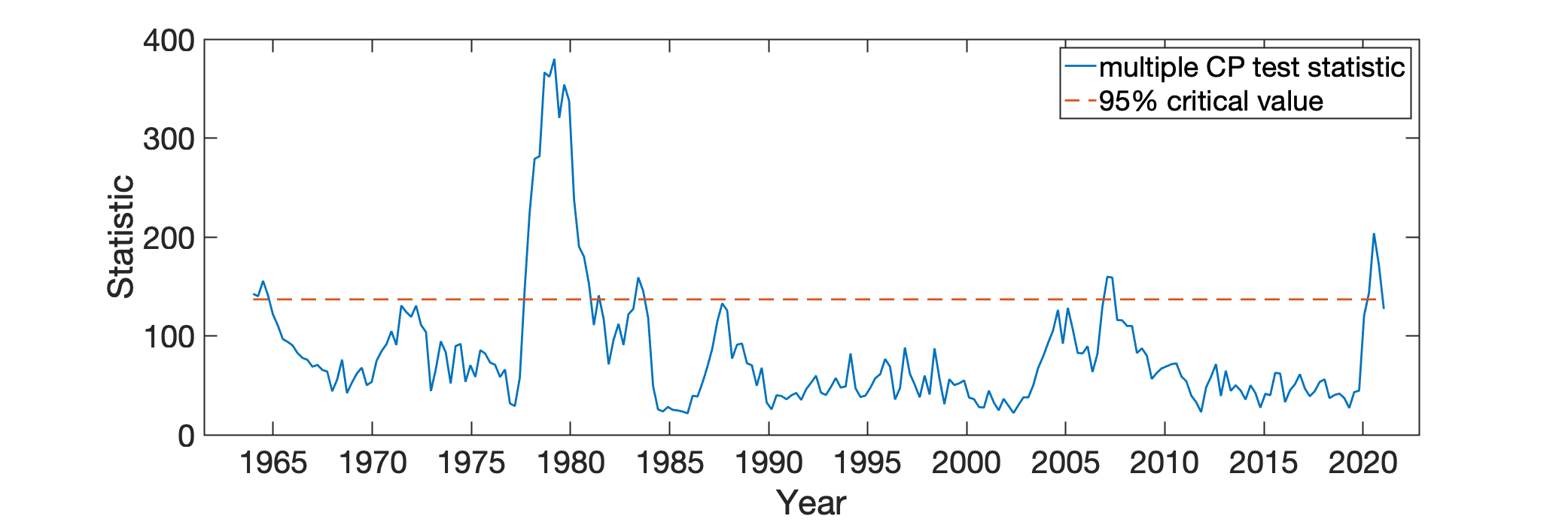}
    \caption{Multiple change-point test statistics}
\end{subfigure}
\bnotefig{This figure shows the test statistics for single and multiple change-point tests for expected shortfall (ES) in the lower 10th percentile for U.S. 10-year zero-coupon bond log returns in different 5-year time windows.}
\label{fig:empir_b10_5y}
\end{figure}

\begin{figure}[htbp]
\tcapfig{Change-Point Tests for U.S. 10-Year Bond Weekly Returns with 10-Year Windows}
\centering
\begin{subfigure}[b]{\textwidth}
    \includegraphics[width=\linewidth]{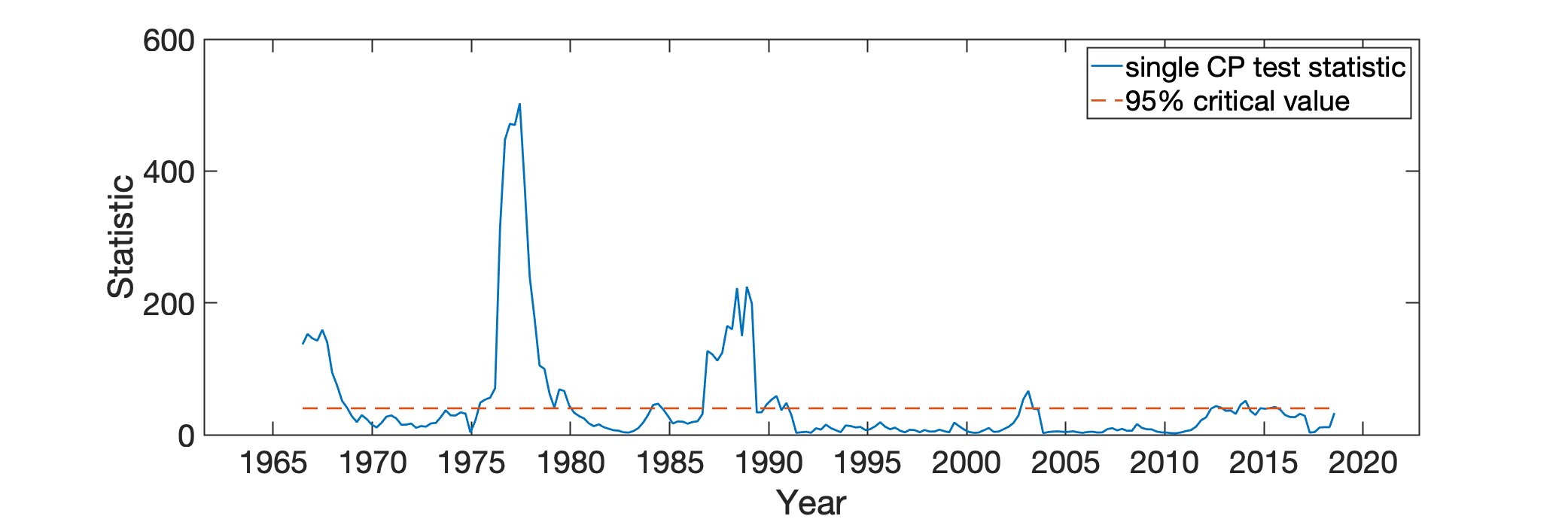}
    \caption{Single change-point test statistics}
\end{subfigure}
\begin{subfigure}[b]{\textwidth}
    \includegraphics[width=\linewidth]{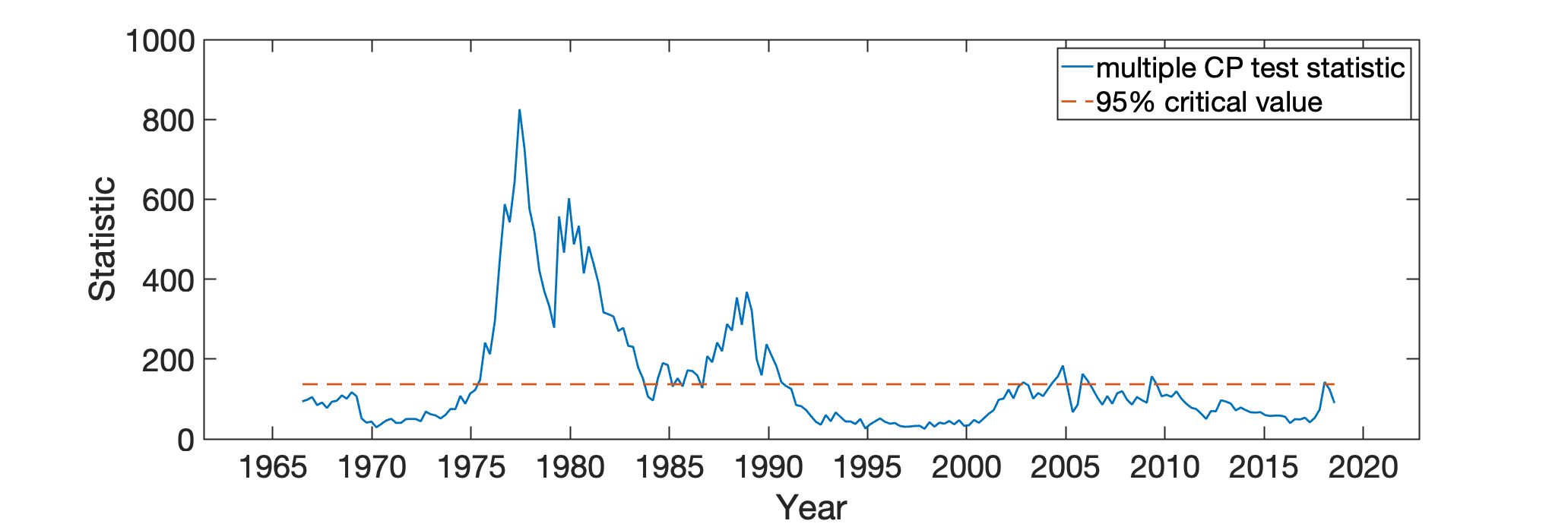}
    \caption{Multiple change-point test statistics}
\end{subfigure}
\bnotefig{This figure shows the test statistics for single and multiple change-point tests for expected shortfall (ES) in the lower 10th percentile for U.S. 10-year zero-coupon bond log returns in different 10-year time windows.}
\label{fig:empir_b10_10y}
\end{figure}

\newpage
\begin{figure}[H]
\tcapfig{Expected Shortfall of Market Returns over Time}
\centering
\minipage{\textwidth}
  \includegraphics[width=\linewidth]{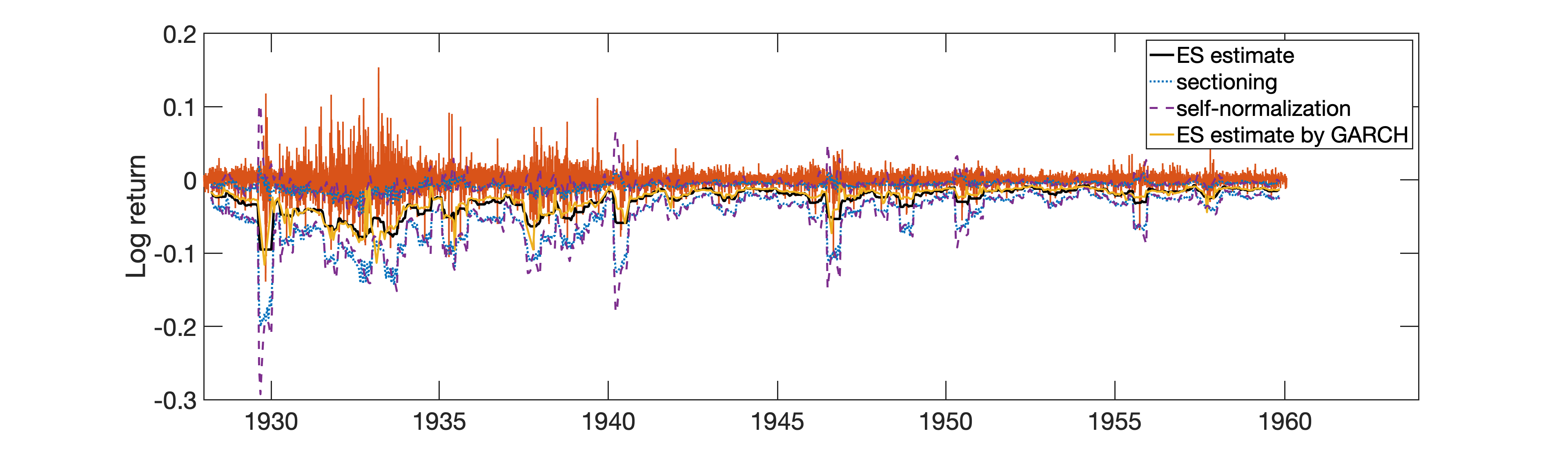}
\endminipage
\\
\minipage{\textwidth}
  \includegraphics[width=\linewidth]{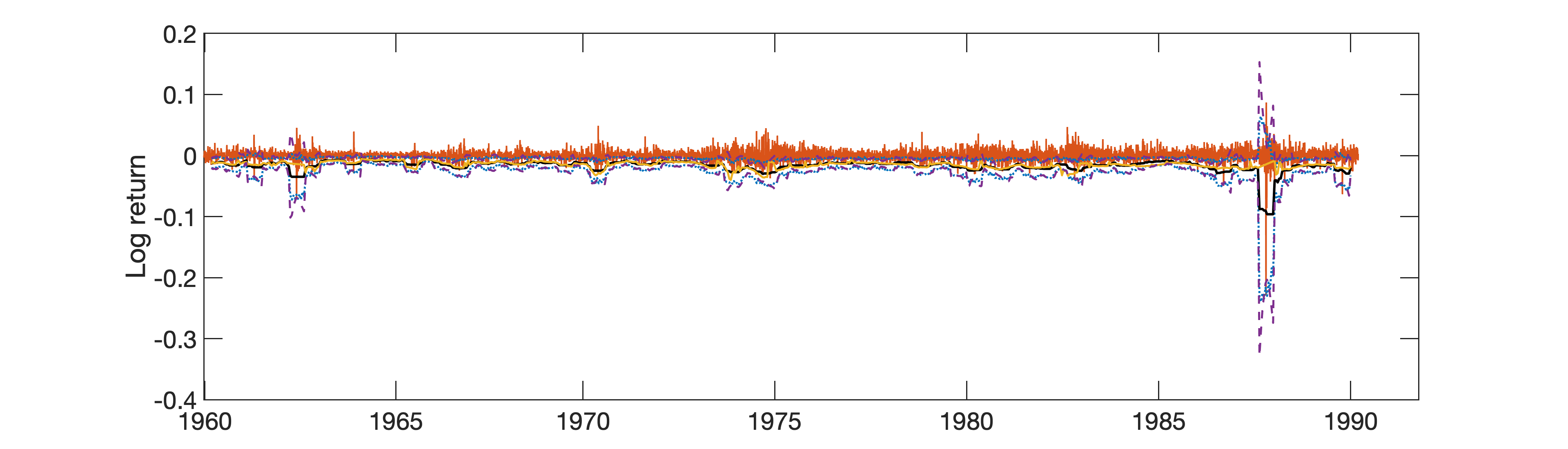}
\endminipage
\\
\minipage{\textwidth}
  \includegraphics[width=\linewidth]{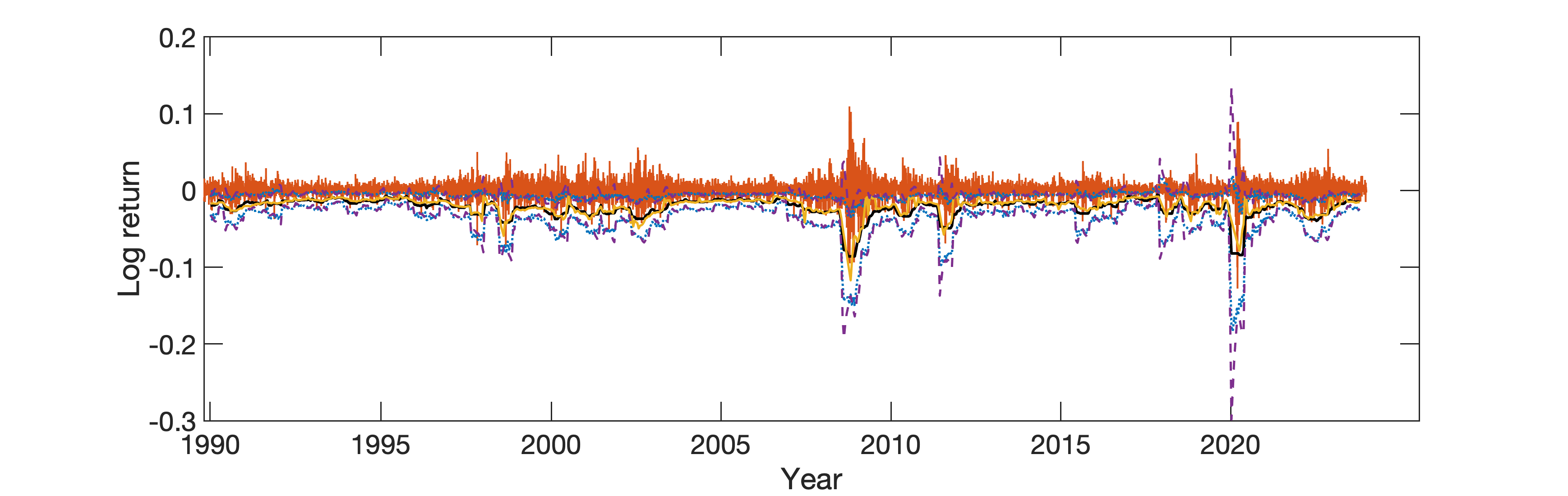}
\endminipage
\bnotefig{This figure shows the daily log returns and expected shortfall with confidence intervals for the S\&P 500 index that approximates a market return time-series. We plot the ES estimate and 95\% confidence bands for the lower 5th percentile. ES is computed using a rolling window of 100 days with 10 day shifts.}
\label{fig:stockGARCH}
\end{figure}

\begin{figure}[H]
\tcapfig{Expected Shortfall of Bond Returns over Time}
\centering
\begin{subfigure}[b]{\textwidth}
    \includegraphics[width=\linewidth]{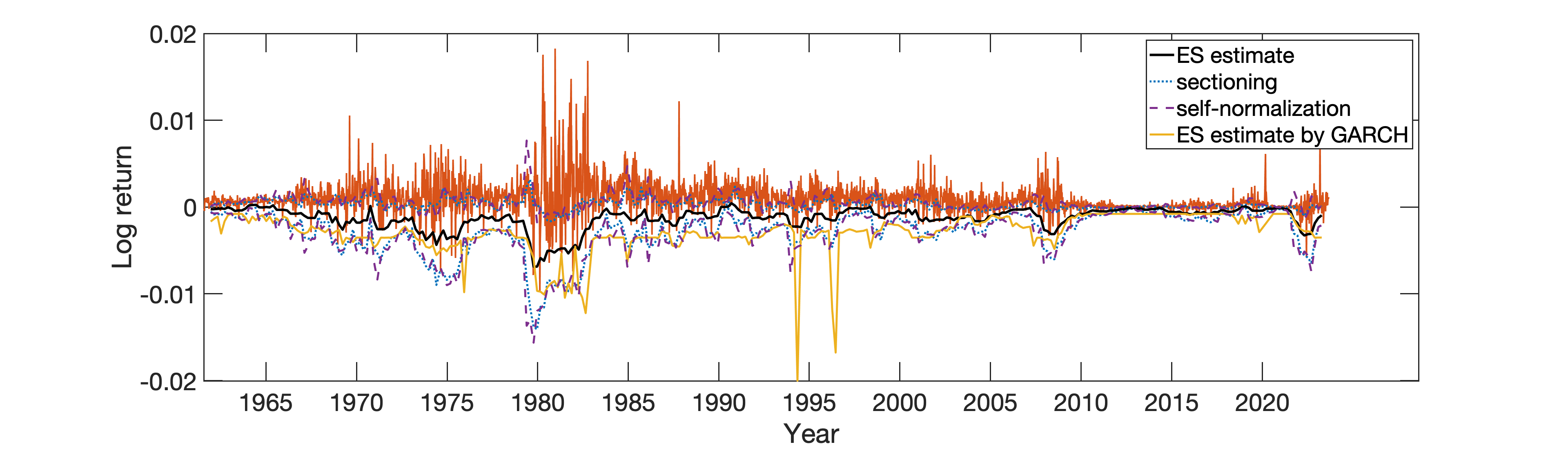}
    \caption{U.S. 1-year zero-coupon bond}
\end{subfigure}
\begin{subfigure}[b]{\textwidth}
    \includegraphics[width=\linewidth]{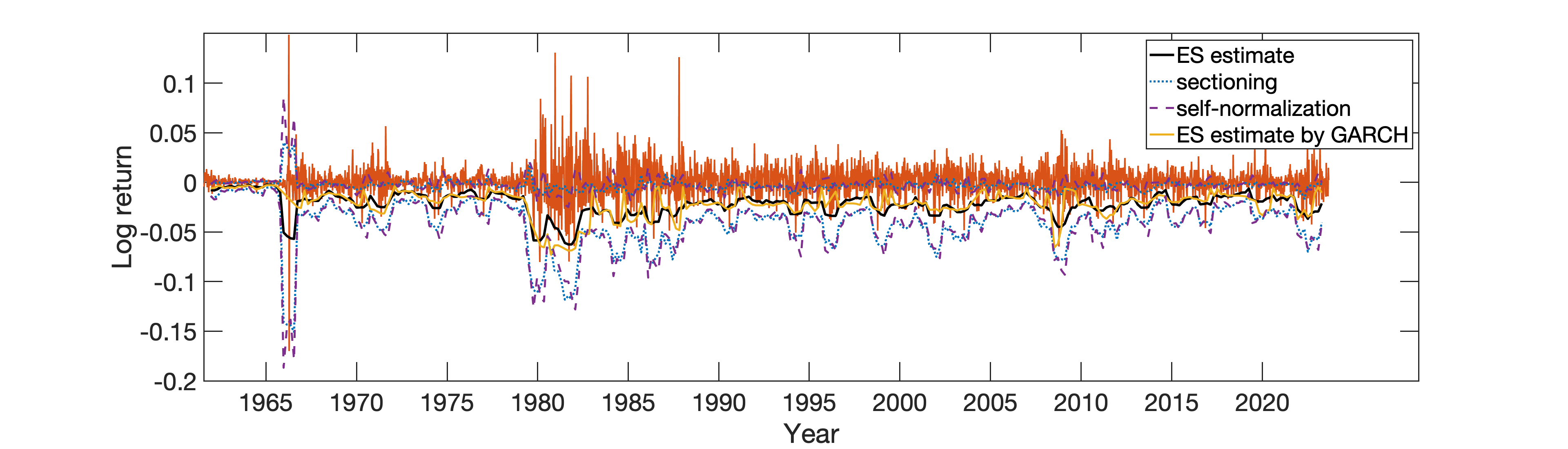}
    \caption{U.S. 10-year zero-coupon bond}
\end{subfigure}
\bnotefig{This figure shows the weekly log returns and expected shortfall with confidence intervals for long- and short-maturity zero-coupon bonds. We plot the ES estimate and 95\% confidence bands for the lower 10th percentile of log returns. ES is computed using a rolling window of 40 weeks with 10 week shifts.}
\label{fig:bondGARCH}
\end{figure}

\begin{table}[H]
\tcapfig{B-Y Multiple Testing Adjustments for Change-Point Tests for Bond Returns}
\label{tab:empirical_change_points_bond_benjamini_yekutieli}
\centering
\begin{tabular}{ccccc}
\toprule
\multicolumn{1}{c}{} 
& \multicolumn{2}{c}{single CP test} & \multicolumn{2}{c}{multiple CP test} \\
\cmidrule(rl){2-5}
time window & statistic & p-value & statistic & p-value \\
\midrule
    \multicolumn{5}{c}{\textbf{US 1-year bond weekly log returns}}\\
    \midrule
    03/21/1976 - 03/08/1981 &300.3 & \hspace{-2.8mm}$<$0.001 &380.7 & \hspace{-2.8mm}$<$0.001 \\
    09/16/1979 - 08/27/1989 &264.3 & \hspace{-2.8mm}$<$0.001 &451.5 & \hspace{-2.8mm}$<$0.001 \\ 
    \cdashline{1-5} 
    09/14/1980 - 09/01/1985 &342.3 & \hspace{-2.8mm}$<$0.001 &310.3 &0.001 \\ 
    10/29/2017 - 10/16/2022 &248.9 & \hspace{-2.8mm}$<$0.001 &308.7 &0.001 \\ 
    \cdashline{1-5}    
    09/27/1970 - 09/07/1980 &18.3 &0.210 &275.6 &0.001 \\
    09/21/1975 - 09/01/1985 &2.3 &0.991 &481.7 & \hspace{-2.8mm}$<$0.001 \\
    03/19/1978 - 02/28/1988 &30.9 &0.090 &285.8 &0.001 \\
    09/14/1980 - 08/26/1990 &146.2 &0.001 &373.0 & \hspace{-2.8mm}$<$0.001 \\
    08/12/2007 - 07/23/2017 &127.6 &0.001 &341.7 & \hspace{-2.8mm}$<$0.001 \\
    11/04/2012 - 10/16/2022 &72.7 &0.011 &273.3 &0.001 \\
    04/29/2018 - 04/16/2023 &98.9 &0.003 &375.6 & \hspace{-2.8mm}$<$0.001 \\
    \midrule
    \multicolumn{5}{c}{\textbf{US 10-year bond weekly log returns}} \\
    \midrule
    06/25/1972 - 06/06/1982 &502.6 & \hspace{-2.8mm}$<$0.001 &825.6 & \hspace{-2.8mm}$<$0.001\\
    06/20/1976 - 06/07/1981 &281.0 & \hspace{-2.8mm}$<$0.001 &361.9 & \hspace{-2.8mm}$<$0.001\\
    03/11/1984 - 02/20/1994 &199.2 &0.001 &321.8 & \hspace{-2.8mm}$<$0.001\\
    \cdashline{1-5}
    07/09/1961 - 06/26/1966 &260.9 & \hspace{-2.8mm}$<$0.001 &142.4 &0.043\\
    \cdashline{1-5}
    03/28/1971 - 03/08/1981 &70.5 &0.012 &295.2 &0.001\\
    06/24/1973 - 06/05/1983 &104.9 &0.002 &422.0 & \hspace{-2.8mm}$<$0.001\\
    06/19/1977 - 05/31/1987 &6.2 &0.581 &270.1 &0.001\\
    06/12/1983 - 05/23/1993 &222.5 &0.001 &354.1 & \hspace{-2.8mm}$<$0.001\\
\bottomrule
\end{tabular}
\bnotefig{
%This table shows the results of the B-Y multiple hypothesis testing procedure applied (separately) to the single and multiple change-point tests, for expected shortfall (ES) in the lower 10th percentile for (separately) U.S. 1-year and U.S. 10-year zero-coupon bond weekly log returns, simultaneously over the rolling time windows used in the Figures \ref{fig:empir_b1_5y}-\ref{fig:empir_b10_10y}. 
%(Recall the setup for these windows discussed in Section \ref{empirical_change_point_detection}.) 
In this table, we consider B-Y multiple hypothesis testing adjustments to single and multiple change-point tests for expected shortfall (ES) in the lower 10th percentile for U.S. 1-year and 10-year zero-coupon bond weekly log returns over the rolling windows in Figures \ref{fig:empir_b1_5y}-\ref{fig:empir_b10_10y}.
Separately for each sub-figure of Figures \ref{fig:empir_b1_5y}-\ref{fig:empir_b10_10y}, we apply the B-Y procedure with FDR control level of 0.05 to all windows (each corresponding to a separate change-point test) within that sub-figure.
(Recall that the setup for these windows was discussed in Section \ref{empirical_change_point_detection}. Also, see below for the details of the B-Y procedure.) 
The top blocks report the windows that the B-Y procedure finds significant for both the single and multiple change-point tests. 
The middle blocks report the windows that the B-Y procedure finds significant for the single change-point test, but not for the multiple change-point test.
The lower blocks report the windows that the B-Y procedure finds significant for the multiple change-point test, but not for the single change-point test. 
For each window, we also report the raw change-point test statistics and corresponding p-values.} 
\end{table}

\noindent \textbf{Benjamini-Yekutieli (B-Y) Procedure \citep{benjamini_yekutieli_2001}} \\
The B-Y procedure for FDR control at level $\alpha \in (0,1)$ consists of the following.
We are given a list of p-values $p_1,\dots,p_m$ and corresponding null hypotheses $h_1,\dots,h_m$.
We first order the p-values in ascending order: $p_{(1)},\dots,p_{(m)}$, with corresponding null hypotheses ordered accordingly: $h_{(1)},\dots,h_{(m)}$.
Next, we find the largest $k$ such that $p_{(k)} \le (k \cdot \alpha)/(m \cdot c_m)$, where $c_m = \sum_{j=1}^m j^{-1}$.
Then, at FDR level $\alpha$, we reject the null hypotheses: $h_{(1)},\dots,h_{(k)}$ (declaring them to be discoveries/significant).

\end{document}